\documentclass[12pt]{article}
\pdfoutput=1

\usepackage{amssymb}
\usepackage{amsmath}
\usepackage{bbold}
\usepackage{graphicx}
\usepackage{cite}
\usepackage{hyperref}
\usepackage{color}

\def\appendix{\par
  \setcounter{section}{0}
  \setcounter{subsection}{0}
  \setcounter{equation}{0}
  \def\thesection{\Alph{section}}}

\expandafter\ifx\csname mathrm\endcsname\relax\def\mathrm#1{{\rm #1}}\fi

\makeatletter

\def\de{\delta}

\def\eps{\epsilon}

\def\refeq#1{\mbox{(\ref{#1})}}

\def\refta#1{\mbox{Table~\ref{#1}}}

\def\refse#1{\mbox{Section~\ref{#1}}}

\def\citeres#1{\mbox{Refs.~\cite{#1}}}

\newcommand{\nn}{\nonumber}

\newcommand{\GeV}{\unskip\,\mathrm{GeV}}

\newcommand{\ri}{{\mathrm{i}}}
\newcommand{\re}{{\mathrm{e}}}
\newcommand{\rd}{{\mathrm{d}}}

\newcommand{\rR}{{\mathrm{R}}}

\newcommand{\rT}{{\mathrm{T}}}

\renewcommand{\L}{\mathcal{L}}

\def\mathswitchr#1{\relax\ifmmode{\mathrm{#1}}\else$\mathrm{#1}$\fi}

\newcommand{\PW}{\mathswitchr W}
\newcommand{\Pw}{\mathswitchr w}
\newcommand{\PZ}{\mathswitchr Z}

\newcommand{\Ph}{\mathswitchr h}
\newcommand{\PH}{\mathswitchr H}

\def\mathswitch#1{\relax\ifmmode#1\else$#1$\fi}

\newcommand{\MW}{\mathswitch {M_\PW}}

\newcommand{\MZ}{\mathswitch {M_\PZ}}
\newcommand{\MH}{\mathswitch {M_\PH}}
\newcommand{\Mh}{\mathswitch {M_\Ph}}

\newcommand{\sw}{\mathswitch {s_\Pw}}
\newcommand{\cw}{\mathswitch {c_\Pw}}

\def\Re{\mathop{\mathrm{Re}}\nolimits}

\newcommand{\MSbar}{{\overline{\mathrm{MS}}}}
\newcommand{\UV}{{\mathrm{UV}}}

\newcommand{\PRTS}{{\mathrm{PRTS}}}
\newcommand{\FJTS}{{\mathrm{FJTS}}}
\newcommand{\ct}{\mathrm{ct}}
\newcommand{\Hct}{\mathrm{Hct}}
\newcommand{\eff}{\mathrm{eff}}
\newcommand{\tree}{\mathrm{tree}}
\newcommand{\SM}{\mathrm{SM}}
\newcommand{\OS}{\mathrm{OS}}
\newcommand{\SESM}{\mathrm{SESM}}
\newcommand{\BSM}{{\mathrm{BSM}}}

\renewcommand{\max}{\mathrm{max}}
\newcommand{\z}{\setbox0\hbox{+}\hbox to \wd0{\hss0\hss}}
\def\limfunc#1{\mathop{\rm #1}}

\def\tr{\limfunc{tr}}
\def\Re{\limfunc{Re}}

\def\slash#1{{\setbox0=\hbox{$#1$}
  \rlap{\ifdim\wd0>.7em\kern.22\wd0\else\kern.1\wd0\fi /}#1}}

\def\braket#1#2{\left\langle #1\vphantom{#2}
  \right. \kern-2.5pt\left| #2\vphantom{#1}\right\rangle }
\def\L{{\cal L}}

\def\oneloop{\text{1-loop}}

\def\rT{{\mathrm{T}}}

\def\rR{{\mathrm{R}}}
\def\rM{{\mathrm{M}}}

\def\barW{\overline{W}}

\newcommand{\eq}[1]{Eq.~\eqref{eq:#1}}
\newcommand{\eqs}[2]{Eqs.~\eqref{eq:#1} and \eqref{eq:#2}}
\newcommand{\eqss}[3]{Eqs.~\eqref{eq:#1}, \eqref{eq:#2}, and \eqref{eq:#3}}
\newcommand{\eqsn}[3]{Eqs.~\eqref{eq:#1}, \eqref{eq:#2}, \eqref{eq:#3}}
\newcommand{\eqsm}[2]{Eqs.~\eqref{eq:#1}\,--\,\eqref{eq:#2}}
\renewcommand{\sec}[1]{Sec.~\ref{se:#1}}

\newcommand{\app}[1]{App.~\ref{app:#1}}

\newcommand{\rcites}[1]{Refs.~\cite{#1}}
\newcommand{\rcite}[1]{Ref.~\cite{#1}}

\newcommand{\ord}{\mathcal{O}}

\DeclareMathAlphabet{\mathbbold}{U}{bbold}{m}{n}
\newcommand{\bbid}{\mathbbold{1}}

\DeclareFontFamily{OT1}{pzc}{}
\DeclareFontShape{OT1}{pzc}{m}{it}{<-> s * [1.20] pzcmi7t}{}
\DeclareMathAlphabet{\mathpzc}{OT1}{pzc}{m}{it}

\allowdisplaybreaks[2]

\begin{document}

  \thispagestyle{empty}
  \def\thefootnote{\fnsymbol{footnote}}
  \setcounter{footnote}{1}
  \null
  \strut\hfill FR-PHENO-2020-010
  \vskip 0cm
  \vfill
  \begin{center}
    {\Large \bf
      \boldmath{Integrating out heavy fields in the path integral
        using the background-field method: general formalism}
      \par} \vskip 2.5em
    {\large
      {\sc Stefan Dittmaier, Sebastian Schuhmacher \\ and Maximilian Stahlhofen}\\[1ex]
      {\normalsize
        \it
        Albert-Ludwigs-Universit\"at Freiburg,
        Physikalisches Institut, \\
        Hermann-Herder-Stra\ss{}e 3,
        D-79104 Freiburg, Germany
      }
    }

    \par \vskip 1em
  \end{center} \par
  \vskip 2cm
\noindent{\bf Abstract:} \\
\par
Building on an older method used to derive non-decoupling effects
of a heavy Higgs boson in the Standard Model, we describe a general
procedure to integrate out heavy fields in the path integral.
The derivation of the corresponding effective Lagrangian including
the one-loop contributions of the heavy particle(s) is particularly
transparent, flexible, and algorithmic.
The background-field formalism allows for a clear separation of tree-level and one-loop effects involving the heavy fields.
Using expansion by regions the one-loop effects are further split into contributions from large and small momentum modes.
The former are contained in Wilson coefficients of effective operators, the latter are reproduced by one-loop diagrams involving effective tree-level couplings.
The method is illustrated by calculating potential non-decoupling effects
of a heavy Higgs boson in a singlet Higgs extension of the Standard Model.
In particular, we work in a field basis corresponding to mass eigenstates and properly take
into account non-vanishing mixing between the two Higgs fields of the model.
We also show that a proper choice of renormalization
scheme for the non-standard sector of the underlying full theory is crucial
for the construction of a consistent effective field theory.
  \par
  \vfill
  \noindent
  \vskip .5cm
  \null
  \setcounter{page}{0}
  \clearpage
  \def\thefootnote{\arabic{footnote}}
  \setcounter{footnote}{0}

  \tableofcontents

\section{Introduction}

After roughly a decade of operation, the Large Hadron Collider (LHC) at CERN
has confirmed the validity of the Standard Model (SM) of particle physics
generically up to energies in the TeV range, without any significant and
convincing deviation from SM predictions.
On the other hand, we know that the SM is incomplete, because it does not
include neutrino masses nor
explain phenomena like Dark Matter or the matter--antimatter asymmetry
in the universe.
To fully exploit the potential of the LHC on its mission to identify the limitations
of the SM and to unravel the structure of potential deviations of experimental
results from SM predictions, a strategy is required that is as model independent
as possible and can be pushed to sufficiently high precision.
Of course, precise SM predictions are the major prerequisite in this task.
However, in order to establish at which accuracy the various sectors of the SM
are tested or to shape observed deviations,
it is necessary to include non-standard effects in analyses.
Besides dedicated analyses in specific models for physics beyond the SM (BSM),
it is desirable to provide, as far as possible, model-independent analyses that
quantify the compatibility of data with the SM before confronting the results
with specific models.

Standard Model Effective Theory (SMEFT)~\cite{Buchmuller:1985jz,Grzadkowski:2010es}
(see also \rcites{Heinemeyer:2013tqa,deFlorian:2016spz,Brivio:2017vri,David:2020pzt}
and references therein)
is such an approach, in which it is assumed
that the SM is the valid theory up to an energy scale $\Lambda$ much larger than
the electroweak (EW) scale $v\approx246\GeV$ and
that new particles have masses of the order
of $\Lambda$. Under this assumption the leading BSM effects are generically suppressed by
powers of $\Lambda$ and can be parametrized in terms of Wilson coefficients
of (local) dimension-5 and dimension-6 effective operators which are added to the SM Lagrangian.
Although there is still a long way towards fully global fits of these Wilson coefficients
to data, larger and larger subsets of operators are being considered (for up-to-date analyses see e.g.\
\rcites{Dawson:2020oco,Ellis:2020unq}), and
SMEFT predictions are being dressed with QCD and EW corrections (for recent calculations see e.g.\ \rcites{Dedes:2019bew,Cullen:2019nnr,Dawson:2019clf,Baglio:2019uty,Degrande:2020evl}).
So far the results of these fits are such that all
Wilson coefficients are still compatible with zero.
Once some coefficients show significant deviations from zero, the question arises
which BSM effects and which new particles cause them.

To answer this question, one evaluates BSM models
at energies well below the mass scale $\Lambda$
of the non-standard particles.
In the effective field theory (EFT) describing this limit the heavy particles are {\it integrated out}, i.e.\ their fields are no longer dynamical degrees of freedom, so that the
particle content is the same as in the SM.
The effects of the BSM particles are reproduced by non-vanishing Wilson coefficients of the higher-dimensional effective operators mentioned above.
There are roughly speaking two types of approaches to compute the relevant Wilson coefficients:
\begin{itemize}
\item For the \textit{diagrammatic matching} a sufficiently large set of Green functions is evaluated both in the underlying BSM model (full theory) and the EFT in terms of Feynman diagrams to a given order in the perturbative (loop) expansion. On the EFT side this requires to first construct a generic basis of operators
(composed of SM fields and respecting the symmetries of the BSM model) to the order of interest in the EFT ($1/\Lambda$) expansion. The Wilson coefficients are then fixed by demanding that corresponding EFT and full-theory amplitudes  match
up to higher orders in $1/\Lambda$.

\item \textit{Functional matching} is based on the path integral defining the
generating functional for Green functions of the full theory. The functional integration over heavy field modes related to the BSM effects is performed  and directly results in a $1/\Lambda$ expanded effective action for the
low-energy (SM) degrees of freedom representing the EFT. No input
on the structure of the EFT operators is required.%
\footnote{In this respect the term ``matching" is actually misleading, because the Wilson coefficients of a generic EFT Lagrangian are not determined by matching EFT and full theory predictions.}
At one-loop order the relevant functional integrals are of Gaussian type and therefore straightforward to carry out. Beyond one loop the feasibility of this method seems unclear.
\end{itemize}

In this paper we describe a generic functional approach to integrate out heavy particles, which is a further development of the method introduced in
\rcites{Dittmaier:1995cr,Dittmaier:1995ee}.
Functional methods for EFT matching have a long history, see e.g.\ \rcites{Gaillard:1985uh,Chan:1986jq,Cheyette:1987qz} for early works.
Following this way, the non-decoupling effects of a heavy SM Higgs boson were computed in \rcites{Dittmaier:1995cr,Dittmaier:1995ee}.
In particular, it was demonstrated there that using
the background-field method (BFM)~\cite{DeWitt:1967ub,DeWitt:1980jv,tHooft:1975uxh,Boulware:1980av,Abbott:1980hw,Denner:1994xt,Denner:2019vbn}
leads to a transparent separation of tree-level and one-loop contributions
in the functional derivation of the effective Lagrangian.
The method of \rcites{Dittmaier:1995cr,Dittmaier:1995ee} was further refined and generalized in \rcite{Fuentes-Martin:2016uol} by employing the expansion by regions~\cite{Beneke:1997zp,Smirnov:2002pj} (see also \rcite{Zhang:2016pja} for a concise review).

In recent years the interest in functional matching has been revived in the context of SMEFT  by \rcite{Henning:2014wua}.
This work initiated the ``Universal One-Loop Effective Action" (UOLEA) program~\cite{Drozd:2015rsp,Henning:2016lyp,Ellis:2016enq,Ellis:2017jns,Kramer:2019fwz,Angelescu:2020yzf,Ellis:2020ivx}
which ultimately aims at deriving a master formula for the one-loop matching of a fully generic BSM model to SMEFT using functional methods.
The basic idea is that the matching in principle only has to be performed once and for all. The relevant Wilson coefficients of the SMEFT operators could then be determined for any specific BSM model via the master formula by replacing the generic with the specific full theory parameters.
So far, however, this goal has not been reached.
While the UOLEA is not yet available for the most general case, for example because couplings of the heavy fields (to be integrated out) involving derivatives are not accounted for, its complexity already suggests that it will be limited to SMEFT operators with dimension $\le6$ in the foreseeable future.
For a review on the current status of the UOLEA program see \rcite{Ellis:2020ivx}.
In order to overcome some of the limitations of the UOLEA approach, there is a
trend towards automation of the matching procedure~\cite{Criado:2017khh,Cohen:2020fcu,Cohen:2020qvb,Fuentes-Martin:2020udw}. Due to its algorithmic nature,
functional matching turns out to be well suited for this purpose, especially at the one-loop level.

In the UOLEA and automation literature quoted above the EFT expansion is essentially based on a power counting of dimensionful quantities, i.e.\ masses and loop momenta of $\ord(\Lambda)$.
However, as noticed already in \rcite{Boggia:2016asg}, some realistic BSM models feature dimensionless parameters like couplings or mixing angles with definite $\Lambda$ scaling, i.e.\ they must be counted as powers of $v/\Lambda$, where $v$ represents a typical SM scale.
For example, the mixing of a BSM-type with as SM-type field to form a heavy and a light mass eigenstate, as it appears in the context of spontaneous symmetry breaking,
requires a mixing angle that is suppressed by powers of $1/\Lambda$
if the heavy field is supposed to decouple.
We will explicitly address this aspect in the present paper using a functional method based on \rcites{Dittmaier:1995cr,Dittmaier:1995ee,Fuentes-Martin:2016uol}.
We will also show that combining background-field gauge invariance with
a non-linear representation of the SM Higgs doublet enables further technical
simplifications, because intermediate manipulations can be
carried out in the unitary gauge, while full gauge invariance is restored
at the end of the calculation.
In this respect we generalize the  matching procedure of \rcites{Dittmaier:1995cr,Dittmaier:1995ee}, where light modes of the heavy field did not contribute in loops.
To account for such contributions we perform the large-mass ($\sim \Lambda$) expansion according to the method of regions~\cite{Beneke:1997zp,Smirnov:2002pj}, which separates heavy and light modes in loop integrals as also proposed in \rcite{Fuentes-Martin:2016uol}.
The loop effects of the heavy modes are encoded in the Wilson coefficients of the
effective Lagrangian, while the loop effects of the light modes
result from insertions of tree-level effective couplings in EFT loop diagrams.
The whole procedure is fully algorithmic and flexible in the sense that
the underlying low-energy need not be specified in advance, i.e.\
the method is also applicable beyond the framework of SMEFT, which assumes
the SM as the leading-order (LO) low-energy theory.

In this article we describe the general framework of our functional matching method and apply it to integrate out a non-standard Higgs boson with large mass $\MH\gg v$ in a Singlet (Higgs) Extension of the SM (SESM), which is defined in different variants in \rcites{Schabinger:2005ei,Patt:2006fw,Bowen:2007ia,Pruna:2013bma,Kanemura:2015fra,Bojarski:2015kra,Altenkamp:2018bcs,Denner:2018opp}.
To keep the presentation transparent, we restrict the calculation here
to the level of non-decoupling effects in the bosonic sector, i.e.\ to
terms of $\ord(\MH^0)$ in the effective Lagrangian, which are non-trivial in the presence of Higgs mixing.
We will deal with the decoupling effects at $\ord(\MH^{-2})$ in a follow-up paper.
A main focus of the present paper will be the
issue of renormalization of the BSM sector of the underlying full theory and its consequences for the EFT.
We will explain how the choice of renormalization and tadpole schemes affects the derivation of the effective Lagrangian, already at the $\ord(\MH^0)$ level.
This aspect, which has mostly been ignored in the existing literature, generally arises
in renormalization schemes where the loop contribution to a renormalization constant
of a BSM parameter and the parameter itself scale differently in the
large-mass limit.
In models with extended Higgs sectors, such effects potentially occur in the
interplay of tadpole renormalization and $\MSbar$ renormalization conditions
(see e.g.\ \rcites{Fleischer:1980ub,Kanemura:2015fra,Krause:2016oke,Denner:2016etu,Denner:2017vms,Altenkamp:2018bcs,Denner:2018opp}).

The low-energy limit of
different SESM variants has been studied repeatedly in the past, see e.g.\ \rcites{Henning:2014wua,deBlas:2014mba,Gorbahn:2015gxa,Chiang:2015ura,Brehmer:2015rna,Egana-Ugrinovic:2015vgy,Boggia:2016asg,Buchalla:2016bse,Ellis:2017jns,Jiang:2018pbd,Haisch:2020ahr,Cohen:2020fcu,Dawson:2021jcl}.%
\footnote{See \rcites{Bilenky:1993bt,delAguila:2016zcb,Zhang:2016pja} for the matching of a SM extension with a charged singlet scalar onto SMEFT.}
In fact, it has become a kind of
test model for different matching techniques as well as to analyze the EFT validity.
In the following we give a brief overview of the most elaborate literature on SESM to SMEFT matching at $\ord(1/\Lambda^2)$ and one-loop level.
We focus on matching calculations that take into account contributions to the Wilson coefficients from both types of loops:
loops that only involve heavy (virtual) particles as well as mixed heavy--light particle loops. The latter were omitted in earlier publications (cf.\ \rcites{Henning:2014wua,Drozd:2015rsp}).
In \rcite{Ellis:2017jns} the matching was performed using functional (UOLEA) methods. While contributions from loops involving fermions were still neglected in \rcite{Ellis:2017jns}, they were included later by a calculation based on Feynman diagrams~\cite{Jiang:2018pbd}.
The results were confirmed by purely diagrammatic matching in \rcite{Haisch:2020ahr} and finally reproduced with the partly automated functional procedure of \rcites{Cohen:2020fcu,Cohen:2020qvb}.
For a recent fit of experimental data to the effective Lagrangian of \rcites{Jiang:2018pbd,Haisch:2020ahr,Cohen:2020fcu} see \rcite{Dawson:2021jcl}.
The two different functional approaches~\cite{Ellis:2017jns,Cohen:2020fcu}
both make use of the BFM in combination with the expansion by regions in order to streamline and simplify the calculations following
 \rcites{Dittmaier:1995cr,Dittmaier:1995ee,Fuentes-Martin:2016uol} (and so does our method).
On the other hand, none of the quoted one-loop matching references~\cite{Ellis:2017jns,Jiang:2018pbd,Haisch:2020ahr,Cohen:2020fcu}
works in a field basis corresponding to mass eigenstates, which is the safest way to consistently
take into account the possibility of
mixing between the SM-type Higgs doublet and the additional scalar field in the (broken phase of the) SESM.
This issue was also addressed
in \rcite{Boggia:2016asg}, albeit using an old-fashioned functional method without the virtues of the BFM and the expansion by regions.
Complications related to SESM renormalization and the treatment of tadpoles in the presence of mixing were avoided there
by choosing a specific on-shell renormalization and tadpole scheme,
while most renormalization procedures in BSM sectors
involve $\MSbar$ conditions to some extent.
In the present article we explore the subtleties arising in different standard renormalization (e.g.\ $\MSbar$) and tadpole schemes.

Our paper is organized as follows:
In \sec{outline} we outline the salient steps and ingredients of the
method and highlight the new features added in this paper.
Section~\ref{se:SESM} describes the Singlet Higgs Extension of the SM
used as test model, the relevant large-mass/low-energy scenario, the formulation of the
model within the BFM, and the non-linear realization of the Higgs sector.
In \sec{largemass} we elaborate on the individual steps of the calculation of the
effective Lagrangian: the separation of heavy and light field modes,
the solution of the functional integral over the heavy quantum field, and the
elimination of the light modes of the heavy Higgs field via its equation of motion.
In \sec{ren} we discuss the renormalization of
the full and the effective theory in detail.
Our conclusions are given in \sec{conclusion}, and \app{FuncDetApp} provides further (pedagogical) details about the functional integration.

\section{Outline of the general method}
\label{se:outline}

The method described in the following is a further development of the method
introduced in \rcites{Dittmaier:1995cr,Dittmaier:1995ee}, where a heavy Higgs
field was integrated out in an SU(2) gauge theory and the SM, respectively,
directly in the path integral.
As already mentioned,
some of the generalizations presented here have already been proposed in
\rcite{Fuentes-Martin:2016uol} (see also \rcites{Zhang:2016pja,Cohen:2020fcu}).
Unlike for several other approaches in the literature, no matching of free parameters
between an ansatz for the effective Lagrangian and explicitly calculated Green functions or
amplitudes is involved.
Furthermore, the use of the BFM yields additional benefits.
Particular strengths of the method are:
\begin{enumerate}
\renewcommand{\labelenumi}{\theenumi}
\renewcommand{\theenumi}{(\roman{enumi})}
\item
a clear separation of tree-level and loop effects of the heavy fields;
\item
the possibility to fix the (background) gauge in intermediate steps of the
calculation and to restore gauge invariance of the effective Lagrangian at the end;
\item
transparency in the sense that at each stage of the calculation it is possible to
identify the origin of all contributions to the effective Lagrangian in terms of
Feynman diagrams;
\item
flexibility due to the fact that no ansatz is made for the effective Lagrangian.
Actually not even the low-energy theory has to be specified in advance, it directly
emerges as part of the result;
\item
An automation of the method is possible, since it is fully algorithmic. In principle,
given a Lagrangian, a large-mass scenario with a corresponding power-counting scheme,
and some details on the renormalization
of the large-mass sector, the actual determination of the effective Lagrangian
at the one-loop level can be carried out by computer algebra.
\end{enumerate}

Since the individual steps in the whole procedure are quite non-trivial
and involve various tricks, we first sketch the different steps and ingredients
before applying the method to a concrete example in the subsequent sections.
This preparatory section will also motivate the splitting of a generic heavy particle field~$H$
into four conceptually different parts,
$H\to \hat H_h+\hat H_l+H_h+H_l$, which is at the heart of the proposed method.
In the course of this brief outline we also explain which generalizations and
optimizations have been made in
\rcite{Fuentes-Martin:2016uol} and are made in this paper
with respect to the original approach
of
\rcites{Dittmaier:1995cr,Dittmaier:1995ee}:
\begin{enumerate}
\item
\textit{Background-field formalism and
non-linear Higgs realization.}
\\
Formulating the theory within the BFM
splits all fields into
background (i.e.\ in some sense semi-classical) and quantum parts.
For a generic heavy field~$H$, this separation reads $H\to \tilde H = \hat H+H$, with $\hat H$ being the background and $H$ the
quantum field.
Diagrammatically this step distinguishes between fields
occurring on tree and loop lines in Feynman graphs.
For tree-level effects, quantum fields are not relevant.
For one-loop corrections, only terms in the Lagrangian
that are bilinear in quantum fields are relevant.
Higher powers of quantum fields only contribute beyond the one-loop level.
Thus, this step determines the terms in the full-theory
Lagrangian that are needed in the subsequent derivation of the EFT Lagrangian.
\\
Employing a non-linear representation
of the scalar sector, it is possible to absorb
all background Goldstone-boson fields into the background
gauge fields by a straightforward
Stueckelberg transformation~\cite{Stueckelberg:1938zz,Stueckelberg:1957zz,Kunimasa:1967zza,Lee:1972yfa},
which reduces the algebraic amount of work
in the subsequent steps considerably.
At the same time this framework remains appropriate also for cases in which heavy Higgs
bosons may not decouple completely.
\item
\textit{Separation of hard and soft field modes.}
\\
Considering all fields consistently in momentum space, it
is possible to additively split the quantum parts~$H$ of the
heavy fields into field modes with small or large
momenta, which we dub ``light (soft) modes"~$H_l$ and
``heavy (hard) modes"~$H_h$,
respectively, i.e.\ $H=H_l+H_h$.
Diagrammatically this splitting expresses the
large-mass expansion of Feynman graphs using the
method of regions~\cite{Beneke:1997zp,Smirnov:2002pj}
in the framework of dimensional
regularization.
Each one-loop diagram
with at least one internal heavy particle
line is decomposed into a part with small and a part with large
loop momentum $q$ (carried by $H_l$ and $H_h$, respectively).
In the large-mass expansion the former contribution
arises from a Taylor expansion of the loop integrand in $q/M,p_i/M,m_i/M \to 0$,
where $\{p_i\}$ are the external momenta, $m_i$ the small masses in the theory,
and $M$ represents the heavy particle mass.
In the region of large loop momenta one expands the integrand only
in $p_i/M,m_i/M \to 0$
(but not in $q/M$)
and is thus left with vacuum-type integrals.
In the EFT the small-momentum regions are reproduced by loop diagrams
with insertions of (higher-dimensional) effective operators,
while the large-momentum contributions are contained in the loop
corrections to the Wilson coefficients of these operators.\\
At one loop, the splitting of loop diagrams into two integration
domains of small and large momenta can be interpreted as a splitting
of the path integral into two functional integrals extending over
light and heavy field modes.
The consistent mode separation according to the method of regions
is a conceptual generalization of the procedure of
\rcites{Dittmaier:1995cr,Dittmaier:1995ee},
where only heavy
field modes appeared in the calculation of the
non-decoupling effects in the leading-order Lagrangian at one loop.
This mode separation has also been suggested in
\rcite{Fuentes-Martin:2016uol} (and applied in  \rcites{Zhang:2016pja,Cohen:2020fcu}) within
 a procedure to calculate the
different parts in the effective Lagrangian, but we consider our
formulation in terms of heavy and light modes of background and quantum fields
and their different treatments in the path integral
conceptually more transparent.

\item
\textit{Integrating out the hard modes of the heavy quantum fields in the path integral.}
\\
Since the part of the Lagrangian that is relevant at one-loop order
is only quadratic in the quantum fields, the path integral over the
heavy field modes~$H_h$ of the
heavy quantum field is of Gaussian type and can be done
analytically. The major complication in this step is the fact that there are
also terms that are linear in the heavy quantum fields~$H_h$.
As we show below, these terms can be removed by a field redefinition of the hard quantum field modes
(of the light particles)
in a fully algorithmic manner.
This means that the resulting part of the Lagrangian quadratic in the heavy quantum field can be directly identified based again on a simple power-counting argument.
This algorithmic handling, which has also been realized in \rcites{Fuentes-Martin:2016uol} (see also \rcite{Zhang:2016pja} and, for a slightly different approach, \rcite{Cohen:2020fcu}),
establishes an important technical improvement
over the procedure described in \rcites{Dittmaier:1995cr,Dittmaier:1995ee},
where the ``diagonalization'' of the Lagrangian was performed via a non-trivial series of individual field shifts.
\\
The result of the straightforward (Gaussian) path integration is a
functional determinant that is expanded for $M\to\infty$.
The terms emerging from this expansion are exactly the
vacuum-type integrals from the large-momentum regions in the
large-mass expansion of the Feynman graphs described above
and produce the one-loop contributions to the
Wilson coefficients of the local effective operators.

\item
\textit{Equations of motion for the soft modes of the
heavy fields and renormalization.}
\\
After the heavy modes~$H_h$
of the heavy field have been integrated out, the effective
Lagrangian still involves the light modes $H_l$ and $\hat H_l$
of the quantum and background
fields of the heavy particle.
As their momenta are much smaller than their mass~$M$,
they do not represent
dynamical degrees of freedom of the EFT.
In fact, they can conveniently be removed from the effective Lagrangian by
applying their equations of motions (EOMs) in the large-mass
expansion.
This procedure can be viewed as a saddle-point
approximation in the path integral over the light modes of the heavy quantum field combined with a large-mass expansion.
It expresses the light modes of the heavy field in terms of all other light fields.
The actual effect of these modes is revealed at a later stage in the perturbative evaluation of Green functions when effective tree-level couplings
are inserted into EFT loop diagrams.
\\
Like the previous one, this step requires a proper power-counting of all
parameters and fields in the limit $M\to\infty$.
We emphasize that in order to obtain a consistent effective Lagrangian the large-mass expansion must be carefully performed taking into account that the full-theory renormalization constants
may have a different scaling behaviour for
$M\to\infty$ than the corresponding renormalized quantities.
\\
As mentioned above, in \rcites{Dittmaier:1995cr,Dittmaier:1995ee}
the light mode~$H_l$ of the heavy SM
Higgs field was irrelevant and ignored, i.e.\
the insertion of effective tree-level vertices into loops did not
occur at the considered order in the heavy-mass expansion.

\item
\textit{Final form of the effective Lagrangian.}
\\
The effective Lagrangian resulting from the previous steps
only involves light background and quantum fields,
but none of the modes of the heavy fields~$H$.
The Lagrangian consists of four different types of contributions:
\begin{enumerate}
\renewcommand{\labelenumii}{\theenumii}
\renewcommand{\theenumii}{(\roman{enumii})}
\item
a tree-level part that depends only on light background fields;
\item a tree-level part that depends both on light background and
quantum fields;
\item a part involving renormalization constants and
light background fields;
\item a part involving the one-loop corrections (from heavy loops) to the effective operators built from light background fields.
\end{enumerate}
Parts~(i) and (ii) combine to a single effective
Lagrangian at lowest order in the coupling constants, which can be used to evaluate tree-level amplitudes at different orders in the large-mass
expansion and one-loop contributions resulting from insertions of effective
vertices in loop diagrams (reproducing the soft momentum regions of loops in the full theory).
Parts~(iii) and (iv) combine to the one-loop correction to the
effective Lagrangian, i.e.\ all NLO contributions to the
Wilson coefficients of the effective operators (reproducing the hard momentum regions of
loops in the full theory).\\
To obtain a more transparent and compact form of the final effective
Lagrangian two further steps are useful. Firstly, the
EOMs of the light fields might be used to eliminate
redundant effective operators that only influence
off-shell Green functions, but no physical scattering amplitudes. This step
is, in particular, necessary to bring the effective Lagrangian into standard SMEFT form.
Secondly, corrections to operators already present
in the
(light-particle sector of the) underlying full theory can be eliminated by absorbing their effect into renormalization constants of the low-energy
theory as far as possible. In a decoupling scenario this means
that the final effective Lagrangian differs from the SM only by operators
with dimensions higher than four.
\end{enumerate}

\section{Heavy Higgs boson in a Higgs singlet extension of the Standard Model}
\label{se:SESM}

\subsection{The singlet Higgs extension}

For the formal description of the singlet Higgs extension of the SM (SESM),
which was formulated in slightly different
versions in \rcites{Schabinger:2005ei,Patt:2006fw,Bowen:2007ia,Pruna:2013bma,Kanemura:2015fra,Bojarski:2015kra,Altenkamp:2018bcs,Denner:2018opp},
we follow the notation and conventions of
\rcites{Altenkamp:2018bcs,Denner:2018opp} and employ a matrix-valued non-linear representation
of the Higgs doublet as defined in \rcites{Dittmaier:1995cr,Dittmaier:1995ee},
\begin{align}
\Phi=\frac{1}{\sqrt{2}}\left(v_{2}+h_{2}\right)U, \qquad
U=\exp\left(2\ri\frac{\varphi}{v_{2}}\right), \qquad
\varphi=\frac{1}{2}\varphi_a\tau_a,
\label{eq:doublet}
\end{align}
where $ \tau_a $ are the Pauli matrices
and the usual convention for the summation over repeated indices is used
throughout the paper.
Here, $h_{2}$ denotes the field of the physical Higgs boson and $v_2$ the corresponding
vacuum expectation value (vev).
The real Goldstone fields $ \varphi_a $ are related
to their counterparts ($\phi^\pm$, $\chi$)
in the linear representation (as used in
\rcites{Denner:1991kt,Denner:1994xt,Denner:2019vbn}) by
\begin{equation}
\phi^\pm=\frac{1}{\sqrt{2}}\left(\varphi_2\pm\ri\varphi_1\right),\qquad\chi=-\varphi_3.
\label{eq:Goldstonefields}
\end{equation}
The covariant derivative of $\Phi$
(and analogously of $U$) reads
\begin{equation}
D_{\mu}\Phi=\partial_{\mu}\Phi-\ri g_{2}W_{\mu}\Phi-\ri g_{1}\Phi\frac{\tau_{3}}{2}B_{\mu},
\end{equation}
with $ W_\mu=W_\mu^a \tau_a/2$ and $g_2$ denoting the SU(2) gauge field and coupling, respectively, and $B^\mu$, $g_{1}$ the U(1) gauge field and coupling.
The conventions in the SM part of the SESM follow
\rcites{Dittmaier:1995ee,Denner:1994xt,Denner:1991kt,Denner:2019vbn}.
The Higgs sector of the SESM Lagrangian is given by
\begin{align}
\L_\mathrm{Higgs}={}&
\frac{1}{2}\tr\left[\left(D_{\mu}\Phi\right)^\dagger\left(D^{\mu}\Phi\right)\right]
+\frac{1}{2}\mu_{2}^2\tr\left[\Phi^\dagger\Phi\right]
-\frac{1}{16}\lambda_{2}\tr\left[\Phi^\dagger\Phi\right]^2
\nn\\
&+\frac{1}{2}\left(\partial_{\mu}\sigma\right)\left(\partial^{\mu}\sigma\right)+\mu_{1}^2\sigma^2-\lambda_{1}\sigma^4-\frac{1}{2}\lambda_{12}\sigma^2\tr\left[\Phi^\dagger\Phi\right],
\end{align}
where a new real scalar field $ \sigma $ is introduced which transforms as a singlet under
the SM gauge groups.
The field $\sigma$ is split into its vev $v_1$ and its
field excitation $h_1$ according to
\begin{align}
\sigma=v_1+h_1.
\label{eq:singlet}
\end{align}
A $ \mathbb{Z}_2 $ symmetry under the transformation $\sigma\rightarrow-\sigma$
is assumed, so that only three new parameters, namely the mass parameter
$\mu_1^2$, the self-coupling parameter $ \lambda_1 $, and the mixed coupling parameter $ \lambda_{12} $ occur.
In analogy to the Higgs sector of the SM the mass parameters fulfill $ \mu_{1,2}^2>0 $, and the coupling parameters are constrained by the vacuum stability conditions
\begin{equation}
\lambda_{1}>0,\qquad\lambda_{2}>0,\qquad\lambda_{1}\lambda_{2}-\lambda_{12}^2>0.
\end{equation}
The Higgs fields $ h,\,H $ corresponding to mass eigenstates are obtained by
a rotation with the mixing angle $\alpha$,
\begin{equation}\label{eq:HfieldRotation}
\begin{pmatrix}H\\h\end{pmatrix} =
\begin{pmatrix}c_\alpha&s_\alpha\\-s_\alpha&c_\alpha\end{pmatrix}
\begin{pmatrix}h_1\\h_2\end{pmatrix},
\end{equation}
where $s_\alpha \equiv \sin(\alpha)$ and $c_\alpha\equiv\cos(\alpha)$.
The Higgs-boson masses expressed in terms of the original parameters are
\begin{equation}
M_\mathrm{h}^2=\frac{1}{2}v_2^2\lambda_2-2v_1v_2\lambda_{12} \frac{s_\alpha}{c_\alpha},\qquad M_\mathrm{H}^2=\frac{1}{2}v_2^2\lambda_{2}
+2v_1v_2\lambda_{12}\frac{c_\alpha}{s_\alpha},
\end{equation}
where we enforce the mass hierarchy $M_\mathrm{H}>M_\mathrm{h}$
without loss of generality by choosing the range for the mixing angle according to
\begin{equation}
0\leq\alpha<\frac{\pi}{2}\quad\text{for }\lambda_{12}\geq0,
\quad\text{and}\quad-\frac{\pi}{2}<\alpha<0\quad\text{for } \lambda_{12} < 0.
\end{equation}
This leaves us with a SM-like Higgs field $ h $ with mass $ M_\mathrm{h} $ and an additional heavier Higgs field $ H $ with mass $ M_\mathrm{H}>M_\mathrm{h} $.%
\footnote{In principle, it is also possible to identify the heavier state $\PH$ with the
observed Higgs particle of mass $125\GeV$, but we do not consider this (experimentally
disfavoured) possibility here, because we want to analyze the heavy-mass limit of the second Higgs
boson of the SESM.}
The reparametrization of the doublet and singlet fields in \eqs{doublet}{singlet}
leads to the tadpole terms $ t_{\mathrm{h}}h $ and $t_{\mathrm{H}}H$  with
\begin{align}
t_\Ph &{} = c_\alpha t_2 - s_\alpha t_1,&
t_\PH &{} = s_\alpha t_2 + c_\alpha t_1,
\end{align}
in the Lagrangian, where
\begin{align}
t_1 &{} = v_1 (2 \mu_1^2 - v_2^2 \lambda_{12} - 4 v_1^2 \lambda_1),&
t_2 &{} = \frac{v_2}{4} (4\mu_2^2 - 4v_1^2 \lambda_{12} - v_2^2 \lambda_2).
\end{align}
At the bare (tree) level two parameters of the theory (here $v_{1,2}$) are
fixed by requiring $ t_{\mathrm{h}}=t_{\mathrm{H}}=0 $.
At loop level the tadpole terms play an important role
in the course of renormalization as described in \sec{ren}.

In the definition of a specific scenario for the limit $ M_\mathrm{H}\rightarrow\infty $,
i.e.\ in defining the scaling behaviour of the BSM parameters, it is
useful to introduce a (dimensionless) power counting parameter
$ \zeta\sim M_\mathrm{H}/M_\mathrm{h}\rightarrow\infty $
that keeps track of the scaling with the heavy mass~$\MH$.
Also, it will be more transparent from now on to work with phenomenologically motivated
input parameters rather than the fundamental parameters of the Lagrangian,
i.e.\ we express the BSM parameters
$ \{\mu_1^2,\lambda_1,\lambda_{12}\}$
in terms of
$\{M_\mathrm{H},s_\alpha,\lambda_{12}\} $.
Note that $s_\alpha$ is most directly related
to the measured {\it signal strengths} of
Higgs production cross sections and decay widths, which are defined by ratios of measured
quantities and SM predictions.
Before we define a specific large-$\MH$ scenario of the SESM, we introduce the scaling powers $a,l$ for $s_\alpha$ and $\lambda_{12}$
as follows,
\begin{equation}
s_\alpha\sim \zeta^{-a},\qquad\lambda_{12}\sim \zeta^{-l}.
\label{eq:scalingpowers}
\end{equation}
The effect of this rescaling on the fundamental parameters of the theory can be calculated from their relations to $\{M_\mathrm{H},s_\alpha,\lambda_{12}\} $,
as given in Eq.~(2.15) of \rcite{Altenkamp:2018bcs}, leading to
\begin{align}
v_1&\sim \zeta^{2-a+l},&&
\mu_1^2\sim \zeta^{\max\{2,-2a,-l\}},&&
\lambda_1\sim \zeta^{\max\{2a-2l-2,-2l-4\}}.
\nn\\
v_2&\sim \zeta^0,&&
\mu_2^2\sim \zeta^{\max\{4-2a+l,2-2a,0\}},&&
\lambda_2\sim \zeta^{\max\{2-2a,0\}},
\end{align}
In the following,
we consider the scenario $ a=1,\, l=0$, i.e.\
\begin{equation}
v_2, \lambda_{1}, \lambda_{2}, \lambda_{12} \sim \zeta^0, \qquad
v_1^2, \mu_1^2, \mu_2^2 \sim \zeta^2,
\label{eq:weak-coupling-scenario}
\end{equation}
in which all mass parameters of the scalar sector are considered to
be large, with the exception of the vev $v_2$, which is tied to the
known W-boson mass.
Self-consistency of the scaling can be checked by applying \eq{weak-coupling-scenario} to
the relation
\begin{align}
  \frac{s_{2\alpha}}{c_{2\alpha}} = \frac{8 v_1 v_2 \lambda_{12}}{16 v_1^2 \lambda_1-v_2^2 \lambda_2},
\end{align}
following from the diagonalization of the Higgs mass matrix~\cite{Altenkamp:2018bcs}.
This shows that
$s_\alpha$ is naturally suppressed according to $s_\alpha\sim\zeta^{-1}$ in agreement with \eq{scalingpowers} for $a=1$, see also \rcite{Boggia:2016asg}.
This is a weakly coupled scenario,
providing the minimal suppression that is required to still deliver a viable description of Higgs data,
which show that the above-mentioned signal strengths
are close to one, i.e.\ $s_\alpha$ has to be small.
In particular, due to the $\zeta^{-1}$ suppression of the mixing angle, $h$ equals the SM-type Higgs field $h_2$ at leading order in the large-mass expansion.

Other physically interesting limits are conceivable, such as the strong-coupling scenario
\begin{equation}
v_1, v_2 \sim \zeta^0, \qquad
\mu_1^2, \mu_2^2, \lambda_{1}, \lambda_{2}, \lambda_{12} \sim \zeta^2,
\end{equation}
in which $s_\alpha\sim \zeta^0$, so that (for $s_\alpha\ne0$)
the low-energy theory does not coincide with the SM in this case. We will not consider such scenarios in this paper,
although the proposed method would be capable of handling also such
scenarios as long as perturbativity is guaranteed. For a tree-level study of the low-energy limit of the SESM in a non-decoupling scenario see e.g.\ \rcite{Buchalla:2016bse}.

\subsection{Background-field formulation and non-linear realization}

Applying the background-field transformation splits each field $ \phi $ into a classical background field $ \hat{\phi} $ and a quantum field $ \phi $. Gauge and physical Higgs fields are split additively,
\begin{equation}
\phi\rightarrow\tilde{\phi}=\hat{\phi}+\phi,
\end{equation}
but the non-linearly parametrized matrix of the Goldstone-boson fields splits
multiplicatively as~\cite{Dittmaier:1995cr,Dittmaier:1995ee}
\begin{equation}
U\rightarrow \tilde{U}=\hat{U}U.
\end{equation}
Owing to the unitarity of $\tilde U $, the combined Lagrangian of the gauge and Higgs sectors of the SESM can be
written as
\begin{align}
\L_\mathrm{gauge+Higgs}
=&-\frac{1}{2}\tr\left[\tilde{W}_{\mu\nu}\tilde{W}^{\mu\nu}\right]
-\frac{1}{4}\tilde{B}_{\mu\nu}\tilde{B}^{\mu\nu}
+\frac{1}{4}\left(v_{2}+\tilde{h}_{2}\right)^2
\tr\left[\left(\tilde{D}_{\mu}\tilde{U}\right)^\dagger\left(\tilde{D}^{\mu}\tilde{U}\right)\right] \nn\\
&+\frac{1}{2}\left(\partial_{\mu}\tilde{h}_{2}\right)\left(\partial^{\mu}\tilde{h}_{2}\right)+\frac{1}{2}\mu_{2}^2\left(v_{2}+\tilde{h}_{2}\right)^2-\frac{1}{16}\lambda_{2}\left(v_{2}+\tilde{h}_{2}\right)^4 \nn\\
&+\frac{1}{2}\left(\partial_{\mu}\tilde{h}_{1}\right)\left(\partial^{\mu}\tilde{h}_{1}\right)+\mu_{1}^2\left(v_{1}+\tilde{h}_{1}\right)^2-\lambda_{1}\left(v_{1}+\tilde{h}_{1}\right)^4 \nn\\
&-\frac{1}{2}\lambda_{12}\left(v_{2}+\tilde{h}_{2}\right)^2\left(v_{1}+\tilde{h}_{1}\right)^2,
\end{align}
where the Goldstone fields occur only in their kinetic term, but not in the Higgs potential. In the BFM, separate gauge choices can be made for background and quantum fields. To eliminate the background Goldstone fields, the unitary gauge is chosen for the background fields which can be achieved by a generalized Stueckelberg transformation~\cite{Stueckelberg:1938zz,Stueckelberg:1957zz,Kunimasa:1967zza,Lee:1972yfa}
\begin{equation}
\hat{W}_{\mu}\rightarrow\hat{U}\hat{W}_{\mu}\hat{U}^\dagger
+\frac{\ri}{g_{2}}\hat{U}\partial_{\mu}\hat{U}^\dagger, \qquad
W_{\mu}\rightarrow\hat{U}W_{\mu}\hat{U}^\dagger, \qquad
\hat{B}_{\mu}\rightarrow\hat{B}_{\mu}, \qquad
B_\mu\rightarrow B_{\mu}.
\label{eq:Stueckelberg}
\end{equation}
This transforms the covariant derivative and the field-strength tensor according to
\begin{equation}
\left(\tilde{D}_{\mu}\tilde{U}\right)\rightarrow\hat{U}\left(\tilde{D}_{\mu}U\right),\qquad
\tilde{F}_{\mu\nu}\rightarrow\hat{U}\tilde{F}_{\mu\nu}\hat{U}^\dagger.
\end{equation}
The inversion of this Stueckelberg transformation, which restores the background
Goldstone fields, is straightforward (see e.g.\ Sec.~5 of \rcite{Dittmaier:1995ee}).
The gauge of the quantum fields is fixed as in the SM by an explicit
gauge-fixing term. We choose~\cite{Dittmaier:1995cr,Dittmaier:1995ee}
\begin{equation}
\L_{\mathrm{fix}}=-\frac{1}{\xi_W}\tr\left[\left(\hat{D}_{W}^{\mu}W_{\mu}+\frac{1}{2}\xi_W g_{2}v_{2}\hat{U}\varphi\hat{U}^\dagger\right)^2\right]-\frac{1}{2\xi_B}\left(\partial^{\mu}B_{\mu}+\frac{1}{2}\xi_B g_{1}v_{2}\varphi_{3}\right)^2,
\end{equation}
where
\begin{equation}
\hat{D}_{W}^{\mu}\phi=\partial^\mu\phi-\ri g_{2}\bigl[\hat{W}^\mu,\phi\bigr],
\end{equation}
for any field $\phi$ in the adjoint representation of SU(2).
The two gauge parameters for the fields $ W^\mu $ and $ B^\mu $ are set equal, $ \xi_{W}=\xi_{B}=\xi, $ to avoid mixing in the tree-level propagators.

\section{Integrating out the heavy Higgs boson}
\label{se:largemass}

\subsection{Separation of hard and soft modes of the heavy Higgs field}
\label{se:modeseparation}

Although usually formulated in terms of momentum domains in Feynman integrals,
the expansion by regions~\cite{Beneke:1997zp,Smirnov:2002pj}
 provides the ideal framework for separating light and heavy field modes.
At the level of a one-loop Feynman integral $I$ with loop momentum $p$,
the idea is to divide the integral domain into
two disjunct parts containing small ($p\sim \Mh$) or large ($p\sim\MH\gg \Mh$) momenta,
\begin{equation}
I = \int\rd^D p\, f(p) = \int\limits_{p\sim \Mh} \!\!\rd^D p\, f(p) \;+ \int\limits_{p\sim\MH}\!\!\rd^D p\, f(p),
\label{eq:I}
\end{equation}
where $f$ denotes an arbitrary integrand.
In four dimensions ($D=4$), a momentum cutoff $\MH \gg \Lambda \gg \Mh$ has to
be introduced to sharply separate the two integration domains.
In dimensional regularization ($D\ne4$), however, following the method of
regions the separation is effectively implemented by a strict Taylor expansion of the
loop integrand in $1/\zeta \sim \Mh/\MH$
before integration, where
$p\sim\zeta^0$ and $p\sim\zeta^1$ in the domains of small and large momenta, respectively.
To formalize this, we introduce the Taylor operators
$\mathcal{T}_{l}(p)$ and $\mathcal{T}_{h}(p)$ by
\begin{align}
\mathcal{T}_{l}(p) \, f(p,p_i,\MH,m_i,c_i)
&{}= \left[ \exp\biggl(\frac{\partial}{\partial \xi_l}\biggr) \, f(\xi_l p,\xi_l p_i,\MH,\xi_l m_i, \xi_l c_i)
\, \right]_{\xi_l\to0},
\nn\\
\mathcal{T}_{h}(p) \, f(p,p_i,\MH,m_i,c_i)
&{}= \left[ \exp\biggl(\frac{\partial}{\partial \xi_h}\biggr) \, f(p,\xi_h p_i,\MH,\xi_h m_i, \xi_h c_i)
\, \right]_{\xi_h\to0},
\label{eq:Tops}
\end{align}
where $p_i^\mu$ generically represents any small external momenta
and $m_i\ll\MH$ stands for any small masses.
In view of our functional approach, c.f.\ \eq{L1loopeff}, we have generalized the integrand here to include additional quantities $c_i \sim \zeta^n$ with $n\le0$
like light background fields and their derivatives.
The integral $I$ of \eq{I}, thus, reads
\begin{equation}
I = \int\rd^D p\, \mathcal{T}_{l}(p)\,f(p) + \int\rd^D p\, \mathcal{T}_{h}(p)\,f(p).
\end{equation}
Formally, the operators can be interpreted as orthogonal projectors, since they obey
the relations
\begin{align}
\left[\mathcal{T}_{l}(p)\right]^k &{}= \mathcal{T}_{l}(p), \qquad
\left[\mathcal{T}_{h}(p)\right]^k = \mathcal{T}_{h}(p), \qquad k \in \mathbb{N},
\nn\\
\mathcal{T}_{l}(p) + \mathcal{T}_{h}(p) &{}=1 ,
\nn\\
\mathcal{T}_{l}(p) \, \mathcal{T}_{h}(p) &{}=
\mathcal{T}_{h}(p) \, \mathcal{T}_{l}(p) = 0,
\label{eq:Toprel}
\end{align}
where the last relation holds, because the successive application of
$\mathcal{T}_{l}(p)$ and $\mathcal{T}_{h}(p)$ (or vice versa) produces
scaleless integrals which vanish in dimensional regularization.

Carrying the concept over to a generic quantum
field variable $\phi(p)$ in momentum space, we define
\begin{equation}
\phi_l(p) = \mathcal{T}_{l}(p)\,\phi(p), \qquad
\phi_h(p) = \mathcal{T}_{h}(p)\,\phi(p),
\end{equation}
so that $\phi(p)=\phi_l(p)+\phi_h(p)$.
We stress that this definition only makes sense
if the momentum $p$ is eventually
integrated over in $D$ dimensions (in the course of a loop calculation).
The separation into light and heavy modes in momentum space
can be translated to
the field $\phi(x)$ in position space via Fourier transformation
(with unit Jacobian determinant in the path integral),
\begin{equation}
\phi_l(x) = \mu^{4-D}\int\frac{\rd^D p}{(2\pi)^D}\,\re^{\ri px}\,\phi_l(p), \qquad
\phi_h(x) = \mu^{4-D}\int\frac{\rd^D p}{(2\pi)^D}\,\re^{\ri px}\,\phi_h(p),
\end{equation}
so that $\phi(x)=\phi_l(x)+\phi_h(x)$.
The parameter $\mu$ denotes the arbitrary reference scale of dimensional regularization,
which is introduced to keep the mass dimensions of quantities at the same values
as for $D=4$.
This additive separation implies a factorization of the path-integral
measure of $\phi$ into factors corresponding to light and heavy modes,
\begin{equation}
\int \mathcal{D}\phi = \int \mathcal{D}\phi_l \int \mathcal{D}\phi_h.
\end{equation}

Finally, we apply this mode separation to the heavy Higgs field
$\tilde H = H+\hat H$ in the BFM,
\begin{equation}
\tilde H(x) = H(x) + \hat H(x) = H_l(x) + H_h(x) + \hat H_l(x) + \hat H_h(x),
\end{equation}
which gets decomposed into four contributions.
Since we apply the EFT only for energies well below $\MH$, the  tree lines in Feynman diagrams only carry small momenta and
we effectively have $\hat{H}_h=0$.

For transparency, we will split the the effective Lagrangian $\L_\eff$
into the {\it tree-level effective Lagrangian} $\L_\eff^\tree$,
which contains all tree-level effects of the heavy field (resulting from $\hat{H}_{l}$)
and provides the effective couplings to be inserted in loops (via $H_{l}$),
and the
{\it one-loop effective Lagrangian} $\delta\L^{\text{1-loop}}_\eff$, which
contains all the (local) one-loop effects of the heavy field at large momentum transfer
(via $H_{h}$).
Another, third part of the Lagrangian, $\delta\L_\eff^{\mathrm{ct}}$
emerges in the course of renormalization, see \sec{ren}.

\subsection{Path integral over the hard modes of the heavy quantum Higgs field}
\label{se:PI}

\subsubsection{Relevant terms in the Lagrangian}

The goal of this section is to carry out the path integral over the
quantum field $H_h(x)$ at the one-loop level.
To this end, we first isolate
all terms in the full-theory
Lagrangian that are bilinear in the
quantum fields and call the resulting part of the Lagrangian $\L^{\text{1-loop}}$,
\begin{align}
\L^{\text{1-loop}}=
&-\frac{1}{2}h_{2}\Delta_{h_{2}}h_{2}
-\frac{1}{2}h_{1}\Delta_{h_{1}}h_{1}
+\tr\left[\barW_{\mu}\Delta_{\barW}^{\mu\nu}\barW_{\nu}\right]
+\frac{1}{2}A_{\mu}\Delta_{A}^{\mu\nu}A_{\nu}
-\tr\left[\varphi\Delta_{\varphi}\varphi\right] \nn\\
&+h_{1}X_{h_{1}h_{2}}h_{2}+h_{2}\tr\left[X_{h_{2}\barW}^\mu \barW_{\mu}\right]
+h_{2}\tr\left[X_{h_{2}\varphi}\varphi\right] \nn\\
&+\tr\left[A_{\mu}X_{A\barW}^{\mu\nu}\barW_{\nu}\right]
+\tr\left[\barW_{\mu}X_{\barW\varphi}^\mu\varphi\right]
+\tr\left[A_{\mu}X_{A\varphi}^\mu\varphi\right]
+\L^{\text{1-loop}}_{\mathrm{ghost}},
\label{eq:L1loop-1}
\end{align}
where
\begin{equation}
\barW^\mu=\frac{1}{2}\left(W_{1}^\mu\tau_{1}+W_{2}^\mu\tau_{2}+Z^\mu\tau_{3}\right)
\end{equation}
and $\L^{\text{1-loop}}_{\mathrm{ghost}}$
comprises all relevant terms containing Faddeev--Popov ghost fields.
Since we are not interested in Green functions with external ghost fields,
$\L^{\text{1-loop}}_{\mathrm{ghost}}$ consists of monomials with exactly two
quantum ghost fields and any additional background fields.
The Lagrangian $\L^{\text{1-loop}}_{\mathrm{ghost}}$ will play no role when
integrating out the heavy quantum field $H(x)$, because ghost fields and $H$ fields
can never appear in the loop part of the same one-loop diagram.

The relevant $\Delta$- and $X$-operators in \eq{L1loop-1} are given by
\begin{align}
\Delta_{h_{1}}&=
\Box-2\mu_{1}^2
+12\lambda_{1}\bigl(v_{1}+\hat{h}_{1}\bigr)^2
+\lambda_{12}\bigl(v_{2}+\hat{h}_{2}\bigr)^2,
\nn\\
\Delta_{h_{2}}&=
\Box-\mu_{2}^2
+\frac{3}{4}\lambda_{2}\bigl(v_{2}+\hat{h}_{2}\bigr)^2
+\lambda_{12}\bigl(v_{1}+\hat{h}_{1}\bigr)^2
-\frac{1}{2}g_{2}^2\tr\bigl[\hat{C}^2\bigr],
\nn\\
\Delta_{\varphi}&=
\hat{D}_{\mu}\biggl(1+\frac{\hat{h}_{2}}{v_{2}}\biggr)^{\!2}\hat{D}^\mu
+g_{2}^2\biggl(1+\frac{\hat{h}_{2}}{v_{2}}\biggr)^{\!2}\hat{C}^2
+\xi M_{W}^2\biggl(1+\frac{\sw^2}{\cw^2} P_{3}\biggr),
\nn\\
\Delta_{\barW,0}^{\mu\nu}&=
 g^{\mu\nu}\Box+\frac{1-\xi}{\xi}\partial^\mu\partial^\nu+g^{\mu\nu}M_{W}^2
\biggl(1+\frac{\sw^2}{\cw^2}P_{3}\biggr),
\nn\\
X_{h_{1}h_{2}}&=
-2\lambda_{12}\bigl(v_{1}+\hat{h}_{1}\bigr)\bigl(v_{2}+\hat{h}_{2}\bigr),
\nn\\
X_{h_{2}\barW}^\mu&=
2g_{2}\biggl(1+\frac{\hat{h}_{2}}{v_{2}}\biggr)M_{W}\hat{C}^\mu
\biggl(1+\frac{1-\cw}{\cw}P_{3}\biggr),
\nn\\
X_{h_{2}\varphi}&=
2g_{2}\biggl(1+\frac{\hat{h}_{2}}{v_{2}}\biggr)\bigl(
\ri g_{1}\hat{B}_{\mu}\tau_{3}\hat{W}^\mu-\hat{C}_{\mu}\partial^\mu
\bigr),
\nn\\
X_{\barW\varphi,0}&= -g_2 \hat{h}_2 \biggl(2+\frac{\hat{h}_{2}}{v_{2}}\biggr) \biggl(1+\frac{1-\cw}{\cw}P_{3}\biggr)\partial^\mu,
\label{eq:Xh2phi}
\end{align}
where (we suppressed $2\times2$ unit matrices for compactness and)
$\hat{C}^2 = \hat{C}^\mu \hat{C}_\mu$ with
\begin{align}
\hat{C}^\mu =
\hat{W}^\mu + \frac{\sw}{\cw}\hat{B}^\mu \frac{\tau_3}{2}
=
\frac{1}{2}\biggl(\hat{W}_{1}^\mu\tau_{1}+\hat{W}_{2}^\mu\tau_{2}
+\frac{1}{\cw}\hat{Z}^\mu\tau_{3}\biggr)
.
\end{align}
Moreover, we have introduced the operators $P_a$ projecting any $2\times2$ matrix $M$
onto the Pauli matrix $\tau_a$,
\begin{equation}
P_{a} M =\frac{\tau_{a}}{2}\tr\left[\tau_{a}M\right].
\end{equation}
Note that we have not given $\Delta_{A}$,
$X_{A\barW}^{\mu\nu}$, and $X_{A\varphi}^\mu$
explicitly, because
they will not be needed for our purposes as will become clear below.
Likewise, for $\Delta_{\barW}^{\mu\nu}$ and $X_{\barW\varphi}$ we show only the leading-order terms indicated by the subscript ``0", because the rest will not be needed
in the derivation of the effective Lagrangian to $\ord(\zeta^{-2})$.
In fact, for the purpose of the present paper, where we only aim for the $\ord(\zeta^0)$  effective Lagrangian, $\Delta_{\barW,0}^{\mu\nu}$ and $X_{\barW\varphi,0}$ are not required either.

Since we want to integrate out the heavy field $H$, we have to express $\L^{\text{1-loop}}$
in terms of the Higgs fields corresponding to mass eigenstates
as defined in Eq.~\eqref{eq:HfieldRotation},
\begin{align}
\L^{\text{1-loop}}= &
-\frac{1}{2}H\Delta_{H}H
-\frac{1}{2}h\Delta_{h}h
+\tr\left[\barW_{\mu}\Delta_{\barW}^{\mu\nu}\barW_{\nu}\right]
+\frac{1}{2}A_{\mu}\Delta_{A}^{\mu\nu}A_{\nu}
-\tr\left[\varphi\Delta_{\varphi}\varphi\right]
\nn\\
& {} +HX_{Hh}h+H\tr\left[X_{H\barW}^\mu\barW_{\mu}\right]
+h\tr\left[X_{h\barW}^\mu\barW_{\mu}\right]
+H\tr\left[X_{H\varphi}\varphi\right]
+h\tr\left[X_{h\varphi}\varphi\right]
\nn\\
& {}
+\tr\left[A_{\mu}X_{A\barW}^{\mu\nu}\barW_{\nu}\right]
+\tr\left[\barW_{\mu}X_{\barW\varphi}^\mu\varphi\right]
+\tr\left[A_{\mu}X_{A\varphi}^\mu\varphi\right]
+\L^{\text{1-loop}}_{\mathrm{ghost}},
\label{eq:L1loophH}
\end{align}
where
\begin{align}
\Delta_{H}&{}= s_{\alpha}^2\Delta_{h_{2}}+c_{\alpha}^2\Delta_{h_{1}}-2c_{\alpha}s_{\alpha}X_{h_{1}h_{2}},
\qquad
\Delta_{h}= c_{\alpha}^2\Delta_{h_{2}}+s_{\alpha}^2\Delta_{h_{1}}+2c_{\alpha}s_{\alpha}X_{h_{1}h_{2}},
\nn\\
X_{Hh}&= c_{\alpha}s_{\alpha}\left(\Delta_{h_{1}}-\Delta_{h_{2}}\right)+\left(c_{\alpha}^2-s_{\alpha}^2\right)X_{h_{1}h_{2}},
\nn\\
X^\mu_{H\barW}&{}= s_{\alpha}X^\mu_{h_{2}\barW}, \qquad
X^\mu_{h\barW} {}= c_{\alpha}X^\mu_{h_{2}\barW}, \qquad
X_{H\varphi}{}= s_{\alpha}X_{h_{2}\varphi}, \qquad
X_{h\varphi}{}= c_{\alpha}X_{h_{2}\varphi}.
\label{eq:hphi}
\end{align}

Up to this point we have not yet split the quantum fields into light and heavy modes
in the Lagrangian $\L^{\text{1-loop}}=\L^{\text{1-loop}}(H,\phi_i)$,
where $\phi_i$ denotes any quantum field other than $H$.
At the one-loop level, we can simply write
\begin{align}
\L^{\text{1-loop}}(H,\phi_i) =
\L^{\text{1-loop}}(H_h,\phi_{i,h}) + \L^{\text{1-loop}}(H_l,\phi_{i,l}),
\end{align}
because one-loop diagrams split into two contributions corresponding to large and small
loop momenta. This can also be understood from \eq{Toprel}.
At the same time recall that all background fields should be interpreted as
light modes, because momenta on external and on tree lines in diagrams
are assumed to be small.

\subsubsection{Diagonalization and functional integration}

Our next step is to express the Lagrangian
$\L^{\text{1-loop}}(H_h,\phi_{i,h})$ for the heavy modes of the quantum fields
in the (``diagonal")  form
\begin{equation}
\L^{\text{1-loop}}(H_h,\phi_{i,h})
= -\frac{1}{2} H_h\tilde{\Delta}_{H}H_h+\L^{\text{1-loop}}_{\mathrm{rem}}(\phi_{i,h}),
\label{eq:L1loopdiag}
\end{equation}
where $\L^{\text{1-loop}}_{\mathrm{rem}}(\phi_{i,h})$ does not depend on $H_h(x)$, but only
on the other quantum fields generically denoted $\phi_{i,h}$.
This can be achieved by a suitable linear
field redefinition of the $\phi_{i,h}$ as demonstrated below.
Of course, both $\tilde{\Delta}_{H}$ and $\L^{\text{1-loop}}_{\mathrm{rem}}$ also
depend on all background fields including $\hat H_l(x)$.
The diagonalization of $\L^{\text{1-loop}}(H_h,\phi_{i,h})$ w.r.t.\ $H_h$
transforms the functional integral over $H_h(x)$
into an integral of Gaussian type, which can be evaluated as
\begin{align}
&\int\mathcal{D}H_h\,\exp\left\{-\frac{\ri}{2}\int\rd^4 x\,H_h\tilde{\Delta}_H H_h\right\}
 {} \,\propto\,
\left\{ \mathpzc{Det}_h\Bigl[ \delta(x-y) \tilde{\Delta}_H(x,\partial_x)
\Bigr]\right\}^{-\frac{1}{2}}
\nn\\
&=\exp \left\{ -\frac12
\mathpzc{Tr}_h \Bigl[ \ln\Bigl(
\delta(x-y) \tilde{\Delta}_H(x,\partial_x) \Bigr) \Bigr]
\right\}
 \,=\,
\exp\left\{\ri\,\mu^{D-4} \int\rd^D x\, \delta\L^{\text{1-loop}}_\eff \right\}.
\label{eq:funcdet}
\end{align}
Dropping an irrelevant constant contribution
the one-loop effective Lagrangian
$\delta\L^{\text{1-loop}}_\eff$
describing the hard loop contributions of $H_h(x)$ can thus be obtained
by translating the functional determinant $\mathpzc{Det}_h$
of the differential operator $\tilde{\Delta}_H(x,\partial_x)$
into a functional trace $\mathpzc{Tr}_h$,
which in turn can be evaluated in terms of a hard momentum-space integral.
The subscript $h$ of $\mathpzc{Det}_h$ and  $\mathpzc{Tr}_h$ indicates the restriction to the subspace of large-momentum modes.
We explain the details of the corresponding functional manipulations in \app{FuncDetApp}
and proceed with the well-known result
\begin{align}
\delta\L^{\text{1-loop}}_\eff & {}= \frac{\ri}{2}\,\mu^{4-D}\int\frac{\rd^D p}{(2\pi)^D}\,
\mathcal{T}_{h}(p) \,
\ln\left( \tilde{\Delta}_H(x,\partial_x+\ri p) \right),
\label{eq:L1loopeff}
\end{align}
where
$\mathcal{T}_{h}(p)$ is defined in \eq{Tops}.
Unlike $\tilde{\Delta}_H$ the term $\L^{\text{1-loop}}_{\mathrm{rem}}$ in \eq{L1loopdiag} is (by construction) independent of any hard scale $\sim \MH$.
Diagonalizing it w.r.t.\ the $\phi_{i,h}$ and performing the corresponding functional integrations in analogy to
\eq{L1loopeff} therefore yields scaleless
large-momentum integrals which vanish in dimensional regularization.
The part $\L^{\text{1-loop}}_{\mathrm{rem}}(\phi_{i,h})$
is therefore irrelevant for the derivation of $\delta\L^{\text{1-loop}}_\eff$.
Note that the $x$ dependence of $\tilde{\Delta}_H$ is only due to background fields.
Thus,
$\partial_x$ only acts on background fields, which all carry small momenta,
 and therefore scales like $\zeta^{-1}$ relative to the large momentum~$p$.

At this point, we need the explicit form of the differential operator
$\tilde{\Delta}_H(x,\partial_x)$ which results from the diagonalization of the
Lagrangian $\L^{\text{1-loop}}(H_h,\phi_{i,h})$ in \eq{L1loopdiag}.
We first formulate this diagonalization in a generic way and subsequently
specialize the result to our model Lagrangian.
Considering \eq{L1loophH} and suppressing the subscripts ``$h$''
indicating heavy modes in the following,
$\L^{\text{1-loop}}(H,\phi_i)$ has the generic form
\begin{align}
\L^{\text{1-loop}}(H,\phi_i) &{}=
-\frac{1}{2}H\Delta_H H + H \mathcal{X}_{Hi} \phi_i
-\frac{1}{2}\phi_i \mathcal{A}_{ij} \phi_j
\label{eq:L1loopschem}
\end{align}
with implicit summations over the labels $i,j$ of the light fields
$\phi_i$, $\phi_j$ which are assumed to be real ($\phi_k=\phi_k^\dagger$).
Taking into account the hermiticity of $\L^{\text{1-loop}}$, the generic
operators $\Delta_H$, $\mathcal{A}_{ij}$, and $\mathcal{X}_{Hi}$
can be assumed to obey the relations
\begin{equation}
\Delta_H {} = \Delta_H^\dagger, \qquad
\mathcal{X}_{Hi} {} = \mathcal{X}_{iH}^\dagger, \qquad
\mathcal{A}_{ij} {} = \mathcal{A}_{ji}^\dagger.
\end{equation}
The following shifts of the light quantum fields,
\begin{equation}
\phi_i \,\to\, \phi_i + \left( \mathcal{A}^{-1} \right)_{ij}\, \mathcal{X}_{jH}\,H,
\end{equation}
which are inspired by the
field transformations described in \rcites{Dittmaier:1995cr,Dittmaier:1995ee},
have unit Jacobian in the functional
integral and change the Lagrangian $\L^{\text{1-loop}}(H,\phi_i)$ only
by terms containing~$H$,
\begin{equation}
\L^{\text{1-loop}}(H,\phi_i) \,\to\,
-\frac{1}{2}H\tilde\Delta_H H
-\frac{1}{2}\phi_i \mathcal{A}_{ij} \phi_j
\end{equation}
with
\begin{equation}
\tilde\Delta_H = \Delta_H - \mathcal{X}_{Hi} \left( \mathcal{A}^{-1} \right)_{ij}\, \mathcal{X}_{jH}.
\label{eq:DelHtilde}
\end{equation}
To evaluate the inverse $\big(\mathcal{A}^{-1}\big)_{ij}$, we split the operators
$\mathcal{A}_{ij}$ into large and small contributions
$\mathcal{D}_{ij}$ and $\mathcal{X}_{ij}$, respectively, in the sense that
all $\mathcal{X}_{ij}$ are suppressed w.r.t.\ all diagonal parts
$\mathcal{D}_{ii}$ at least by one power of $1/\zeta$,
\begin{align}
\mathcal{A}_{ij} = \mathcal{D}_{ij} - \mathcal{X}_{ij}, \qquad
\mathcal{X}_{ii} = 0,
\end{align}
so that $\mathcal{D}_{ij}$ is invertible (but not necessarily fully diagonal).%
\footnote{This is always possible, because the leading terms of $\mathcal{A}_{ij}$ for large $p$ correspond to the inverse propagators of the light fields.}
Without loss of generality, we take the diagonal parts of
the $\mathcal{X}_{ij}$ to vanish.
Below we will relate the $\mathcal{D}_{ij}$ to the $\Delta_{u}$
and the $\mathcal{X}_{ij}$ to the $X_{uv}$ of \eq{L1loophH}
(with $u,v = \barW,\varphi,H,h,\ldots$).
The scaling assumption holds because, upon the replacement $\partial_x \to \partial_x+\ri p$ according to \eq{L1loopeff}, the kinetic terms in the $\Delta_{u}$ contain at least one
power of the large momentum $p \sim \MH$ more than the interaction terms $X_{ij}$.
As realized also in \rcite{Fuentes-Martin:2016uol}, the inverse
of the operator of $\mathcal{A}_{ij}$,
can then be expressed as ($\mathcal{D}^{-1}$ times) a Neumann series,
\begin{align}
\left( \mathcal{A}^{-1} \right)_{ij} &{} = \left( (\mathcal{D}-\mathcal{X})^{-1} \right)_{ij}
= \left( \mathcal{D}^{-1} + \mathcal{D}^{-1}\mathcal{X}\mathcal{D}^{-1}+\mathcal{D}^{-1}\mathcal{X}\mathcal{D}^{-1}\mathcal{X}\mathcal{D}^{-1}+\dots \right)_{ij}
\label{eq:InvA}
\\
&{} = \left(\mathcal{D}^{-1}\right)_{ij}
+ \left(\mathcal{D}^{-1}\right)_{ik} \mathcal{X}_{kl} \left(\mathcal{D}^{-1}\right)_{lj}
+ \left(\mathcal{D}^{-1}\right)_{ik} \mathcal{X}_{kl} \left(\mathcal{D}^{-1}\right)_{lm}
\mathcal{X}_{mn} \left(\mathcal{D}^{-1}\right)_{nj}
+ \ldots\,.
\nn
\end{align}
Here the intermediate field indices such as $k,l,\dots$
(but not the  external indices $i,j$) are summed over.
This implies that the Lorentz and internal symmetry group indices of adjacent $\big(\mathcal{D}^{-1}\big)_{ij}$ and $\mathcal{X}_{ij}$ factors are properly contracted (except for the left- and rightmost indices).

The application of this generic diagonalization procedure to our model
requires a careful identification of the operators
$\mathcal{D}_{ij}$, and $\mathcal{X}_{ij}$
with their concrete realizations in \eq{L1loophH}.
Explicitly writing out also the adjoint SU(2) and Lorentz indices
we have the following assignments
 for the non-vanishing
$\mathcal{D}_{ij}$, which are diagonal in the field type (but not in the Lorentz and SU(2) indices),
\begin{align}
\mathcal{D}_{\barW_\mu^a\barW_\nu^b}
&{}= -2\tr\Bigl[\textstyle\frac{\tau_a}{2} \Delta^{\mu\nu}_{\barW} \frac{\tau_b}{2}\Bigr],
&
\mathcal{D}_{A_\mu A_\nu} &{}= -\Delta^{\mu\nu}_A,
\nn\\
\mathcal{D}_{\varphi_a\varphi_b}
&{}= 2\tr\Bigl[\textstyle\frac{\tau_a}{2} \Delta_{\varphi} \frac{\tau_b}{2}\Bigr],
&
\mathcal{D}_{hh} &{}= \Delta_h.
\label{eq:Delexpr}
\end{align}
For our model the relevant $\Delta_{u}$
expressions on the r.h.s.\ are given in \eqs{Xh2phi}{hphi}.
The inverse $\big(\mathcal{D}(x,\partial_x+\ri p)^{-1}\big)_{ij}$ required in \eq{InvA} can now be easily obtained, again in terms of a Neumann series,
by realizing that its leading-order term in the $\zeta$ expansion
is the usual momentum-space propagator of the respective light field with momentum $p$.
Accordingly, for $u=\barW,\varphi$ the $\Delta_{u}$ are, at leading order in $1/\zeta$, proportional to the unit matrix with fundamental SU(2) indices.
Hence, we can also compute%
\footnote{Note that defining $\mathcal{A}_{ij}$, $\mathcal{D}_{ij}$, $\mathcal{X}_{ij}$
in such a way that their indices are individual SU(2) components rather than complete SU(2) multiplets ($\barW$, $\varphi$)  makes it unnecessary to project onto the subspace of SU(2) generators when inverting $\Delta_{u}$ as was done in (Sections~3 of)	\rcites{Dittmaier:1995cr,Dittmaier:1995ee}.}
\begin{equation}
  \big(\mathcal{D}^{-1}\big)_{\varphi_a\varphi_b}
  = 2\tr\Bigl[\textstyle\frac{\tau_a}{2} \Delta_{\varphi}^{-1} \frac{\tau_b}{2}\Bigr],
  \qquad
  \big(\mathcal{D}^{-1}\big)_{\barW_\mu^a\barW_\nu^b}
  = -2\tr\Bigl[\textstyle\frac{\tau_a}{2} \big(\Delta_{\barW}^{-1}\big)^{\mu\nu} \frac{\tau_b}{2}\Bigr].
\end{equation}
Corresponding expansions to the order required in this work are given below.
For the non-vanishing non-diagonal parts $\mathcal{X}_{ij}$
($= \mathcal{X}_{ji}^\dagger$) we have
\begin{align}
\mathcal{X}_{A_\mu\barW_{a,\nu}} &{}= \tr\Bigl[\textstyle\frac{\tau_a}{2} X_{A\barW}^{\mu\nu}\Bigr],
&
\mathcal{X}_{\barW_{a,\mu}\varphi_b} &{}=
\tr\Bigl[\textstyle\frac{\tau_a}{2} X_{\barW\varphi}^\mu\frac{\tau_b}{2}\Bigr],
&
\mathcal{X}_{A_\mu\varphi_a} &{}= \tr\Bigl[\textstyle X_{A\varphi}^\mu \frac{\tau_a}{2} \Bigr],
\nn\\
\mathcal{X}_{Hh} &{}= X_{Hh},
&
\mathcal{X}_{H\barW_{a,\mu}} &{}=
\tr\Bigl[\textstyle X_{H\barW}^\mu\frac{\tau_a}{2}\Bigr],
&
\mathcal{X}_{h\barW_{a,\mu}} &{}=
\tr\Bigl[\textstyle X_{h\barW}^\mu\frac{\tau_a}{2}\Bigr],
\nn\\
\mathcal{X}_{h\varphi_a} &{}=
\tr\Bigl[\textstyle X_{h\varphi}\frac{\tau_a}{2}\Bigr],
&
\mathcal{X}_{H\varphi_a} &{}=
\tr\Bigl[\textstyle X_{H\varphi}\frac{\tau_a}{2}\Bigr].
&
\label{eq:Xijexpr}
\end{align}

\subsubsection{Large-mass expansion}

Aiming at a final effective Lagrangian $\delta\L^{\text{1-loop}}_\eff$
that includes all
(non-decoupling) effects of $\ord(\MH^0)$, we need
for the calculation of $\ln \big(\tilde\Delta_H(x,\partial_x+\ri p )\big)$ in
\eq{L1loopeff}, where $p\sim\MH\sim\zeta$, all terms of order $\zeta^{-4}$.
Since
\begin{align}
\tilde\Delta_H(x,\partial_x+\ri p) = -(p^2-\MH^2)+\Pi(x,p,\partial_x)
\label{eq:tildeDeltaH-decomp}
\end{align}
with $\Pi(x,p,\partial_x)$ at most of $\ord(\zeta^1)$, the
operator $\tilde\Delta_H(x,\partial_x+\ri p)$ is required to
$\ord(\zeta^{-2})$.
The scaling behaviour of the individual operators
$X_{uv}(x,\partial_x+\ri p)$ and $\Delta_{u}(x,\partial_x+\ri p)$ (and hence $\mathcal{X}_{ij}$ and $\mathcal{D}_{ij}$) can be easily determined from \eqs{Xh2phi}{hphi} and is summarized in \refta{tab:scaling-Delta-X}.%
\begin{table}[t]
\centerline{
\begin{tabular}{l|cccccccccc}
\hline
Operator & $\Delta^{-1}_{u=\barW,A,\varphi,h,H}$ &
$X_{A\barW}$ &
$X_{\barW\varphi}$ &
$X_{A\varphi}$ &
$X_{Hh}$ &
$X_{H\barW}$ &
$X_{h\barW}$ &
$X_{H\varphi}$ &
$X_{h\varphi}$
\\
\hline
Scaling & $\zeta^{-2}$ &
$\zeta^1$ & $\zeta^1$ &
$\zeta^0$ & $\zeta^1$ & $\zeta^{-1}$ & $\zeta^0$ &
$\zeta^0$ & $\zeta^1$
\\
\hline
\end{tabular} }
\caption{Scaling behaviour of the operators $\Delta_{u}^{-1}(x,\partial_x+\ri p)$ and $X_{uv}(x,\partial_x+\ri p)$
according to \eqs{Xh2phi}{hphi}
in the hard-momentum region where $p\sim\MH\sim\zeta$, $\hat{H}\sim s_a \sim \zeta^{-1}$.
}
\label{tab:scaling-Delta-X}
\end{table}%
\footnote{Here we anticipate that $\hat{H}_l\sim \zeta^{-1}$. This scaling behaviour is confirmed by the explicit result for the equation of motion of $\hat{H}_l$ in \sec{LeffLO}, but can also be directly understood from the scaling of the heavy Higgs propagator: $\langle H_lH_l \rangle \sim 1/\MH^2$.}
From \eqsm{DelHtilde}{Delexpr} we thus obtain
\begin{align}
\tilde\Delta_H(x,\partial_x+\ri p) \,={}&\, \Delta_H
- X_{Hh} \, \Delta_h^{-1} \, X_{hH}
\nn\\
&
- \tr\left[\textstyle X_{H\varphi}\frac{\tau_a}{2}\right] \,
  2\tr\left[\textstyle\frac{\tau_a}{2} \Delta_{\varphi}^{-1} \frac{\tau_b}{2}\right] \,
  \tr\left[\textstyle X_{\varphi H}\frac{\tau_b}{2}\right]
\nn\\
&
-2\, \tr\left[\textstyle X_{H\varphi}\frac{\tau_a}{2}\right] \,
  2\tr\left[\textstyle\frac{\tau_a}{2} \Delta_{\varphi}^{-1} \frac{\tau_b}{2}\right] \,
  \tr\left[\textstyle X_{\varphi h}\frac{\tau_b}{2}\right] \,
  \Delta_h^{-1} \, X_{hH}
\nn\\
&
- X_{Hh} \, \Delta_h^{-1} \,
  \tr\left[\textstyle X_{h\varphi}\frac{\tau_a}{2}\right] \,
  2\tr\left[\textstyle\frac{\tau_a}{2} \Delta_{\varphi}^{-1} \frac{\tau_b}{2}\right] \,
  \tr\left[\textstyle X_{\varphi h}\frac{\tau_b}{2}\right] \,
  \Delta_h^{-1} \, X_{hH}
\nn\\
&
+ \ord(\zeta^{-3}),
\label{eq:tildeDeltaH-expand}
\end{align}
where the fourth term on the r.h.s.\ actually represents two equal contributions,
corresponding to the two different orders
$\mathcal{X}_{H\varphi_a}\dots \mathcal{X}_{hH}$ and
$\mathcal{X}_{Hh}\dots \mathcal{X}_{\varphi_a H}$
of the operator chain.

The operators $\Delta_{u}(x,\partial_x+\ri p)$ and
$X_{uv}(x,\partial_x+\ri p)$ appearing in
\eq{tildeDeltaH-expand} can be directly read from
\eqs{Xh2phi}{hphi}
to the needed order in $1/\zeta$.
With these ingredients the individual contributions $\Pi^{(\kappa)}$
of order $\zeta^{-\kappa}$
to $\Pi(x,p,\partial_x)$, as defined in \eq{tildeDeltaH-decomp},
follow in a straightforward way, and we can evaluate
$\ln \bigl(\Delta_H(x,\partial_x+\ri p) \bigr)$ as series expansion,
\begin{align}
\ln\left(\tilde{\Delta}_H\left(x,\partial_{x}+\ri p\right)\right)
= \ln(-p^2+\MH^2)
- \sum_{n=1}^\infty \frac{1}{n} \,\biggl(\frac{\Pi}{p^2-\MH^2}\biggr)^{\!n},
\end{align}
where the $n$th term of the sum contributes at most at order $\zeta^{-n}$.
After that we can drop the $\mathcal{T}_{h}(p)$ operator in \eq{L1loopeff}.
Taking into account that odd powers of $p^\mu$
integrate to zero and dropping an irrelevant constant we arrive at%
\footnote{
  In the corresponding diagrammatic calculation the loop integrands, which are expanded in $\zeta$ depend only quadratically on $\MH$ and $s_\alpha$ (upon eliminating $\hat{H}$ using its EOMs). It is therefore intuitively clear that $\Pi^{(\kappa)}$ with odd $\kappa$ is proportional to odd powers of $p^\mu$.
  This can be easily verified with the explicit expressions given in \eq{tildeDeltaH-expand} and \eqsm{DeltaH}{XHhDhXhpDpXphDhXhH}.
}
\begin{align}
\delta\L^{\text{1-loop}}_\eff =
\frac{\ri}{2}\,\mu^{4-D}\int\frac{\rd^D p}{(2\pi)^D}\,\left[
-\frac{\Pi^{(0)}+\Pi^{(2)}}{p^2-\MH^2}
-\frac{(\Pi^{(0)})^2}{2(p^2-\MH^2)^2} \right]
+ \,\ord(\zeta^{-2}).
\label{eq:LeffPi}
\end{align}
Note that the $p^\mu$-even terms $\propto \Pi^{(-1)} \Pi^{(1)}$ and  $\propto \Pi^{(-1)} \Pi^{(0)} \Pi^{(-1)} (+ \mathrm{perm.})$ with $\Pi^{(-1)}=2\ri p\!\cdot\!\partial_x$ vanish in \eq{LeffPi} like total derivatives or because there is no background field for the partial derivative to act on.
The relevant terms of the $\Delta^{-1}_y(x,\partial_x+\ri p)$ read
\begin{align}
\Delta_h^{-1} ={}&  -\frac{1}{p^2} - \frac{1}{p^4} \biggl[
2\ri p\!\cdot\!\partial_x+\Box_x+\Mh^2
+\frac{3\Mh^2}{v_2}\hat h
+\frac{\MH^2 s_\alpha}{v_2}\hat H
+\frac{3(\Mh^2+\MH^2s_\alpha^2)}{2v_2^2} \hat h^2
\nn\\
& \quad
-\frac{g_2^2}{2}  \tr\bigl[\hat{C}^2\bigr] \biggr]
+ 4 \frac{(p\!\cdot\!\partial_x)^2}{p^6}
+ \,\ord(\zeta^{-5}),
\label{eq:Delhinv}\\
\Delta_\varphi^{-1} ={} & -\frac{1}{p^2} \biggl( 1+\frac{\hat h}{v_2} \biggr)^{\!\!\!-2}
\bbid
+ \,\ord(\zeta^{-3}),
\label{eq:Delphiinv}
\end{align}
where $\bbid$ is the $2\times2$ unit matrix.
Furthermore we have according to \eqs{Xh2phi}{hphi}
\begin{align}
\Delta_H={}& {-}p^2 + \MH^2
+2\ri p\!\cdot\!\partial_x + \Box_x
\nn\\
& {}
-\frac{g_2^2 s_\alpha^2}{2} \tr\bigl[\hat{C}^2\bigr]
+ \left[\frac{3 s_\alpha^2  \left(\Mh^2+ s_\alpha^2\MH^2\right)}{2 v_2^2}
+\frac{6 \lambda_{12}^2 v_2^2}{\MH^2}+\lambda_{12}
\left(1-6 s_\alpha^2\right)\right] \hat{h}^2
\nn\\
& {}
+ \left[2 \lambda_{12} v_2
\left(s_\alpha^2-\frac{3 \Mh^2}{\MH^2}-2\right)+\frac{s_\alpha^2 \left(\Mh^2-\MH^2 \left(s_\alpha^2-2\right)\right)}{v_2}\right] \hat{h}
+\frac{6  \lambda_{12}^2 v_2^2}{\MH^2 s_\alpha^2} \hat{H}^2
\nn\\
& {}
+\left[\frac{6\lambda_{12} v_2}{s_\alpha} \left(\frac{\Mh^2}{\MH^2}-s_\alpha^2+1\right)+\frac{3 \MH^2 s_\alpha^3}{v_2}\right]\hat{H}
+\left(6 \lambda_{12} s_\alpha-\frac{12 \lambda_{12}^2 v_2^2}{\MH^2 s_\alpha}\right) \hat{h} \hat{H}
\nn\\
& {}
+\ord\bigl(\zeta^{-3}\bigr),
\label{eq:DeltaH}
\end{align}
which corresponds to the contribution from loops involving heavy Higgs modes only.
With \eqs{Delhinv}{Delphiinv} the remaining terms in \eq{tildeDeltaH-expand} are
\begin{align}
X_{Hh} \, \Delta_h^{-1} \, X_{hH} ={}&
\frac{g_2^2 \MH^4 s_\alpha^2}{2 p^4 v_2^2} \tr\bigl[\hat{C}^2\bigr]\hat{h}^2
+\frac{g_2^2 \MH^2 s_\alpha^2}{p^2 v_2} \tr\bigl[\hat{C}^2\bigr]\hat{h}
-\frac{3 \MH^4 s_\alpha^2}{2 p^4 v_2^4}
\left(\Mh^2+\MH^2 s_\alpha^2\right) \hat{h}^4
\nn\\
&
-\frac{3 \MH^2 s_\alpha^2}{p^4 v_2^3} \Bigl[\Mh^2 \left(\MH^2+p^2\right)+p^2 \left(\MH^2 s_\alpha^2-2 \lambda_{12} v_2^2\right)\Bigr] \hat{h}^3
\nn\\
&
+\frac{\MH^2 s_\alpha^2}{p^4 v_2^2}
\Bigl[\bigl(2 s_\alpha^2-1\bigr)p^2 \MH^2
-4 \lambda_{12}p^2 v_2^2-\Mh^2 \bigl(\MH^2+4 p^2\bigr)\Bigr]\hat{h}^2
\nn\\
&
-\frac{\MH^2  s_\alpha}{p^4 v_2^3}
\left(\MH^4 s_\alpha^2+4 \lambda_{12} p^2 v_2^2\right) \hat{h}^2 \hat{H}
+\frac{4  \MH^2 s_\alpha}{p^2} \biggl(2 \lambda_{12}-\frac{\MH^2 s_\alpha^2}{v_2^2}\biggr)\, \hat{h} \hat{H}
\nn\\
&
-\frac{ \MH^4 s_\alpha^2 }{p^6 v_2^2}
\,\left[
p^2 \hat{h}\Box \hat{h}
+2 \ri p^2\, \hat{h}\, p^\mu \partial_\mu\hat{h}
-4 \hat{h}(p^\mu \partial_\mu)^2\hat{h} \right]
+\ord\bigl(\zeta^{-3}\bigr),
\\[.5em]
\rlap{$\tr\left[\textstyle X_{H\varphi}\frac{\tau_a}{2}\right]
2\tr\left[\textstyle\frac{\tau_a}{2} \Delta_{\varphi}^{-1} \frac{\tau_b}{2}\right]
\tr\left[\textstyle X_{\varphi H}\frac{\tau_b}{2}\right]
= \displaystyle  -\frac{2 g_2^2 s_\alpha^2}{p^2}
\,p^\mu p^\nu \tr\bigl[\hat{C}_\mu\hat{C}_\nu\bigr]
+\ord\bigl(\zeta^{-3}\bigr),$}
\phantom{X_{Hh} \, \Delta_h^{-1} \, X_{hH} =}
\label{eq:XHhDhXhH}
\\[.5em]
\rlap{$
2 \tr\left[\textstyle X_{H\varphi}\frac{\tau_a}{2}\right]
  2\tr\left[\textstyle\frac{\tau_a}{2} \Delta_{\varphi}^{-1} \frac{\tau_b}{2}\right]
  \tr\left[\textstyle X_{\varphi h}\frac{\tau_b}{2}\right]
  \Delta_h^{-1} X_{hH}$}
\phantom{X_{Hh} \, \Delta_h^{-1} \, X_{hH} =}
\label{eq:XHpDpXphDhXhH}
\nn\\
={}&
-\frac{4 g_2^2 \MH^2 s_\alpha^2 }{p^4 v_2}
\,p^\mu p^\nu \tr\bigl[\hat{C}_\mu\hat{C}_\nu\bigr] \hat{h}
+\ord\bigl(\zeta^{-3}\bigr),
\\[.5em]
\rlap{$
X_{Hh} \,\Delta_h^{-1}
  \tr\left[\textstyle X_{h\varphi}\frac{\tau_a}{2}\right]
  2\tr\left[\textstyle\frac{\tau_a}{2} \Delta_{\varphi}^{-1} \frac{\tau_b}{2}\right]
  \tr\left[\textstyle X_{\varphi h}\frac{\tau_b}{2}\right]
  \Delta_h^{-1}  X_{hH}$}
\phantom{X_{Hh} \, \Delta_h^{-1} \, X_{hH} =}
\nn\\
={}&
-\frac{2 g_2^2 \MH^4 s_\alpha^2 }{p^6 v_2^2}
\,p^\mu p^\nu \tr\bigl[\hat{C}_\mu\hat{C}_\nu\bigr] \hat{h}^2
 +\ord\bigl(\zeta^{-3}\bigr).
 \label{eq:XHhDhXhpDpXphDhXhH}
\end{align}
At this point the correspondence between the individual
terms in $\delta\L^{\text{1-loop}}_\eff$ \refeq{eq:LeffPi} and Feynman graphs in a diagrammatic
calculation is most obvious: The external lines of the diagrams are uniquely given by
the background fields contained in each monomial of $\delta\L^{\text{1-loop}}_\eff$,
the internal lines of the light fields $\phi_i$ originate from the factors
$\Delta_{\phi_i}$ with $\phi_i=h,\varphi,\dots$, and the heavy internal $H$ lines correspond to the
factors $1/(p^2-\MH^2)$. Note, however, that in general internal loop lines in diagrams lead to
sequences of powers of the corresponding propagators owing to the Taylor expansion for
$p_i\ll p,\MH$, where $p_i$ stands for external momenta represented by $\partial$
operators in $\delta\L^{\text{1-loop}}_\eff$.
Therefore, the terms in $\delta\L^{\text{1-loop}}_\eff$ actually correspond to the
individual terms of the Taylor-expanded Feynman diagrams in the hard momentum region.

Inserting the results of \eqsm{Delhinv}{XHhDhXhpDpXphDhXhH} into \eq{LeffPi}
effectively leads to
\begin{align}
\Pi^{(0)} ={}&
\left(\lambda_{12}+\frac{\MH^4 s_\alpha^2}{p^2 v_2^2}\right) \hat{h}^2
+ \left(\frac{2 \MH^2 s_\alpha^2}{v_2}-4 \lambda_{12} v_2\right) \hat{h}
+\frac{6 \lambda_{12} v_2}{s_\alpha}\, \hat{H}, \\
\Pi^{(2)} ={}&
-\frac{(D\!-\!4) g_2^2  \MH^4s_\alpha^2}{2 D\, p^4 v_2^2} \tr\bigl[\hat{C}^2\bigr] \hat{h}^2
-\frac{(D\!-\!4) g_2^2  \MH^2 s_\alpha^2}{D \,p^2 v_2} \tr\bigl[\hat{C}^2\bigr]\hat{h}
-\frac{(D\!-\!4) g_2^2s_\alpha^2}{2 D} \tr\bigl[\hat{C}^2\bigr]
\nn\\
&
+\frac{3 \MH^4s_\alpha^2 \left(\Mh^2+\MH^2s_\alpha^2\right)}{2 p^4 v_2^4}\, \hat{h}^4
+3 s_\alpha^2\biggl[\frac{ \Mh^2 \MH^4}{p^4 v_2^3}
+\frac{ \MH^2 \left(\Mh^2+\MH^2s_\alpha^2\right)}{p^2 v_2^3}
-\frac{2 \lambda_{12} \MH^2}{p^2
	v_2}\biggr]\hat{h}^3
\nn\\
&
+s_\alpha^2 \biggl[
\frac{\Mh^2	\MH^4}{p^4 v_2^2}
+\frac{4 \Mh^2 \MH^2-\!2\MH^4s_\alpha^2}{p^2 v_2^2}
+3\frac{\Mh^2+\!\MH^2s_\alpha^2}{2v_2^2}
+\lambda_{12} \Bigl(\frac{4 \MH^2}{p^2}-6\Bigr)
+\frac{6 \lambda_{12}^2 v_2^2}{\MH^2s_\alpha^2}\biggr]\hat{h}^2
\nn\\
&
+\left[2 \lambda_{12} v_2 \left(s_\alpha^2-\frac{3
	\Mh^2}{\MH^2}\right)+\frac{s_\alpha^2 \left(\Mh^2-\MH^2s_\alpha^2\right)}{v_2}\right]
\hat{h}
+\frac{6 \lambda_{12}^2 v_2^2}{\MH^2s_\alpha^2}\hat{H}^2
\nn\\
&
+\biggl[6\lambda_{12}v_2 \biggl(\frac{\Mh^2}{\MH^2s_\alpha}-s_\alpha \biggr)
+\frac{3 \MH^2s_\alpha^3}{v_2}\biggr]\hat{H}
+ \biggl(\frac{\MH^6s_\alpha^3}{p^4 v_2^3}+\frac{4 \lambda_{12} \MH^2s_\alpha}{p^2
	v_2}\biggr) \hat{h}^2 \hat{H}
\nn\\
&
+ \left[\frac{4 \MH^4s_\alpha^3}{p^2 v_2^2}+\lambda_{12}
\left(6s_\alpha-\frac{8 \MH^2s_\alpha}{p^2}\right)-\frac{12 \lambda_{12}^2 v_2^2}{\MH^2s_\alpha}\right]\hat{h}\hat{H}
+\frac{(D-4)  \MH^4s_\alpha^2}{D\,p^4 v_2^2}\, \hat{h}\Box \hat{h}\,,
\end{align}
under the integral over $p$
in \eq{LeffPi}, where we have already performed the tensor reduction of the
$p^\mu p^\nu$ terms, which for rank-2 vacuum integrals is achieved by
the replacement
\begin{align}
p^\mu p^\nu \;\to\; \frac{p^2}{D}\,g^{\mu\nu}.
\end{align}
The loop integration over $p$ involves only the very simple vacuum integrals
\begin{align}
I_{ab} &= \frac{(2\pi\mu)^{4-D}}{\ri \pi^2} \!\int \! \rd^D p\; \frac{1}{(p^2 - \MH^2 +\ri 0)^a (p^2 +\ri 0)^b}
\nn\\
&=
(4\pi\mu^2)^{(4-D)/2} \, (-1)^{a+b} \,
\frac{\Gamma \bigl(\frac{D}{2}-b \bigr)
\Gamma \bigl(a+b-\frac{D}{2}\bigr)}{\Gamma (a) \Gamma \bigl(\frac{D}{2}\bigr)}\,
\MH^{D-2a-2 b},
\label{eq:MI}
\end{align}
which obey the useful relations
\begin{align}
I_{0b} {}= 0,
\qquad\quad
I_{11} &{}= I_{10}/\MH^2, &
I_{12} &{}= I_{10}/\MH^4,
\nn\\
I_{21} &{}= I_{20}/\MH^2 - I_{10}/\MH^4, &
I_{22} &{}= I_{20}/\MH^4 - 2I_{10}/\MH^6.
\end{align}
The integrals $I_{0b}$ vanish, because they are scaleless;
the other relations follow from partial fractioning.
We can thus express $\delta\L^{\text{1-loop}}_\eff$ solely in terms of
$I_{10}$ and $I_{20}$ and obtain
\begin{align}
\delta\L^{\text{1-loop}}_\eff ={}
 & \frac{1}{32 \pi^2} \Biggl\{
-\frac{(D-4) g_2^2 s_\alpha^2 }{2 D}I_{10}
\tr\bigl[\hat{C}^2\bigr] \biggl(1+\frac{\hat h}{v_2}\biggr)^2
+\frac{(D-4) s_\alpha^2 }{D\, v_2^2} I_{10}\, \hat{h}\Box\hat{h}
\nn\\
&
+\biggl[\frac{1}{2} \lambda_{12}^2 I_{20}
- \frac{\lambda_{12} s_\alpha^2}{v_2^2} (I_{10} - \MH^2 I_{20})
+ \frac{s_\alpha^2}{2v_2^4}
  \Big(3 \Mh^2  I_{10} + \MH^2 s_\alpha^2  (I_{10} + \MH^2 I_{20} ) \Big)
\biggr] \hat{h}^4
\nn\\
&
- \biggl[
 4\lambda_{12}^2 v_2  I_{20}
+ \frac{2 \lambda_{12} s_\alpha^2}{v_2} (I_{10} + \MH^2 I_{20})
\nn\\
& \qquad
- \frac{s_\alpha^2}{v_2^3} \Big(6\Mh^2 I_{10}
   + \MH^2 s_\alpha^2 (I_{10} + 2\MH^2 I_{20} ) \Big)
\biggr] \hat{h}^3
\nn\\
&
+ \biggl[
 \frac{2 \lambda_{12}^2 v_2^2}{\MH^2} (3 I_{10}+4\MH^2 I_{20})
- \lambda_{12} \Big(8 \MH^2 s_\alpha^2 I_{20} - (1 - 2 s_\alpha^2)I_{10} \Big)
\nn\\
& \qquad
+ \frac{s_\alpha^2}{2v_2^2} \Big(13\Mh^2  I_{10}
  + 4 \MH^4 s_\alpha^2 I_{20}
  +  \MH^2 (2 - s_\alpha^2)I_{10} \Big)
\biggr] \hat{h}^2
\nn\\
&
+ \biggl[ \frac{s_\alpha^2}{v_2} \Big(\Mh^2 + \MH^2 (2 - s_\alpha^2)\Big)
 - \frac{2 \lambda_{12}v_2}{\MH^2} \Big(3\Mh^2 + \MH^2(2- s_\alpha^2)\Big)
\biggr] I_{10} \hat{h}
\nn\\
&
+ \frac{6\lambda_{12}^2 v_2^2 }{\MH^2 s_\alpha^2}
  (I_{10} + 3\MH^2 I_{20} ) \hat{H}^2
+ \frac{3}{v_2} \biggl[ \MH^2 s_\alpha^3
  + \frac{2 \lambda_{12} v_2^2}{\MH^2 s_\alpha}
  \Big(\Mh^2 + \MH^2 (1 - s_\alpha^2)\Big)
\biggr] I_{10} \hat{H}
\nn\\
&
+ \frac{1}{s_\alpha v_2^3 } \Bigl[
6 \lambda_{12}^2 v_2^4  I_{20}
- 2 \lambda_{12}  s_\alpha^2 v_2^2 (I_{10} - 3 \MH^2 I_{20})
+\MH^2 s_\alpha^4 I_{10}
\Bigr] \hat{h}^2 \hat{H}
\nn\\
&
+ \frac{2}{\MH^2 s_\alpha v_2^2} \,
(\MH^2 s_\alpha^2 - 2 \lambda_{12} v_2^2)
  \Bigl[2\MH^2 s_\alpha^2  I_{10}
   + 3\lambda_{12} v_2^2  ( I_{10} + 2 \MH^2 I_{20} ) \Bigr]
 \hat{h} \hat{H} \Biggr\}
\nn\\
&
+\ord\bigl(\zeta^{-2}\bigr),
\label{eq:Leffi}
\end{align}
Upon inserting
\begin{align}
I_{10} = \MH^2(L_\eps+1) + \ord(\eps), \qquad
I_{20} = L_\eps + \ord(\eps),
\label{eq:I10}
\end{align}
with
\begin{equation}
L_\eps = \Delta +\ln \biggl(\frac{\mu^2}{\MH^2} \biggr),
\qquad
\Delta = \frac{1}{\eps}-\gamma_{\mathrm{E}}+ \ln(4\pi),
  \label{eq:Leps}
\end{equation}
and expanding in $\eps=(4-D)/2$ we have
\begin{align}
\delta\L^{\oneloop}_{\eff} &=
\frac{1}{32 \pi^2}\Biggl\{
\frac{1}{4} g_2^2 \MH^2 s_\alpha^2 \tr\bigl[\hat{C}^2\bigr]
\biggl(1+\frac{\hat h}{v_2}\biggr)^2
-\frac{\MH^2 s_\alpha^2}{2 v_2^2}\, \hat{h}\Box\hat{h}
\nn\\
&
+ \bigg[\frac{\lambda_{12}^2}{2}L_\eps
+\frac{3  \Mh^2 \MH^2 s_\alpha^2}{2 v_2^4} (L_\eps+1)
+\frac{ \MH^4 s_\alpha^4}{2 v_2^4} (2 L_\eps+1)
-\frac{\lambda_{12} \MH^2 s_\alpha^2}{v_2^2}
\biggr]\hat{h}^4
\nn\\
&
+ \biggl[
\frac{6 \Mh^2 \MH^2 s_\alpha^2}{v_2^3}(L_\eps+1)
+\frac{\MH^4 s_\alpha^4}{v_2^3}(3 L_\eps+1)
-\frac{2 \lambda_{12} \MH^2 s_\alpha^2}{v_2}(2 L_\eps+1)
-4 \lambda_{12}^2 v_2L_\eps
\biggr] \hat{h}^3
\nn\\
&
+ \biggl[
   \frac{13\Mh^2 \MH^2 s_\alpha^2}{2 v_2^2}  (L_\eps+1)
  +  \frac{\MH^4 s_\alpha^2}{2 v_2^2}
   \left( 3 s_\alpha^2L_\eps + 2L_\eps -s_\alpha^2+2\right)
  \nn\\
  &\quad
  +\lambda_{12} \MH^2 \left(1-10  s_\alpha^2 L_\eps +L_\eps-2 s_\alpha^2\right)
  + 2\lambda_{12}^2 v_2^2 (7 L_\eps+3)
\biggr]\hat{h}^2
\nn\\
&
+\biggl[ \frac{\MH^2 s_\alpha^2}{v_2}  \left(\Mh^2+(2-s_\alpha^2) \MH^2 \right)
 (L_\eps+1)
 -2
  \lambda_{12} v_2 \left(3 \Mh^2 + (2-s_\alpha^2) \MH^2  \right)(L_\eps+1)
  \biggr]\hat{h}
\nn\\
&
+\frac{6\lambda_{12}^2 v_2^2}{s_\alpha^2} (4 L_\eps+1) \hat{H}^2
+ 3\biggl[\frac{2
   \lambda_{12} v_2}{s_\alpha} \left(\Mh^2+ (1-s_\alpha^2) \MH^2 \right)
   +\frac{\MH^4 s_\alpha^3}{v_2}
\biggr](L_\eps+1) \hat{H}
\nn\\
&
+\biggl[
\frac{ \MH^4 s_\alpha^3}{v_2^3} (L_\eps+1)
 +\frac{2 \lambda_{12} \MH^2 s_\alpha}{v_2}  (2 L_\eps-1)
+\frac{6  \lambda_{12}^2 v_2}{s_\alpha} L_\eps
\biggl] \hat{h}^2 \hat{H}
\nn\\
&
+\frac{2}{s_\alpha v_2^2} \, \bigl(\MH^2 s_\alpha^2-2\lambda_{12} v_2^2\bigr)
\Bigl[ 2\MH^2 s_\alpha^2(L_\eps+1)+3\lambda_{12} v_2^2(3 L_\eps+1) \Bigr]
\hat{h} \hat{H}
\Biggr\}
\nn\\
&
 +\ord\bigl(\zeta^{-2} , \eps \bigr).
\label{eq:Leff1loop}
\end{align}
This expression represents the bare effective Lagrangian from integrating out heavy modes at one loop in unitary (background) gauge.
In order to bring $\delta\L^{\text{1-loop}}_\eff$ into a manifestly
gauge-invariant form, we can invert the Stueckelberg transformation in \eq{Stueckelberg} by replacing
\begin{equation}
\hat{C}_\mu \rightarrow \frac{\ri}{g_{2}}\hat{U}^\dagger \bigl(\hat{D}_\mu \hat{U}\bigr).
\end{equation}
We emphasize that the (seemingly non-decoupling) $\delta\L^{\text{1-loop}}_\eff$ must be properly renormalized, taking into account full-theory as well as EFT counterterms,
 before it can be used to compute physical observables.
We will come back to this point in \sec{ren}.

\subsection{Heavy Higgs equation of motion and lowest-order effective Lagrangian}
\label{se:LeffLO}

At the end of \refse{se:modeseparation}, we have already outlined how the final
effective Lagrangian $\L_\eff$
breaks up into different parts,
\begin{align}
\L_\eff = \L_{\mathrm{SM}}(\hat\phi_i,\phi_i) +
\delta\L_\eff^\tree(\hat\phi_i,\phi_i) +
\delta\L^{\text{1-loop}}_\eff(\hat\phi_i) +
\delta\L_\eff^{\mathrm{ct}}(\hat\phi_i),
\label{eq:Leffdecomp}
\end{align}
where all field arguments correspond to light field modes.
The arguments of the full SM Lagrangian $\L_{\mathrm{SM}}(\hat\phi_i,\phi_i)$
comprise all background and quantum fields of the SM, since all
SM particles can propagate along tree and loop lines in EFT Feynman diagrams.

The part $\delta\L_\eff^\tree(\hat\phi_i,\phi_i)$ of the
effective Lagrangian quantifies all lowest-order couplings between SM fields
that are induced by exchange of a heavy Higgs boson.
The terms in $\L_\eff^\tree= \L_\SM+\delta\L_\eff^\tree$
built from
background fields $\hat\phi_i$ only are sufficient for the construction of all tree-level diagrams contributing to Green functions up to some target order $\zeta^{-n}$.
The effective couplings in $\delta\L_\eff^\tree$ involving (SM) quantum fields
$\phi_i$ give rise to loop diagrams
that are related to
the small-momentum regions of full-theory loop diagrams
involving the quantum field $H$.
Note that most of the terms in $\delta\L_\eff^\tree$ depend on the
background and quantum fields only via their sum $\tilde\phi_i=\hat\phi_i+\phi_i$
by construction within the BFM.%
\footnote{The only parts of the BFM quantized full-theory Lagrangian that do not depend on the sum $\hat\phi_i+\phi_i$ of background and quantum fields are the gauge-fixing
Lagrangian of the quantum fields and the ghost Lagrangian.}
In this section we derive $\delta\L_\eff^\tree$. To this end,
we eliminate $\hat H_l$ and $H_l$ from the full SESM Lagrangian by solving the
EOM for the $H_l$~field in terms of a series in inverse powers of $\zeta$.
This is possible, since all derivatives $\partial$ acting on light field modes
scale as $\zeta^0$ are therefore $\zeta^{-1}$ suppressed compared to the heavy Higgs mass $\MH$.
The effects of the heavy field modes in hard loops,
where $\partial$ effectively
counts as $\zeta^1$, are contained in $\delta\L^{\text{1-loop}}_\eff(\hat\phi_i)$
constructed in the previous section.
The last contribution to the effective Lagrangian,
$\delta\L_\eff^{\mathrm{ct}}(\hat\phi_i)$,
which accounts for counterterm contributions from the renormalization of the
heavy-H-boson sector in the full and effective theory,
is constructed in the next section.

To derive the EOM for the light modes $\hat H_l$ and $H_l$, we start from the dependence
of the full-theory Lagrangian on light and SM field modes, which we summarize in a Lagrangian
dubbed $\L^\tree(\hat\phi_i,\phi)$. This part is given by
(with $\tilde{H}_{l} \sim \zeta^{-1}$)
\begin{align}
\L^\tree=\L_{\mathrm{SM}}
-\frac{s_{\alpha}^2\MH^2}{8v_2^2}\tilde{h}^4-\frac{1}{2}\MH^2\tilde{H}_{l}^2-\frac{s_{\alpha}\MH^2}{2v_2}\tilde{h}^2\tilde{H}_{l}+\mathcal{O}\left(\zeta^{-2}\right).
\label{eq:Ltree}
\end{align}
Here and in the following we suppress the subscript $l$ of the soft modes of light (SM) particles, which represent the degrees of freedom of the EFT.
Since $\L_{\mathrm{SM}}$ does not depend on $\hat H_l$ and $H_l$, the EOM resulting
from the variation of $H_l$ reads
\begin{align}
0 = \MH^2 \tilde{H}_l
+\frac{s_\alpha \MH^2}{2v_2}\tilde{h}^2+\mathcal{O}\left(\zeta^{-1}\right)
\label{eq:Hl-EOM}
\end{align}
with the straightforward solution
\begin{equation}
\tilde{H}_l=-\frac{s_\alpha}{2v_2}\tilde{h}^2+\mathcal{O}\left(\zeta^{-3}\right).
\label{eq:Hl-solution}
\end{equation}
Note that this result a posteriori confirms our counting $\tilde{H}_{l} \sim \zeta^{-1}$.
Inserting this solution back into $\L^\tree$ given in \eq{Ltree}
leads to
\begin{equation}
\L_\eff^\tree=\L_{\mathrm{SM}}+
\delta\L_\eff^\tree, \qquad
\delta\L_\eff^\tree=\mathcal{O}\left(\zeta^{-2}\right),
\label{eq:Lefftree}
\end{equation}
showing that there are no non-decoupling effects of the SESM with a heavy H~boson
at tree level in the weak-coupling scenario in \eq{weak-coupling-scenario}.
Note, however, that the individual Feynman rules of the
full theory do not all simply turn into their SM versions in this limit.
The non-standard $\tilde{h}^4$ coupling in $\L^\tree$, for instance, is rather
compensated by the leading contribution of the four-point interaction of
$\tilde h$~fields induced by tree-level
$\tilde H$~exchange, when the $\tilde H$~propagator
shrinks to a point and $\tilde H$ is effectively given by
\eq{Hl-solution}.

In order to obtain $\delta\L^{\text{1-loop}}_\eff$ in \eq{Leff1loop}
in terms of SM fields, we have
to eliminate
the light mode $\hat H_l$ of the heavy-Higgs background field, which
proceeds along the same lines as above using the EOM \refeq{eq:Hl-EOM}.
There are, however, two differences. Firstly, the dependence of the solution
on the quantum field $H_l$ is irrelevant and can be discarded at the
one-loop level, because these terms would only contribute as part of a second loop.
Secondly, the term proportional to $\hat H_l$
in $\delta\L^{\text{1-loop}}_\eff$ of \eq{Leff1loop}
has a prefactor scaling like $\zeta^3$, so that the solution for
$\hat H_l$ is needed to order $\zeta^{-3}$, i.e.\ the solution in
\eq{Hl-EOM} has to be supplemented by further terms.
This task is straightforward and yields
\begin{align}
\hat{H}_{l}={} &
-\frac{s_{\alpha}}{2v_2}\hat{h}^2
+\frac{s_{\alpha}}{2v_2\MH^2}\left(\Box-2\Mh^2
-2\lambda_{12}v_2^2+s_{\alpha}^2\MH^2\right)\hat{h}^2
\nn\\
&{}
-\frac{s_{\alpha}}{2v_2^2\MH^2} \left(\Mh^2-s_{\alpha}^2\MH^2+2\lambda_{12}v_2^2 \right)\hat{h}^3
-\frac{s_{\alpha}\lambda_{12}}{4v_2\MH^2}\hat{h}^4
\nn\\
& {}
+\frac{g_2^2 s_{\alpha}}{2\MH^2}\tr\bigl[\hat{C}^2\bigr](v_2+\hat{h})
+\mathcal{O}\bigl(\zeta^{-5}\bigr).
\label{eq:Hl-EOM-ord3}
\end{align}

For later convenience we also derive the EOM for the light
Higgs field $\tilde h$,
\begin{align}
0 ={} & \left(\Box+\Mh^2\right)\tilde h
- \frac{g_2^2}{2} \tr\bigl[\tilde C^2\bigr]  (v_2+\tilde h)
+ \frac{3\Mh^2}{2v_2} \,\tilde h^2
+ \frac{\Mh^2+\MH^2 s_\alpha^2}{2v_2^2}\, \tilde h^3
\nn\\
& {}
+ \frac{\MH^2 s_\alpha}{v_2}\, \tilde h \tilde H_l
+\ord(\zeta^{-2})
\nn\\
={} & \left(\Box+\Mh^2\right)\tilde h
- \frac{g_2^2}{2} \tr\bigl[ \tilde C^2\bigr]  (v_2+\tilde h)
+ \frac{3\Mh^2}{2v_2} \,\tilde h^2
+ \frac{\Mh^2}{2v_2^2} \,\tilde h^3
+\ord(\zeta^{-2}),
\label{eq:EOMh}
\end{align}
where the solution \eq{Hl-solution} has been inserted for
$\tilde H_l$ in the last line.

\section{Renormalization}
\label{se:ren}

\subsection{Renormalization of the SM}
\label{se:SMren}

\begin{sloppypar}
Of course, the one-loop renormalization of the SM is by now standard,
both in the conventional quantization formalism and in the BFM
(see e.g.\ \rcites{Denner:1994xt,Denner:1991kt,Denner:2019vbn}
and references therein).
As shown in the previous section,
the SM coincides with the EFT describing the large-$\MH$ limit of the SESM at
tree level in the leading order of the large-$\MH$ expansion.
To prepare ourselves for the renormalization of the SESM and
the  EFT, it is therefore instructive to first recall some aspects of the SM renormalization.
In the formulation below, we closely follow \rcite{Denner:2019vbn}
both conceptually and concerning notation and conventions for
field-theoretical quantities.

Before renormalization, the defining ``bare'' Lagrangian depends on
parameters whose physical meaning is obscure when they
are used to parametrize physical observables.
Likewise, the fields
occurring in the bare Lagrangian are, in general, not canonically
normalized. In order to introduce parameters
and fields with clear meaning and well-defined normalization, respectively,
the original ``bare'' quantities are split into
renormalized quantities and renormalization constants.
Denoting all bare quantities with subscript ``0'', we write
\begin{align}
c_{i,0} = c_i + \delta c_i, \qquad
\hat\phi_{i,0} =
\left(\delta_{ij} + \textstyle\frac{1}{2}\delta Z_{ij}\right) \hat\phi_j,
\label{eq:bareSMparams}
\end{align}
for generic parameters $c_i$ and background fields $\hat\phi_i$.
The renormalized parameters are denoted by $c_i$ and
the corresponding renormalization constants by $\delta c_i$.
The renormalization constants $\delta c_i$
are fixed by renormalization conditions in order to tie the
renormalized parameters to measurable quantities, which in turn
give them their precise physical meaning.
The choice of the field renormalization constants $\delta Z_{ij}$, on the other hand, is only a matter of convenience.
The matrix structure of the field renormalization constants $\delta Z_{ij}$
is conveniently determined
by demanding that (at least) at some specific momentum transfer the renormalized fields $\hat\phi_i$ do not mix.
The renormalization of the (virtual) quantum fields $\phi_i$ is not necessary.
\end{sloppypar}

Specifically, we perform the ``renormalization transformations"
for the relevant physical parameters in the SM as follows:
\begin{align}
e_0 = (1+\delta Z_e)e,
\quad
s_{\Pw,0} = \sw + \delta\sw,
\quad
M_{\PW,0}^2 = \MW^2 + \delta\MW^2,
\quad
M_{\Ph,0}^2 = \Mh^2 + \delta\Mh^2.
\end{align}
This fixes the renormalization of the gauge couplings $g_1,g_2$
and the parameters $\cw,\MZ,v_2$, which are related to the
gauge-boson masses by%
\footnote{Note that we write the parameters of the SM Higgs potential here with a subscript ``2", i.e.\ $v_2$, $\mu_2$, $\lambda_2$, whereas the Higgs field $h$ has no subscript in order to match the notation for the SM-like part of the SESM in view of the next sections.}
\begin{align}
g_1 = \frac{e}{\cw}, \qquad
g_2 = \frac{e}{\sw}, \qquad
\cw^2 = 1-\sw^2 = \frac{\MW^2}{\MZ^2}, \qquad
v_2 = \frac{2\MW}{g_2}.
\label{eq:SMparamrel}
\end{align}
These relations are valid for bare and renormalized quantities.
This, in particular, implies
\begin{align}
\delta v_2 ={} &
v_2\biggl( \frac{\delta\MW^2}{2\MW^2}+\frac{\delta\sw}{\sw}-\delta Z_e \biggr).
\label{eq:dv2}
\end{align}
The renormalization of the parameters $\mu_2^2$ and $\lambda_2$
of the Higgs potential depends on the scheme that is employed to
treat the SM tadpole parameter
\begin{align}
t_{\Ph,0}  = v_{2,0} \biggl(\mu_{2,0}^2 - \frac{1}{4} \lambda_{2,0}v_{2,0}^2 \biggr).
\label{eq:th-rentrafo}
\end{align}
The SM tadpole term $t_{\Ph,0} \tilde h_0$ in $\L_\mathrm{SM}$
is the term linear in the bare Higgs field
$\tilde h_0$, while
$v_{2,0}/\sqrt{2}$ is the constant contribution from the
bare Higgs doublet field.
The (squared) bare Higgs-boson mass is given by
\begin{align}
M^2_{\Ph,0} = -\mu_{2,0}^2+\frac{3}{4}\lambda_{2,0}v_{2,0}^2,
\label{eq:Mh0}
\end{align}
and for the renormalized Higgs parameters we adopt the renormalization conditions
\begin{align}
\mu_2^2 = \frac{\Mh^2}{2}, \qquad
\lambda_2 = \frac{2\Mh^2}{v_2^2}.
\end{align}
In order to determine the renormalization constants $ \delta\mu_2^2$ and $\delta\lambda_2$ in
\begin{align}
\mu_{2,0}^2 = \mu_2^2 + \delta\mu_2^2, \qquad
\lambda_{2,0} = \lambda_2^2 + \delta\lambda_2^2, \qquad
\end{align}
we still have to fix the tadpole parameter $t_\Ph$.
Similar to the descriptions of \rcite{Denner:2019vbn}
we use two different prescriptions in parallel:%
\footnote{Our description differs from the procedure described in
Sect.~3.1.6 of \rcite{Denner:2019vbn} by introducing the bare vev $v_{2,0}$.
In the FJTS our $v_{2,0}$ effectively plays the same role
as the parameter $v_0$ in \rcite{Denner:2019vbn} for the
FJTS; in the PRTS our $v_{2,0}$ corresponds to the PRTS parameter $\bar v$
of \rcite{Denner:2019vbn}.
The formal treatment described here seems somewhat more generic, but the
PRTS and FJTS schemes are fully equivalent to the ones of
\rcite{Denner:2019vbn}.}
\begin{itemize}
\item
{\it Parameter-renormalized tadpole scheme}
(PRTS)~\cite{Denner:1991kt}:
Demanding
that the renormalized vev $v_2$ corresponds to the true (corrected)
minimum of the Higgs potential implies that the renormalized tadpole
parameter vanishes,
\begin{align}
t_\Ph =t_{\Ph,0}-\delta t_\Ph = 0.
\end{align}
The tadpole renormalization constant
$\delta t_\Ph$ is then simply given by the bare tadpole parameter $t_{\Ph,0}$ in \eq{th-rentrafo},
\begin{align}
\delta t_\Ph  = t_{\Ph,0}
= v_{2,0} \left(\mu_{2,0}^2 - \frac{1}{4}\lambda_{2,0} v_{2,0}^2 \right)
= v_2 \left(\delta \mu_2^2
- \frac{1}{4}\delta \lambda_2 v_2^2
- \frac{1}{2}\lambda_2 v_2\delta v_2 \right).
\end{align}
Together with \eq{Mh0} this fixes $\delta\mu_2^2$ and $\delta\lambda_2$ in terms of
$\delta\Mh^2$, $\delta v_2$, and $\delta t_\Ph$.
\item
{\it Fleischer--Jegerlehner tadpole scheme}
(FJTS)~\cite{Fleischer:1980ub}:
The bare tadpole parameter is consistently set to zero, $t_{\Ph,0}=0$,
so that, according to \eq{th-rentrafo} $v_{2,0}=2\sqrt{\mu_{2,0}^2/\lambda_{2,0}}$,
 and no renormalization
of the tadpole parameter is performed.
The bare Higgs-boson mass is thus given by
\begin{align}
M^2_{\Ph,0} = 2\mu_{2,0}^2 = \frac{1}{2}\lambda_{2,0}v_{2,0}^2.
\end{align}
This directly fixes $\delta\mu_2^2$ and $\delta\lambda_2$, in terms of
$\delta\Mh^2$ and $\delta v_2$.
A tadpole counterterm
\begin{align}
\delta t_\Ph  = - \Mh^2\Delta v_\Ph
\label{eq:Deltav2}
\end{align}
is effectively generated by a field shift $\hat h\to\hat h+\Delta v_\Ph$ in the
Lagrangian, which does not affect physical observables.
\end{itemize}
In both schemes there is a term $\delta t_\Ph\hat h$
in the counterterm Lagrangian, and $\delta t_\Ph$ is chosen to
compensate explicit tadpole diagrams in Green functions, i.e.\
\begin{align}
\delta t_\Ph  = - T^{\hat h},
\end{align}
where $T^{\hat h}$ ($= \Gamma^{\hat h}$) denotes the
unrenormalized one-point vertex function
of the background Higgs field at one loop.
The tadpole renormalization constant $\delta t_\Ph$ also enters
many other contributions in the counterterm Lagrangian.
These terms depend on the tadpole scheme.
For the sake of compact notation we introduce the expressions
$\delta t_\Ph^\PRTS$ and $\delta t_\Ph^\FJTS$,
where $\delta t_\Ph^\PRTS$ equals $\delta t_\Ph$ only in the PRTS and
is zero in the FJTS, and vice versa.

The field renormalization can either be performed in
the basis of the gauge multiplets $\hat W_\mu, \hat B_\mu, \hat \Phi$
or in the basis spanned by the fields $\hat W^\pm_\mu, \hat A_\mu, \hat Z_\mu, \hat h$
that correspond to mass eigenstates.
For our purposes, the gauge field renormalization will not play a role.
In the following, the only relevant
field renormalization
transformation is the one
of the Higgs field,
which we formulate directly for~$\hat h$:
\begin{align}
\displaystyle
\hat h_0 = \left(1+\frac{1}{2}\delta Z_{\hat h\hat h}\right) \hat h.
\end{align}

The part of the counterterm Lagrangian $\delta\L_{\SM}^{\ct}$ that results
from the SM Higgs sector by the renormalization
transformations described above
is denoted $\delta\L_{\SM}^{\PH\ct}$ and
(in compact notation for both schemes)
given by
\begin{align}
\delta\L_{\SM}^{\PH\ct} ={} &
-\frac{1}{2}\delta\Mh^2\, \hat h^2 \biggl(
1 + \frac{\hat h}{v_2} + \frac{\hat h^2}{4 v_2^2} \biggr)
+ \delta v_2
\Biggl[ \frac{g_2^2}{2} \tr\bigl[\hat C^2\bigr]  (v_2+\hat h)
+  \frac{\Mh^2}{4 v_2^2} \,\hat h^3\biggl(2+\frac{\hat h}{v_2}\biggr)
\Biggr]
\nn\\
& {}
-\frac{1}{2} \delta Z_{\hat h\hat h} \hat h
\Biggl[ \Box \hat h
- \frac{g_2^2}{2} \tr\bigl[\hat C^2\bigr] (v_2+\hat h)
+ \Mh^2 \hat h
\biggl(1 + \frac{3 \hat h}{2v_2}+ \frac{\hat h^2}{2v_2^2} \biggr)
\Biggr]
+\delta t_\Ph \,\hat h
\nn\\
& {}
-\delta t_\Ph^\PRTS \,\frac{\hat h^3}{2 v_2^2}
\biggl( 1 + \frac{\hat h}{4 v_2} \biggr)
+\delta t_\Ph^\FJTS
\Biggl[
\frac{\hat h^2}{2 v_2} \biggl(3+\frac{\hat h}{v_2} \biggr)
- \frac{g_2^2}{2\Mh^2} \tr\bigl[\hat C^2\bigr] (v_2+\hat h)
\Biggr].
\label{eq:LSMct}
\end{align}

In the SM, the mass parameters for the W, Z, and Higgs bosons
are usually defined as on-shell (OS) masses, which determine the
locations of the poles in the respective propagators.
This fixes the mass renormalization constants according to
\begin{align}
\delta \MW^2 = \Sigma^{\hat W}_\rT(\MW^2), \qquad
\delta \MZ^2 = \Sigma^{\hat Z\hat Z}_\rT(\MZ^2), \qquad
\delta \Mh^2 = \Sigma^{\hat h}(\Mh^2),
\end{align}
where $\Sigma^{\cdots}_{(\rT)}(p^2)$ denotes the corresponding self-energy
(with ``T'' indicating its transverse part) for momentum transfer~$p$.
Following the conventions of \rcite{Denner:2019vbn},
at the one-loop
level $\Sigma^{\cdots}_{(\rT)}(p^2)$ includes the contributions from
one-particle-irreducible (1PI) loop diagrams,
explicit tadpole diagrams, as well as tadpole
coun\-ter\-terms, but no contributions from other
renormalization constants.
Note that according to \eq{SMparamrel} fixing $\delta\MW^2$ and $\delta\MZ^2$ also fixes $\delta \sw$.

We complement these OS renormalization conditions by the
OS condition for the electric charge $e$, where $\delta Z_e$
is fixed by requiring that $e$ does not receive any correction
in the Thomson limit, where a physical charged particle interacts
with a photon of vanishing momentum.
The explicit form of $\delta Z_e$,
which involves only loops of charged particles in the $AA$ and
$AZ$ propagators, will not be needed in the following, because
neutral Higgs bosons do not contribute to $\delta Z_e$
at one loop.
The explicit form of the field renormalization constants,
which we assume to be fixed in the OS renormalization scheme,
will not be required either.
Only their scaling properties
in the considered large-mass limit of the SESM will be relevant and are quoted below.

\subsection{Renormalization of the SESM}
\label{se:SESMren}

Renormalization schemes for the SESM were worked out in
\rcites{Kanemura:2015fra,Bojarski:2015kra,Denner:2017vms,Altenkamp:2018bcs,Denner:2018opp}
in different variants. We follow the proposals of
\rcites{Altenkamp:2018bcs,Denner:2018opp} which
employ the parameters $\MH$, $s_\alpha$, and $\lambda_{12}$
(or alternatively $\lambda_{1}$)
as independent parameters in the BSM sector of the model.
We apply the renormalization transformations
\begin{align}
M_{\PH,0}^2 = \MH^2 + \delta\MH^2, \qquad
s_{\alpha,0} = s_\alpha + \delta s_\alpha, \qquad
\lambda_{12,0} = \lambda_{12} + \delta\lambda_{12},
\end{align}
which are supplemented by the renormalization transformations
of the SM-like parameters described in \sec{SMren}.
In \rcites{Altenkamp:2018bcs,Denner:2018opp}
several conceptually different renormalization schemes
for the (sine of the) mixing angle ($s_\alpha$) are discussed:
\begin{itemize}
\item
$\MSbar$ renormalization~\cite{Altenkamp:2018bcs}
with the PRTS or FJTS for treating tadpoles,
\item
OS renormalization~\cite{Denner:2018opp}
based on the ratio of amplitudes with external $\Ph/\PH$ bosons
with the PRTS or FJTS for treating tadpoles,
\item
symmetry-inspired renormalization~\cite{Denner:2018opp}
based on rigid (global) and BFM gauge
invariance of the model.
\end{itemize}
The benefits and drawbacks of these schemes for the renormalization of $s_\alpha$ are discussed in \rcite{Denner:2018opp} in detail.
In the present paper, we focus on $\MSbar$ and OS renormalization.
In the OS scheme, the renormalization constant $\delta s_\alpha$
can be calculated from the field renormalization constants of the
$\hat h/\hat H$ system, which are introduced below, using
Eq.~(3.13) of \rcite{Denner:2018opp}.
The result for $\delta s_\alpha$ in the $\MSbar$ scheme can be obtained
from $\delta s_\alpha$ in the OS scheme upon taking only its ultraviolet (UV) divergent parts.
Explicitly, we have in these schemes
\begin{align}
\delta s_\alpha ={} &
\left\{\begin{array}{ll}
\ord(\zeta^{-1}) & \;\mbox{for the OS/PRTS and $\MSbar/\PRTS$ schemes,}
\\[.2em]
\displaystyle
-\frac{s_\alpha}{\Mh^2 v_2}\,T^{\hat h} +
\ord(\zeta^{-1}) & \;\mbox{for the OS/FJTS scheme,}
\\[.8em]
\displaystyle
-\frac{s_\alpha}{\Mh^2 v_2}\,T^{\hat h}\big|_\UV +
\ord(\zeta^{-1}) & \;\mbox{for the $\MSbar/\FJTS$ scheme,}
\end{array}\right.
\label{eq:dsa}
\end{align}
where the ``UV'' label
indicates that only UV-divergent parts proportional to
$\Delta$, as given in \eq{Leps},
are absorbed into $\delta s_\alpha$.
The $\MSbar$ renormalization constants can be deduced from the
corresponding OS counterparts upon dropping the UV-finite parts.
Here and in the following we only give explicit expressions for the terms in the
large-$\MH$ expansion that will be relevant for the final effective Lagrangian to $\ord(\zeta^0)$.

The tadpole contributions from the large-momentum region of all relevant one-loop tadpole diagrams
can be directly read off the linear Higgs field terms in \eq{Leffi}.
At leading order in the large-mass expansion the explicit expressions are
\begin{align}
T^{\hat h} =
-\frac{2\lambda_{12}v_2^2 - \MH^2 s_\alpha^2}{16 \pi^2 v_2} \, I_{10}
+ \ord(\zeta^0),
\qquad
T^{\hat H} =
\frac{3\lambda_{12}v_2}{16 \pi^2 s_\alpha} \,I_{10}
+ \ord(\zeta^1),
\label{eq:tadpoles}
\end{align}
for the background light ($\hat h$) and heavy Higgs ($\hat{H}$)
fields, respectively,
where $I_{10} =\ord(\zeta^2)$ is given by \eq{I10}.
The soft-momentum regions contribute to \eq{tadpoles}
only at $\ord(\zeta^0)$.
Like in the SM,
the SESM tadpole counterterms are fixed by
\begin{align}
\delta t_\Ph = -T^{\hat h}, \qquad
\delta t_\PH = -T^{\hat H},
\label{eq:dtSESM}
\end{align}
which applies both in the PRTS and FJTS.

In all SESM renormalization schemes considered here,
the mass $\MH$ is on-shell renormalized, and the
coupling parameter $\lambda_{12}$ (or $\lambda_{1}$) with the $\MSbar$ prescription.
Explicit results for $\delta\MH^2$ and $\delta\lambda_{12}$ can be
obtained in a straightforward way
(see also the explicit results in
\rcite{Altenkamp:2018bcs}), but for our purpose we actually only need
their scaling behaviour as $\MH\to\infty$, namely
\begin{equation}
\delta\MH^2 = \Sigma^{\hat H}(\MH^2) = \ord(\zeta^2),
\qquad
\delta\lambda_{12} = \ord(\zeta^0),
\label{eq:dMH}
\end{equation}
both in the PRTS and FJTS.

The renormalization constants of the SM-like parameters are
obtained in full analogy to their counterparts in the SM.
The required leading terms in the large-$\MH$ limit are
\begin{align}
\delta\Mh^2 ={} &
\frac{\MH^2 s_\alpha^2+\lambda_{12}v_2^2}{16\pi^2 v_2^2} I_{10}
+ \frac{3}{v_2}\, \delta t_\Ph^\FJTS
+\frac{s_\alpha}{v_2}\, \delta t_\PH^\FJTS
+\ord(\zeta^0),
\nn\\
\delta\MW^2 ={} &
\frac{g_2^2 v_2}{2\Mh^2}\,\delta t_\Ph^\FJTS
+ \ord(\zeta^0),
\qquad
\delta\sw = \ord(\zeta^0),
\qquad
\delta Z_e = \ord(\zeta^0).
\label{eq:SESMparrenconsts}
\end{align}
The first three terms of $\delta\Mh^2$ and the first term of $\delta\MW^2$ can also be directly read off from the $\hat{h}^2$ and $\tr[\hat{C}^2]$ terms of \eqs{Leffi}{LSMct}.

As for the field renormalization, only the two Higgs fields and their
mixing are of interest in the following,
\begin{equation}\label{eq:fieldren}
\begin{pmatrix} \hat H_0 \\ \hat h_0 \end{pmatrix}
= \begin{pmatrix} 1+\frac{1}{2}\delta Z_{\hat H\hat H} & \delta Z_{\hat H\hat h} \\
    \delta Z_{\hat h\hat H} &  1+\frac{1}{2}\delta Z_{\hat h\hat h} \end{pmatrix}
\begin{pmatrix} \hat H \\ \hat h\end{pmatrix}.
\end{equation}
For practical calculations, these field renormalization constants are fixed
by OS conditions, which guarantee that the particle residues in the diagonal propagators are
equal to one and that different field types do not mix on their mass shells;
the explicit prescription for calculating $\delta Z_{\dots}$ from the Higgs self-energies
can, e.g., be found in Eqs.~(4.8)--(4.11) of
\rcite{Altenkamp:2018bcs} (see also \rcite{Denner:2018opp}).
As a matter of fact, the explicit form of none of the
Higgs field renormalization constants $\delta Z_{ij}$
will be required for the calculation of the final effective Lagrangian
(not even their scaling behaviour as $\MH\to\infty$).
Nevertheless, it is helpful to know some of their leading terms in the large-$\MH$ expansion,
\begin{align}
\label{eq:dZHiggs}
\delta Z_{\hat H\hat H} ={} & \ord(\zeta^0), \qquad
\delta Z_{\hat H\hat h} =
\frac{2s_\alpha}{v_2\Mh^2} \, \delta t_\Ph^\FJTS
+ \ord(\zeta^{-1}),
\\
\delta Z_{\hat h\hat h} ={}& \ord(\zeta^0),
\qquad
\delta Z_{\hat h\hat H} =
-\frac{3\MH^2 s_\alpha}{16\pi^2 v_2^2} \Re\{B_0(\MH^2,0,0)\}
- \frac{2s_\alpha}{v_2\Mh^2} \, \delta t_\Ph^\FJTS
+ \ord(\zeta^{-1}).
\nn
\end{align}
The scalar two-point one-loop integral $B_0$ is defined as in
\rcites{Denner:1991kt,Denner:2019vbn} and given by
\begin{align}
B_0(p_1^2\!=\!\MH^2,0,0) =
L_\eps + 2 + \ri\pi + \ord(\epsilon).
\end{align}
The results in \eq{dZHiggs}
can be easily derived from the one-loop Higgs-boson
self-energies, as e.g.\ described in \rcites{Altenkamp:2018bcs,Denner:2018opp},
and by applying our power-counting in $\zeta$.
They can,
for instance, be used to derive the OS scheme
renormalization constant $\delta s_\alpha$ in \eq{dsa}
as suggested in \rcite{Denner:2018opp}.

Before we turn to the renormalization of the EFT and
the contribution of the SESM counterterm Lagrangian to the
effective Lagrangian, we comment on the use of a
non-linear parametrization of the Higgs doublet $\Phi$
and potential implications on the renormalization procedure.
In fact, great care is mandatory when adopting renormalization
schemes that have been designed for linear realizations of the Higgs doublet.
Note that vertex functions even with the same external field
content in general change by switching from a linear to a non-linear
Higgs realization.
This also concerns the
structure of UV divergences of vertex functions, and the differences
might be quite drastic.
In the non-linear Higgs realization, for instance,
the Higgs self-energy $\Sigma^{\hat h\hat h}(p^2)$,
involves UV-divergent terms proportional to $p^4$,
which cannot appear for linearly realized Higgs bosons.
This, in particular, implies that the ``renormalized" Higgs self-energy
$\Sigma^{\hat h\hat h}_\rR(p^2) = \Sigma^{\hat h\hat h}(p^2)
-\delta\Mh^2+\delta Z_{\hat h\hat h}(p^2-\Mh^2)$
is not UV~finite.
Of course, this does not spoil the UV~finiteness
of S-matrix elements, since the theory with non-linearly realized
Higgs doublet
is still renormalizable.
The compensation of UV~divergences
simply does not happen inside 1PI vertex functions (such as the self-energies)
after renormalization,
but results from a non-trivial conspiracy of the divergences between
different renormalized vertex functions.

As long as the same renormalization transformations in the linearly
and non-linearly realized
theories are used with the same
OS renormalization conditions, the resulting renormalized theories
are fully equivalent, because OS conditions make use of
properties of S-matrix elements that are independent of the nature of the Higgs
field realizations.
Thus, the OS renormalization~\cite{Denner:2018opp}
of the mixing angle $\alpha$ works in the SESM with linear or non-linear
Higgs realizations exactly in the same way.
More care is already needed for $\MSbar$ renormalization, where
the determination of the renormalization constant $\delta s_\alpha^{\MSbar}$
has to be carried out based on S-matrix elements in the
non-linear realization, while it is sufficient to
consider some appropriate 1PI vertex function in the linear
realization, as e.g.\  in \rcite{Altenkamp:2018bcs}.
A safe way
to determine $\delta s_\alpha^{\MSbar}$ is to take
the UV-divergent part of $\delta s_\alpha^{\OS}$.
On the other hand, the translation of the symmetry-inspired
BFM schemes of \rcite{Denner:2018opp} to the non-linearly realized
theory is non-trivial, because these
schemes are based on
properties of the UV structure of specific vertex functions, which
drastically differ from the ones in the non-linear realization.
We, therefore, do not consider these symmetry-inspired renormalization
schemes in this paper.
Of course, one possibility to apply these schemes
would be to integrate out the heavy Higgs field directly starting from the
linearly realized SESM Lagrangian.

\subsection{Renormalization of the EFT}
\label{se:EFTren}

In \sec{PI} we have integrated out the hard modes from the SESM at one-loop order. The result is the contribution to the effective Lagrangian given in \eq{Leff1loop}, which contains $1/\eps$ singularities of UV and infrared (IR) origin.
The UV divergences are absorbed by the ($\zeta$-expanded) SESM renormalization constants of the previous section, while the IR divergences correspond to UV divergences of the EFT (with opposite sign) and can thus be interpreted as part of the counterterms of the EFT.

In \sec{LeffLO} we have worked out the effective Lagrangian
$\L_\eff^\tree$ describing all tree-level effects of the SESM  and
found that to $\ord(\zeta^0)$ it has the same form as the SM Lagrangian, see \eq{Lefftree}.
Using bare parameters and fields in the original (full-theory)
Lagrangian in \eq{Ltree},
this procedure automatically includes the SESM counterterms and yields the effective Lagrangian $\L_\eff^{\tree,0}$.
The ``0" superscript indicates that the parameters and fields of $\L_\eff^{\tree,0}$
 are the bare quantities
of \eq{bareSMparams}, where the $\delta c_i$ and $\delta Z_i$ are the one-loop SESM renormalization constants expanded to sufficiently high order in $1/\zeta$.
As long as the one-loop SESM renormalization constants have the same large-$\MH$ scaling behaviour as
originally assumed for the associated renormalized  quantities,
there is no further contribution from SESM counterterms to the effective Lagrangian.
Note, in particular, that $\delta \MH^2$ in \eq{dMH} is eliminated at $\ord(\zeta^0)$ together with $\MH$ in $\L_\eff^\tree$ upon using the EOM in \eq{Hl-solution}.

In \eqss{dsa}{SESMparrenconsts}{dZHiggs} we have
observed, however, that, depending on the scheme, some of the renormalization constants are enhanced by positive powers of $\zeta$ compared to the scaling assumed for the corresponding renormalized quantities.
The associated SESM counterterms thus give rise to additional contributions to the effective Lagrangian, which we dub $\delta \L_\eff^{\SESM\ct}$. They are derived in the same way as \eq{Lefftree}, i.e.\ employing the
heavy-Higgs EOM, and also comprise the tadpole counterterms
of \eq{dtSESM} in analogy to \eq{LSMct}.
We find
\begin{align}
\delta \L_\eff^{\SESM\ct} ={} &
\delta\Mh^2\, \frac{3s_\alpha^2}{4v_2}
\hat h^3 \left( 1 + \frac{7\hat h}{6 v_2} +\frac{\hat h^2}{3 v_2^2} \right)
\nn\\
& {}
+\delta v_2\,\frac{s_\alpha^2}{4v_2^3}\,\hat h \left[
\hat h\,\Box \hat h^2
- g_2^2 v_2 \tr\bigl[\hat C^2\bigr] (v_2^2-\hat h^2)
-v_2 \Mh^2 \hat h^2 \left(3+ \frac{7\hat h}{v_2} + \frac{3\hat h^2}{v_2^2} \right)
\right]
\nn\\
& {}
+\frac{s_\alpha \delta s_\alpha}{4v_2^2}\, \hat h \left[
-\hat h\,\Box \hat h^2 - 2g_2^2 v_2 \tr\bigl[\hat C^2\bigr] (v_2 + \hat h)^2
+ v_2 \Mh^2 \,\hat h^2\left(6 + \frac{7\hat h}{v_2} + \frac{2\hat h^2}{v_2^2} \right)  \right]
\nn\\
& {}
+ \delta Z_{\hat h\hat H}\, \frac{s_\alpha}{4v_2} \,\hat h^2
\left[ \Box \hat h
- \frac{g_2^2}{2} \tr\bigl[\hat C^2\bigr] (v_2+\hat h)
+ \Mh^2 \hat h \left(1 + \frac{3 \hat h}{2v_2}+ \frac{\hat h^2}{2v_2^2} \right)
\right]
\nn\\
& {}
+\delta t_\Ph^\PRTS \,
\frac{s_\alpha^2}{v_2^2}\,\hat h^3
\left( 1 + \frac{17\hat h}{16 v_2} + \frac{\hat h^2}{4 v_2^2} \right)
\nn\\
& {}
+\delta t_\PH^\PRTS \,\frac{s_\alpha}{\MH^2} \biggr[
\frac{g_2^2}{2} \tr\bigl[\hat C^2\bigr]  (v_2+\hat h)
- \frac{2 \Mh^2 + (1\!-\!s_\alpha^2)\MH^2 + 2 \lambda_{12} v_2^2}{2 v_2} \hat h^2
\nn\\
& \quad {}
- \frac{2 \Mh^2 + (2\!-\!5s_\alpha^2) \MH^2 + 4 \lambda_{12} v_2^2}{4 v_2^2} \hat h^3
- \frac{(1\!-\!8 s_\alpha^2)\MH^2 + 2 \lambda_{12} v_2^2}{8 v_2^3} \hat h^4
+ \frac{\MH^2 s_\alpha^2}{4 v_2^4} \hat h^5
\biggl]
\nn\\
& {}
+\delta t_\Ph^\FJTS \,\frac{s_\alpha^2}{2v_2^2\Mh^2} \Biggl[
\hat h^2 \,\Box \hat h
+ \frac{g_2^2 v_2}{2} \tr\bigl[\hat C^2\bigr] (v_2+\hat h) (v_2 +3\hat h )
\nn\\
&\quad {}
- v_2\Mh^2 \hat h^2\left(
\frac{9}{2} + \frac{7\hat h}{v_2} + \frac{5\hat h^2}{2 v_2^2} \right)
\Biggr]
\;+\; \ord(\zeta^{-2}),
\label{eq:LeffSESMct}
\end{align}
where $\delta v_2$ obeys \eq{dv2} and, according to \eq{SESMparrenconsts}, scales like $\zeta^2$ in the $\FJTS$.
The other renormalization constants and tadpole counterterms in \eq{LeffSESMct} are given in the previous section.
Note that $\delta \L_\eff^{\SESM\ct}$ is independent of
$\delta Z_{\hat H\hat H}$, $\delta Z_{\hat H\hat h}$, and
$\delta t_\PH^\FJTS=-\MH^2\Delta v_\PH$ in analogy to \eq{Deltav2}.
This is true to any order in $1/\zeta$,
because these renormalization constants are connected to field redefinitions
of the heavy field $\hat H$, which is eliminated via its EOM.
In fact, the EOM effectively eliminates
the combination
$\hat H(1+\delta Z_{\hat H\hat H}/2)+\delta Z_{\hat H\hat h}/2\hat h+\Delta v_\PH$.
Similarly, the term in \eq{LeffSESMct} that is
proportional to $\delta Z_{\hat h\hat H}$
can be removed upon
using the EOM \refeq{eq:EOMh} for the field $\hat h$ (again the quantum part $h$ can be dropped here, because it would only contribute at two loops).
Recall that the use of the EOM for $\hat h$ changes off-shell parts of Green functions, but not S-matrix elements, so that predictions
for observables remain unaffected.

With \eqsn{Leff1loop}{Lefftree}{LeffSESMct} and the SESM renormalization constants of \sec{SESMren} we can now write down the complete ``bare'' effective Lagrangian:
\begin{align}
  \L_\eff &= \L_\eff^{\tree,0} + \delta\L^{\oneloop}_{\eff}
  + \delta \L_\eff^{\SESM\ct} \nn\\
  &= \L_\eff^\tree + \delta\L_\eff^{\oneloop,\mathrm{ren}} + \delta\L_\eff^\ct.
  \label{eq:Leff}
\end{align}
In the second line we have reshuffled the terms
contributing to $\L_\eff$ in such a way that
$\L_\eff^\tree$ equals \eq{Lefftree},
$\delta\L_\eff^{\oneloop,\mathrm{ren}}$ is finite,
and $ \delta\L_\eff^\ct$ consists of the one-loop EFT counterterms.
Although, the effective Lagrangian in \eq{Leff} is already suitable
for phenomenological studies at (fixed) NLO in the loop expansion,
this Lagrangian is ``bare'' in the sense that it explicitly includes the
UV-divergent counterterms (containing $1/\eps$ poles) required to render the one-loop corrections to physical observables based on $\L_\eff^{\tree}$ finite.
The one-loop contributions from hard momentum ($\sim \MH$) modes are encoded in
 $\delta\L^{\oneloop}_{\eff}$.

The exact form of $\L_\eff^\tree$, $\delta\L_\eff^{\oneloop,\mathrm{ren}}$, and $\delta\L_\eff^\ct$ is only unique after fixing a
renormalization scheme for the EFT.
In the course of
our derivation, the renormalization scheme for the parameters and fields of
the $\ord(\zeta^0)$ part of $\L_\eff^{\tree}$, which (here) equals $\L^\SM$,
 is initially inherited from the underlying renormalized full theory.
For instance, the masses in the EFT are, according to \sec{SESMren}, initially on-shell renormalized.
Note, however, that the associated counterterms in $\delta\L_\eff^\ct$ differ in general from the
 respective SESM counterterms in $\delta \L_\eff^{\SESM\ct}$,
because the contributions from large momentum modes cancel with terms in $\delta\L^{\oneloop}_{\eff}$.
The (bare) Wilson coefficients of the remaining (BSM-type) effective operators are initially composed of renormalized full-theory parameters.
The renormalization conditions for these Wilson coefficients are in general not predetermined by the chosen full-theory renormalization scheme, because the respective operators are usually not part of the full-theory Lagrangian.

Once the (bare) $\L_\eff$ in \eq{Leff} is derived, we can of course adopt any suitable renormalization scheme for the Wilson coefficients as well as for the SM-type parameters in the EFT by moving finite terms between
$\L_\eff^\tree + \delta\L_\eff^{\oneloop,\mathrm{ren}}$, and $\delta\L_\eff^\ct$.
In particular, to resum large logarithms ($\propto \ln \Mh/\MH$) via renormalization group equations (RGEs) one may want to choose a (modified)
minimal subtraction ($\MSbar$) scheme for the Wilson coefficients and couplings of the EFT.
In that case we write
\begin{equation}
  C_{i,0} = C_i(\mu_\rR) + \delta C_i(\mu_\rR),
\end{equation}
where $C_{i,0}$ is the already determined bare coefficient of some effective operator
and $\mu_\rR$ denotes the renormalization scale on which
the renormalized coefficient as well as its (one-loop) renormalization constant depend.
The boundary (matching) condition of the corresponding one-loop
RGE at a matching scale $\mu_\rM$  is then given by
\begin{align}
C_i(\mu_\rM) = C_{i,0} - \delta C_i(\mu_\rM),
\end{align}
where in the $\MSbar$ scheme $\delta C_i(\mu_\rM)$ equals the
(divergent) terms proportional to $\Delta$ in $C_{i,0}$,
with $\Delta$ as defined in \eq{Leps}.
The $\MSbar$ matching scale is identified with $\mu$ in \eq{Leps} and
$\mu_\rM \equiv \mu \sim \MH$ should be chosen to render the logarithms in \eq{Leff1loop} small.
In this way  a good convergence behaviour of the perturbative expansion of $C_i(\mu_\rM)$ is maintained.
Solving the one-loop RGE for $C_i(\mu_\rR)$ then resums logarithms $\propto (\ln \mu_R/\mu_\rM)^n$ at leading logarithmic order in
renormalization-group-improved perturbation theory.%
\footnote{The corresponding one-loop anomalous dimension matrix for all dimension-6 SMEFT operators was computed in \rcites{Grojean:2013kd,Elias-Miro:2013gya,Elias-Miro:2013mua,Jenkins:2013zja,Jenkins:2013wua,Alonso:2013hga,Alonso:2014zka}.}
In EFT computations of physical observables $\mu_\rR$ is fixed to a typical low-energy scale ($\sim \Mh$).

\subsection{Final form of the effective Lagrangian}
\label{se:FinalL}

Finally, to check the decoupling of all BSM effects, we have to
investigate whether
\begin{equation}
    \delta \L_\eff^\BSM \equiv  \L_\eff - \L_\SM^0 = \ord(\zeta^{-2}),
\end{equation}
where $\L_\SM^0$ denotes the ``bare" SM Lagrangian including appropriate one-loop counterterms such that $\L_\SM^0$ cancels all SM-type operators of $\L_\eff$.%
\footnote{This is, e.g., achieved by choosing the same renormalization scheme(s) for the SM parameters as for the corresponding full theory quantities and posing corresponding renormalization conditions.}
In particular, the part of the SM counterterm Lagrangian relevant here takes
the form of \eq{LSMct}.
Consequently,
$\delta\L_\eff^\BSM$ consists of non-SM-type operators only.

Using \eq{Leff} we can thus write
\begin{align}
  \delta \L_\eff^\BSM &=
  \big[\delta\L^{\oneloop}_{\eff} + \delta \L_\eff^{\SESM\ct} \big]^\BSM,
\end{align}
where $[\L]^\BSM$ returns only the non-SM-type operators in $\L$.
Subtracting from $\delta \L_\eff^{\SESM\ct}$ in \eq{LeffSESMct} a SM-type term of the form of \eq{LSMct} with appropriately
adjusted renormalization constants and tadpole counterterms we have
\begin{align}
  \big[\delta \L_\eff^{\SESM\ct} \big]^\BSM ={}&
\biggl(\delta v_2 - \frac{\delta t_\Ph^\FJTS}{\Mh^2}\biggr)
\frac{s_\alpha^2}{4 v_2^2} \,\hat h \biggl[
  -g_2^2 \,\tr\bigl[\hat C^2\bigr]
    (v_2+\hat h) (3 v_2+ \hat h)
  + \Mh^2 \, \hat h^2 \biggl(3-\frac{\hat h^2}{v_2^2}\biggr)
\biggr]
\nn\\
& {}
+ \left(3\delta t_\Ph^\PRTS + \delta t^\PRTS_\PH s_\alpha
        + \delta\Mh^2 v_2\right)
  \frac{s_\alpha^2}{8 v_2} \,
  \hat h^3  \left(6+\frac{7\hat h}{v_2}+\frac{2\hat h^2}{v_2^2}\right)
\nn\\
& {}
- \delta t_\Ph \, \frac{s_\alpha^2}{16 v_2^2} \, \hat h^3
  \left(20+\frac{25\hat h}{v_2}+\frac{8\hat h^2}{v_2^2}\right)
\nn\\
& {}
+ \left( \frac{\delta s_\alpha}{s_\alpha} - \frac{\delta v_2}{v_2} \right)
  \frac{s_\alpha^2}{4v_2^2}
  \left[ -\hat h^2\, \Box \hat h^2
  -2 g_2^2 v_2 \,\tr\bigl[\hat C^2\bigr] \hat h (\hat h + v_2)^2
  \phantom{\left(\frac{7\hat h}{v_2}\right)} \right.
\nn\\
& \qquad \left. {}
  + \Mh^2 v_2 \,\hat h^3
  \left(6+\frac{7\hat h}{v_2} + \frac{2\hat h^2}{v_2^2} \right)
  \right]
\;+\; \ord(\zeta^{-2}).
\label{eq:LHcteff}
\end{align}
Here we have eliminated the terms involving the
field renormalization constant $\de Z_{\hat h\hat H}$ by using the
EOM \refeq{eq:EOMh} for $\hat h$ as described below \eq{LeffSESMct}.
The remaining (SESM) renormalization constants appearing on the r.h.s.\
of \eq{LHcteff} are grouped
in a way that makes the simultaneous use of the PRTS and FJTS
particularly simple. The first term in \eq{LHcteff} does not contribute
at the considered order ($\zeta^0$) at all, because in $\delta v_2$
as derived from \eqs{dv2}{SESMparrenconsts}
only the term $\delta t_\Ph^\FJTS/\Mh^2$ contributes
at $\ord(\zeta^2)$ which is cancelled by the explicit
$\delta t_\Ph^\FJTS$ term.
In the second term on the r.h.s.\ of \eq{LHcteff},
the explicit PRTS tadpole terms and the FJTS tadpoles implicitly contained in
$\delta\Mh^2$ combine with the 1PI parts of $\delta\Mh^2$ exactly
in the same way,
so that the overall contribution of the second line
is independent of the tadpole scheme.
The third term is given by the same
tadpole term in the PRTS and FJTS.
Only the last term on the r.h.s.\ of \eq{LHcteff},
which involves $\delta s_\alpha$, depends on the tadpole
scheme as well as on the
renormalization scheme chosen for the (sine of the) mixing angle $\alpha$ in the SESM.
Combining $\delta v_2$ with the scheme-dependent results for $\delta s_\alpha$
given in \eq{dsa}, we find that
$(\delta s_\alpha/s_\alpha - \delta v_2/v_2)s_\alpha^2$
vanishes at $\ord(\zeta^0)$ in all but the $\MSbar/\FJTS$ scheme,
where it is proportional to the UV-finite part of
$s_\alpha^2 T^{\hat h}/(\Mh^2 v_2)$.

To simplify the final step towards the effective Lagrangian at
$\ord(\zeta^0)$, we now insert the explicit ($\zeta$-expanded) expressions
for the SESM renormalization constants and tadpole counterterms into \eq{LHcteff}
everywhere but in the last term to obtain
\begin{align}
\big[\delta \L_\eff^{\SESM\ct} \big]^\BSM ={} &
- \frac{\MH^2 s_\alpha^2}{64\pi^2 v_2^3}
\left(\MH^2 s_\alpha^2 - 2\lambda_{12} v_2^2\right) \,
\hat h^3 \left(1+\frac{3\hat h}{4v_2}\right)
\left(L_\eps + 1 \right)
\nn\\
& {}
+ \left( \frac{\delta s_\alpha}{s_\alpha} - \frac{\delta v_2}{v_2} \right)
  \frac{s_\alpha^2}{4v_2^2}
  \left[ -\hat h^2\, \Box \hat h^2
  -2 g_2^2 v_2 \,\tr\bigl[\hat C^2\bigr] \hat h (\hat h + v_2)^2
  \phantom{\left(\frac{7\hat h}{v_2}\right)} \right.
\nn\\
& \qquad \left. {}
  + \Mh^2 v_2 \,\hat h^3
  \left(6+\frac{7\hat h}{v_2} + \frac{2\hat h^2}{v_2^2} \right)
  \right]
\;+\; \ord(\zeta^{-2}).
\label{eq:LHcteff2}
\end{align}
Now, adding $\delta\L^{\oneloop}_{\eff}$ of \eq{Leff1loop} and again dropping terms
that can be absorbed in the SM counterterm in \eq{LSMct}, we end up with
\begin{align}
  \delta \L_\eff^\BSM ={}&
  \left( \frac{\delta s_\alpha}{s_\alpha} - \frac{\delta v_2}{v_2} \right)
  \frac{s_\alpha^2}{4v_2^2}
  \Bigg[ -\hat h^2\, \Box \hat h^2
  -2 g_2^2 v_2 \,\tr\bigl[\hat C^2\bigr] \hat h (\hat h + v_2)^2
\nn\\
  &+ \Mh^2 v_2 \,\hat h^3
  \left(6+\frac{7\hat h}{v_2} + \frac{2\hat h^2}{v_2^2} \right)
  \Bigg]
\;+\; \ord(\zeta^{-2}),
\end{align}
with the SESM renormalization constants $\delta s_\alpha$ as given in \eq{dsa} and
\begin{equation}
  \delta v_2 =  v_2 \frac{\delta\MW^2}{2\MW^2} + \ord(\zeta^0)
  = \frac{\delta t_\Ph^\FJTS}{\Mh^2} + \ord(\zeta^0)
\end{equation}
according to \eqs{dv2}{SESMparrenconsts}.
Hence, we observe decoupling of the heavy Higgs boson~H in
the SESM for $\MH\to\infty$, i.e.
\begin{align}
\delta\L_\eff^\BSM ={} & \ord(\zeta^{-2}), \quad
\mbox{for the schemes OS/PRTS, OS/FJTS, and $\MSbar$/PRTS},
\end{align}
but non-decoupling in the $\MSbar$/FJTS scheme,
\begin{align}
\delta\L_\eff^\BSM \big|_{\MSbar/\FJTS} ={} &
\frac{\MH^2 s_\alpha^2(2\lambda_{12}v_2^2-\MH^2 s_\alpha^2)}{64\pi^2 \Mh^2v_2^4} \,
\left[ \ln\left(\frac{\mu_\rM^2}{\MH^2}\right)+1\right]
\nn\\
& \quad {} \times
  \left[ \hat h^2\, \Box \hat h^2
  +2 g_2^2 v_2 \,\tr\bigl[\hat C^2\bigr]\hat h (\hat h + v_2)^2
  - \Mh^2 v_2 \,\hat h^3
  \left(6+\frac{7\hat h}{v_2} + \frac{2\hat h^2}{v_2^2} \right)
  \right]
\nn\\
& {}
+ \ord(\zeta^{-2}).
\label{eq:LeffMSFJTS}
\end{align}
Here we have used \eqs{tadpoles}{dtSESM},
and identified the reference scale $\mu$
of dimensional regularization with the  $\MSbar$ renormalization scale of the full theory, which is in turn interpreted as the matching scale $\mu_\rM$ of the EFT (not to be confused with the EFT renormalization scale $\mu_R$).
Accordingly, the  $\MSbar$ renormalized mixing angle of the SESM depends on this scale, i.e.\ $s_\alpha\equiv \bar s_\alpha(\mu_\rM^2)$ in \eq{LeffMSFJTS}.

At first sight, the explicit appearance of the renormalization scale $\mu_\rM$
in $\delta\L_\eff|_{\MSbar/\FJTS}$ of \eq{LeffMSFJTS} seems odd, because it
potentially appears in NLO corrections to observables without being compensated
by some implicit $\mu_\rM$ dependence in LO contributions, since the
tree-level effective Lagrangian is just $\L_\SM$ to $\ord(\zeta^0)$
and thus independent of $s_\alpha$.
To resolve this puzzle, we have to remember that our weak-coupling
scenario in \eq{weak-coupling-scenario} assumes that
$s_\alpha\equiv \bar s_\alpha(\mu_\rM^2)=\ord(\zeta^{-1})$.
Here we have emphasized that the renormalized parameter $s_\alpha$, in which the
perturbative expansion works, is a running parameter $\bar s_\alpha(\mu_\rM^2)$
tied to the SESM renormalization scale $\mu_\rM$.
The running of $\bar s_\alpha(\mu_\rM^2)$ follows from the
$\mu_\rM$ independence of the bare parameter $s_{\alpha,0}$ and the
UV divergence of the renormalization constant in the SESM,
\begin{equation}
\frac{\partial \bar s_\alpha(\mu_\rM^2)}{\partial\ln\mu_\rM^2} = \beta_{s_\alpha},
\label{eq:saRGE}
\end{equation}
where the one-loop $\beta$-function
of $s_\alpha$ in the $\MSbar$/FJTS scheme can be directly read off \eq{dsa}:
\begin{equation}
\beta_{s_\alpha}\big|_{\MSbar/\FJTS} =
\frac{\partial \delta s_\alpha}{\partial\Delta} =
\frac{\MH^2 s_\alpha(2\lambda_{12}v_2^2-\MH^2 s_\alpha^2)}{16\pi^2 \Mh^2v_2^2}
+ \ord(\zeta^{-1}).
\end{equation}
Solving  \eq{saRGE} iteratively to NLO in the loop ($\lambda_{12} \sim g_2^2$)
expansion for $\MH\to\infty$, we find
\begin{equation}
\bar s_\alpha(\mu_\rM^2)\big|_{\MSbar/\FJTS}
={}
\bar s_\alpha(\hat\mu_\rM^2)
+ \frac{\MH^2\bar  s_\alpha(\hat\mu_\rM^2)
\bigl(2\lambda_{12}v_2^2-\MH^2 \bar s_\alpha(\hat\mu_\rM^2)^2\bigr)}{16\pi^2 \Mh^2v_2^2}
\ln\biggl(\frac{\mu_\rM^2}{\hat\mu_\rM^2}\biggr)
+ \ord(\zeta^{-1}),
\label{eq:NLOsol}
\end{equation}
where $\hat\mu_\rM$ can take an arbitrary value different from $\mu_\rM$.
From \eq{NLOsol} we see that if we start from a specific renormalization scale $\mu_\rM$
for which the assumption $\bar s_\alpha(\mu_\rM^2)=\ord(\zeta^{-1})$ holds,
this assumption is not fulfilled for $\bar s_\alpha(\hat\mu_\rM^2)$ anymore
if $\mu_\rM\ne\hat\mu_\rM$.
Thus, if we want to change the SESM renormalization scale,
we have to take into account the terms in $\L^\tree_\eff$ that are promoted from $\ord(\zeta^{-2})$ to $\ord(\zeta^0)$, when substituting \eq{NLOsol} for $s_\alpha$ and counting the one-loop correction to $s_\alpha(\mu_\rM^2)$ as $\ord(\zeta^1)$. Note that two-loop terms have to be dropped consistently after the replacement.
The relevant terms in $\L^\tree_\eff$
are derived in a straightforward
way following \sec{LeffLO} and read (before the substitution)
\begin{align}
\delta\L_\eff^\tree ={} &
-\frac{\bar s_\alpha(\mu_\rM^2)^2}{8v_2^2}
  \left[ \hat h^2\, \Box \hat h^2
  +2 g_2^2 v_2 \,\tr\bigl[\hat C^2\bigr]\hat h (\hat h + v_2)^2
  - \Mh^2 v_2 \,\hat h^3
  \left(6+\frac{7\hat h}{v_2} + \frac{2\hat h^2}{v_2^2} \right)
  \right]
  \nn\\
&+\ldots,
\end{align}
where the ellipses refer to terms that  are of $\ord(\zeta^{-2})$ even after incorporating the enhanced one-loop correction of \eq{NLOsol}.
Combining $\delta\L_\eff^\tree$ with the
one-loop part of $\delta\L_\eff$ given in \eq{LeffMSFJTS},
we can now write the final effective Lagrangian to $\ord(\zeta^0)$ in the form
\begin{align}
\delta\L_\eff^\BSM\big|_{\MSbar/\FJTS} ={} &
\frac{\bar s_\alpha(\MH^2)^2}{8v_2^2}
\left[ -1 + \frac{\MH^2 \bigl(2\lambda_{12}v_2^2-\MH^2 \bar s_\alpha(\MH^2)^2\bigr)}{8\pi^2 \Mh^2v_2^2} \right]
\nn\\
& \quad {} \times
  \left[ \hat h^2\, \Box \hat h^2
  +2 g_2^2 v_2 \,\tr\bigl[\hat C^2\bigr]\hat h (\hat h + v_2)^2
  - \Mh^2 v_2 \,\hat h^3
  \left(6+\frac{7\hat h}{v_2} + \frac{2\hat h^2}{v_2^2} \right)
  \right]
\nn\\
& {}
+ \ord\big(\zeta^{-2}\big),
\label{eq:LeffMSFJTS2}
\end{align}
which is renormalization/matching scale independent at
one-loop order.
To compactify the result, we have set $\hat\mu_\rM=\MH$.
Note that in \eq{LeffMSFJTS2}, and accordingly in \eq{NLOsol}, the loop and large-mass expansions are intertwined in the sense that one should treat
$\bar s_\alpha(\hat\mu_\rM^2)^n$ to be of $\ord(\zeta^{2-n})$ or $\ord(\zeta^{-n})$ when it appears in a tree-level or a one-loop term, respectively.
This non-uniform scaling behaviour of $s_\alpha$ in the $\MSbar$/FJTS scheme continues at higher loop orders and is particularly problematic when it comes to resummation in the EFT (for quantities where $s_\alpha$ appears in the anomalous dimension).
Concerning \eq{LeffMSFJTS2}, we can loosely speaking say
that adding the one-loop corrections to the effective Lagrangian
in the $\MSbar$/FJTS scheme effectively changes the scale at which $\bar s_\alpha$ is evaluated from $\mu_\rM$, where $\bar s_\alpha(\mu_\rM^2)$ is strongly suppressed,
to $\hat\mu_\rM=\MH$, where $\bar s_\alpha(\MH^2)$ is enhanced by one-loop corrections of $\ord(\zeta^1)$.

Using the EOM of $\hat h$,
given in \eq{EOMh}, but with $\hat h$ interpreted as SM Higgs field (i.e.\ dropping again the quantum part of $\tilde h$),
and absorbing some terms into the SM renormalization constants in $\L^{\Hct}_\SM$,
the operator appearing in \eqs{LeffMSFJTS}{LeffMSFJTS2} can be rewritten as
\begin{equation}
\delta\L_\eff^\BSM\big|_{\MSbar/\FJTS} =
C_{\Phi\Box}(\mu_\rM^2) \, Q_{\Phi\Box}
\,+\, \ord(\zeta^{-2}),
\label{eq:LeffMSFJTS3}
\end{equation}
where $Q_{\Phi\Box}$ is one of the SMEFT operators in the Warsaw
basis~\cite{Grzadkowski:2010es} usually written as
\begin{equation}
Q_{\Phi\Box} =
\frac{1}{4}\big(\tr\bigl[ \hat\Phi_\SM^\dagger\hat\Phi_\SM\bigr] \big)\Box
\big(\tr\bigl[ \hat\Phi_\SM^\dagger\hat\Phi_\SM\bigr] \big)
=
\big(\hat\phi_\SM^\dagger\hat\phi_\SM\big)\Box
\big(\hat\phi_\SM^\dagger\hat\phi_\SM\big).
\end{equation}
Here $\hat\Phi_\SM$ denotes the matrix-valued SM background Higgs field
and $\hat\phi_\SM$ is
 the corresponding two-component SM background
Higgs doublet field in the linear realization, i.e.
\begin{equation}
\hat\phi_\SM =
\begin{pmatrix} \hat\phi^+ \\ (v_2+\hat h+\ri\hat \chi)/\sqrt{2} \end{pmatrix}.
\end{equation}
Note that we write $\hat h$ instead of $\hat h_2$ for the SM Higgs field.
The background Goldstone-boson fields $\hat\phi^\pm$ and $\hat\chi$ are
defined as in \eq{Goldstonefields}.
According to \eq{LeffMSFJTS2} the Wilson coefficient $C_{\Phi\Box}(\mu_\rM^2)$ in
\eq{LeffMSFJTS3} is
\begin{align}
C_{\Phi\Box}(\mu_\rM^2) = {} &
-\frac{\bar s_\alpha(\mu_\rM^2)^2}{2v_2^2} +
\frac{\MH^2 \bar s_\alpha(\mu_\rM^2)^2
\left(2\lambda_{12}v_2^2-\MH^2 \bar s_\alpha(\mu_\rM^2)^2\right)}{16\pi^2 \Mh^2v_2^4} \,
\left[ \ln\left(\frac{\mu_\rM^2}{\MH^2}\right)+1\right]
+ \ord(\zeta^{-2})
\nn\\
= {} &
\frac{\bar s_\alpha(\MH^2)^2}{2v_2^2} \left[ -1 +
\frac{\MH^2\left(2\lambda_{12}v_2^2-\MH^2 \bar s_\alpha(\MH^2)^2\right)}{8\pi^2 \Mh^2v_2^2} \right]
+ \ord(\zeta^{-2}).
\end{align}
In this simple example, no particular effort was needed to bring the final
result into SMEFT form. In more complicated cases, we first have to
translate the final effective Lagrangian into a basis of gauge-invariant
operators upon inverting the Stueckelberg transformation in \eq{Stueckelberg}
as described in \rcites{Dittmaier:1995cr,Dittmaier:1995ee}.
In a second step EOMs for the light fields, in our case the SM fields,
can be used to bring all occurring operators in the effective Lagrangian
into canonical form, which is the SMEFT basis in our case.
We note that despite $Q_{\Phi\Box}$ being a dimension-six operator it formally contributes, because of its Wilson coefficient $C_{\Phi\Box}$, to the effective Lagrangian at $\ord(\zeta^0)$ in the $\MSbar$/FJTS scheme.
On the other hand, re-expressing $\bar s_\alpha(\MH^2)$ in terms of the on-shell renormalized $s_\alpha$ and consistently re-expanding to one-loop order renders $C_{\Phi\Box}$ to be of $\ord(\zeta^{-2})$.

\begin{sloppypar}
Comparing the different renormalization schemes, we have to conclude that the
$\MSbar$/FJTS scheme does not reflect the true nature of the $\MH\to\infty$ limit
of the SESM in a sound way. The decoupling behaviour observed
in the other schemes and at tree level
is broken at the one-loop level or, more precisely, decoupling at one loop
only happens
at the fine-tuned scale $\mu_\rM=\MH/\re$, with $\re$ being Euler's constant, where
$C_{\Phi\Box}(\mu_\rM^2)$ is of $\ord(\zeta^{-2})$.
The origin of this odd behaviour is the fact that the renormalization constant
$\delta s_\alpha$ and thus the one-loop contribution to $\bar s_\alpha(\mu_\rM^2)$
does not scale in the same way as
initially assumed for the bare (tree-level) parameter
$s_{\alpha,0}$ in the heavy-mass limit.
As a result, some NLO corrections in the SESM
tend to get unnaturally large in
the $\MSbar$/FJTS scheme for large $\MH$, so that this scheme
is not recommendable for the SESM with large $\MH$.
In particular, this scheme does not allow to retain $\bar s_\alpha(\mu_\rM^2)$
(without loop-order dependent fine-tuning)
as a parameter in the EFT describing the large-mass limit in a consistent way, because it obscures the power counting.%
\footnote{
    In \rcite{Boggia:2016asg} an on-shell renormalization scheme  was adopted for the SESM (with linear Higgs realization), so that the results are independent of the chosen (``$\beta_h$'') tadpole scheme.
    Hence, there are no subtleties connected to tadpole renormalization like the spurious $\ord(\zeta^0)$ non-decoupling terms in $\MSbar$/FJTS.
}
\end{sloppypar}

Finally, we note that the unpleasant behaviour of the $\MSbar$/FJTS scheme
is certainly not tied to the specific case of the SESM.
Artificially large corrections in the $\MSbar$/FJTS scheme for mixing angles
have, for instance, also been found in scenarios of the Two-Higgs-Doublet Models
with large Higgs-boson masses in \citeres{Krause:2016oke,Denner:2016etu,Altenkamp:2017ldc,Altenkamp:2017kxk}.
In general, this typically occurs when the full-theory loop expansion of an EFT parameter has non-uniform scaling behaviour in the heavy-mass limit.
Nevertheless a case-by-case study is always recommended in order to analyse the decoupling behaviour of each renormalization scheme.

\section{Conclusion}
\label{se:conclusion}

Building on earlier work, in this article we have
described a general procedure to integrate out heavy fields
directly in the path integral and to derive an effective
Lagrangian at the one-loop level.
The method is based on the {\it background-field formalism}, which
implies a natural separation of tree-level and loop effects of the
heavy fields, and on the {\it expansion by regions},
which further separates loop effects into contributions from
large and small momentum modes.
Combining these concepts, together with additional technical
tricks (non-linear Higgs realization, field redefinitions, EOMs, etc.),
lends the method some particular strengths:

\begin{itemize}
\item
{\it Transparency:}
The clear separation of tree-level and loop effects of the heavy fields
and the further decomposition of field modes into light and heavy
degrees of freedom render the procedure very transparent.
At every stage of our calculation it is possible to
identify the origin of all contributions to the effective Lagrangian
in terms of (combinations of)
Feynman diagrams.

\item
{\it Flexibility:}
The method for integrating out heavy fields is fully flexible in the sense
that no preknowledge of the low-energy effective theory is needed, i.e.\
no ansatz for the effective Lagrangian is made in advance.
The fields in the full theory just have to be divided into
sets of light fields, providing the dynamical degrees of freedom
at low energies, and heavy fields, which will be integrated out and
the effects of which will appear in effective operators composed of the
light fields. Besides that, given a large mass scale $\Lambda$,
a proper definition of the large-mass scenario
has to be specified by a power-counting scheme for all model parameters
in the limit $\Lambda\to\infty$.
\item
{\it Gauge invariance:}
In the the background-field method the gauge of the background fields,
which correspond to the fields on tree lines in Feynman diagrams,
can be fixed independent of the gauge of the quantum fields,
which are the fields appearing inside loops.
This feature can, for instance, be exploited to simplify the explicit calculation
of the effective Lagrangian, by choosing a specific background gauge
in intermediate steps and restoring gauge invariance at the end.
This proves particularly powerful in combination with a non-linear realization of the SM Higgs sector.
\item
{\it Algorithmic organization:}
The method is fully algorithmic and suitable for automation.
Given a properly defined large-mass scenario and
some details on the renormalization
of the large-mass sector, the actual determination of the effective Lagrangian
at the one-loop level can, in principle, be carried out by computer algebra.
Recently, some steps in this procedure have already been automated, see \rcites{Cohen:2020qvb,Fuentes-Martin:2020udw}.
\end{itemize}

Compared to other related approaches described in the literature (such as the
UOLEA approach) for integrating out
heavy fields, our presentation might seem somewhat lengthy, but to a large extent this is
due to the fact that our formulation is very close to the actual NLO machinery used in
precision calculations in SM extensions. In the first place this means that we work in a field
basis corresponding to mass eigenstates by diagonalizing mass matrices involving heavy
fields before integrating out the heavy degrees of freedom. This procedure does not only
avoid doubts on the consistent treatment of mixing effects raised in the literature w.r.t.\
other approaches, it also very naturally prepares an appropriate framework to include
renormalization prescriptions that are designed for phenomenological analyses
(e.g.\ by taking mixing angles as independent parameters).

We have illustrated the method by considering
a singlet Higgs extension of the Standard Model in which a heavy Higgs boson~H
exists in addition to the known Standard-Model-like Higgs boson~h,
which is experimentally investigated at the LHC.
To be precise, we have calculated potential non-decoupling effects of H
in the limit $\MH\to\infty$, assuming a weak coupling scenario
in which the mixing angle $\alpha$ between H and the singlet scalar of the model
is suppressed by a factor $\sim\Mh/\MH$.
We have carried out our calculation in a field basis corresponding to
mass eigenstates, in order to avoid issues
in the mixing
between fields of light and heavy particles.
In the course of the calculation we have emphasized the issue of
renormalization of the non-standard sector of the theory---an
aspect that is widely ignored in the literature on the
construction of effective Lagrangians for heavy-particles effects.
Non-trivial contributions connected to renormalization
appear whenever model parameters and the corresponding
renormalization constants scale differently in the heavy-mass limit.
Spontaneously broken gauge theories with extended scalar sectors
are particularly prone to such issues,
because heavy Higgs-boson
masses often enhance scalar self-couplings.
In this context, the renormalization of vacuum expectation values
and corresponding tadpole contributions in the full SESM deserve particular care.
In the specific model with the heavy Higgs singlet~H we
observe for example full decoupling for $\MH\to\infty$ using an on-shell
renormalization scheme for the Higgs mixing angle $\alpha$. For commonly used $\MSbar$ renormalization schemes for $\alpha$, on the other hand,
we find decoupling or non-decoupling depending on the treatment
of tadpole contributions.
In the latter case the construction of a consistent EFT is problematic.

Owing to its transparent, flexible, and algorithmic structure
the method opens a vast field of applications.
The natural next step is to extend the calculation
of all heavy-Higgs effects associated with the singlet extension considered in this paper to order $1/\MH^2$
and the determination of the corresponding dimension-six SMEFT Lagrangian
for various renormalization schemes.
The effects of integrating out the heavy Higgs field on the fermionic sector of the SESM, neglected here for brevity, also remain to be analyzed in detail.
These tasks and some phenomenological applications will be addressed
in a forthcoming publication.

\section*{Acknowledgements}

We thank Michele Boggia for his collaboration in an early stage of this work.
S.D.\ gratefully acknowledges Giampiero Passarino for fruitful discussions on the
subject of effective field theories.

\bigskip

\appendix
\numberwithin{equation}{section}
\section{Evaluation of the functional determinant }
\label{app:FuncDetApp}

In this appendix we describe the evaluation of the functional
determinant in \eq{funcdet}.
Let us first introduce the Hilbert space version of the heavy-mode
projection operator in \eq{Tops}:
\begin{equation}
  \langle p| T_h |\phi\rangle \equiv \langle p |\phi_h\rangle   \equiv \mathcal{T}_{h}(p) \langle p| \phi\rangle =
  \mathcal{T}_{h}(p)\, \phi(p) = \phi_h(p).
\end{equation}
We now write the differential operator
$\tilde{\Delta}_H(x,\partial_x)$, which operates on functions
$\psi(x)$, $\phi(x)$ in Minkowski space, as matrix elements
of a linear operator $M$ acting on the elements $|\psi\rangle$, $|\phi\rangle$
of the corresponding Hilbert space. In the usual bracket notation, we
thus have
\begin{align}
|\phi\rangle ={}& M\,|\psi\rangle, \qquad \psi(x) = \langle x|\psi\rangle,
\qquad \phi(x) = \langle x|\phi\rangle,
\\
\phi(x) ={}& \int\rd^Dy\, \langle x|M|y\rangle\,\psi(y)
={} \tilde{\Delta}_H(x,\partial_x) \, \psi(x),
\end{align}
so that we can identify
\begin{equation}
\langle x|M|y\rangle
=  \tilde{\Delta}_H(x,\partial_x)\, \delta(x-y)
= \delta(x-y) \, \tilde{\Delta}_H(y,\partial_y).
\label{eq:Melement}
\end{equation}
The last relation is obtained via partial integration under the $y$-integral
and expresses the hermiticity of $\tilde{\Delta}_H$.
With this notation we can
replace the clumsy expression
$\mathpzc{Det}_h \big[\delta(x-y) \tilde{\Delta}_H(x,\partial_x) \big]$
in \eq{funcdet} by a more accurate one: $\mathpzc{Det}_h[M]$, which represents the functional determinant of the suboperator of $M$ that acts only on the subspace of hard-momentum states $|H_h\rangle$.
We then evaluate the 1-loop part of the EFT action as
\begin{align}
\mu^{D-4}\!\!
\int\rd^Dx\; \delta\L^{\text{1-loop}}_\eff
={}& \frac{\ri}{2} \ln\bigl(\mathpzc{Det}_h[M]\bigr)
={} \frac{\ri}{2}\, \mathpzc{Tr}_h\left[\ln(M) \right]
={} \frac{\ri}{2}\, \mathpzc{Tr}\left[T_h \ln(M) \right]
\nn\\
={}& \frac{\ri}{2}\int\!\frac{\rd^Dp}{(2\pi)^D}\, \langle p| T_h \ln(M) |p\rangle
=
\frac{\ri}{2}\int\!\frac{\rd^Dp}{(2\pi)^D} \, \mathcal{T}_{h}(p) \,\langle p| \ln(M) |p\rangle
\nonumber\\
={}& \frac{\ri}{2}\int\!\frac{\rd^Dp}{(2\pi)^D} \, \mathcal{T}_{h}(p) \int\!\rd^Dx \int\!\rd^Dy\,
\langle p|x\rangle\,\langle x |\ln(M)\,|y\rangle\,\langle y|p\rangle
\nonumber\\
={}&
\frac{\ri}{2}\int\!\frac{\rd^Dp}{(2\pi)^D}\, \mathcal{T}_{h}(p)  \int\!\rd^Dx \int\!\rd^Dy\;
\re^{-\ri px}\,\langle x |\ln(M)\,|y\rangle\,\re^{\ri py},
\label{eq:Leff_lnM}
\end{align}
where $\langle x|p\rangle=\re^{ipx}$ is the eigenfunction of the derivative operator $-\ri\partial^\mu_x$ with eigenvalue $p^\mu$ and $|p\rangle$ denotes the corresponding momentum eigenstate.
The matrix element $\langle x| \ln(M)\, |y\rangle$
is evaluated via the usual
power series of the logarithm of the operator~$M$,
which we express in terms of the deviation~$N$ from the unit operator~$\bbid$,
\begin{equation}
M \equiv \bbid-N, \qquad
\ln(M) = -\sum_{k=1}^\infty \frac{N^k}{k}.
\end{equation}
Writing the matrix element of $N$ according to \eq{Melement} as
\begin{equation}
\langle x|N|y\rangle = \delta(x-y) \, n(y,\partial_y), \qquad
\tilde{\Delta}_H=1-n,
\end{equation}
we have
\begin{align}
\langle x| \ln(M)\, |y\rangle &{}=
-\sum_{k=1}^\infty \frac{1}{k} \langle x|N^k\, |y\rangle
\nonumber\\
&{}=
-\sum_{k=1}^\infty  \frac{1}{k} \int\rd^Dx_1 \cdots \int\rd^Dx_{k-1}\,
\langle x|N\, |x_1\rangle \,\langle x_1|N\, |x_2\rangle
\cdots \langle x_{k-1}|N\, |y\rangle
\nonumber\\
&{}=
-\sum_{k=1}^\infty  \frac{1}{k} \int\rd^Dx_1 \cdots \int\rd^Dx_{k-1}\,
\langle x|N\, |x_1\rangle \, \delta(x_1-x_2) \, n(x_2,\partial_{x_2}) \cdots
\nonumber\\
&{}=
-\sum_{k=1}^\infty  \frac{1}{k} \langle x|N\, |y\rangle \, n(y,\partial_y)^{k-1}
= -\delta(x-y) \sum_{k=1}^\infty  \frac{n(y,\partial_y)^k}{k}
\nonumber\\
&{}=
\delta(x-y) \ln\left[1-n(y,\partial_y)\right]
= \delta(x-y) \ln\left[\tilde{\Delta}_H(y,\partial_y)\right].
\end{align}
Inserting this into \eq{Leff_lnM} we obtain for the 1-loop
effective Lagrangian
\begin{align}
\delta\L^{\text{1-loop}}_\eff(x)
={}&
\frac{\ri}{2}\,\mu^{4-D} \int\!\frac{\rd^Dp}{(2\pi)^D}\, \mathcal{T}_{h}(p)  \int\rd^Dy\,
\re^{-\ri px}\, \delta(x-y) \ln\left[\tilde{\Delta}_H(y,\partial_y)\right] \re^{\ri py}
\nonumber\\
={}&
\frac{\ri}{2}\,\mu^{4-D} \int\!\frac{\rd^Dp}{(2\pi)^D}\, \mathcal{T}_{h}(p) \;
\re^{-\ri px}\, \ln\left[\tilde{\Delta}_H(x,\partial_x)\right] \re^{\ri px}
\nonumber\\
={}&
\frac{\ri}{2}\,\mu^{4-D} \int\!\frac{\rd^Dp}{(2\pi)^D}\, \mathcal{T}_{h}(p) \,
\ln\left[\tilde{\Delta}_H(x,\partial_x+\ri p)\right],
\end{align}
which is the result given in \eq{L1loopeff}.

\addcontentsline{toc}{section}{References}
\bibliographystyle{jhep}
\bibliography{largeMH}

\providecommand{\href}[2]{#2}\begingroup\raggedright\begin{thebibliography}{10}

\bibitem{Buchmuller:1985jz}
W.~{Buchm\"uller} and D.~Wyler, {\it {Effective Lagrangian Analysis of New
  Interactions and Flavor Conservation}},  {\em Nucl. Phys. B} {\bf 268} (1986)
  621--653.

\bibitem{Grzadkowski:2010es}
B.~Grzadkowski, M.~Iskrzynski, M.~Misiak, and J.~Rosiek, {\it {Dimension-Six
  Terms in the Standard Model Lagrangian}},  {\em JHEP} {\bf 10} (2010) 085,
  [\href{http://arXiv.org/abs/1008.4884}{{\tt arXiv:1008.4884}}].

\bibitem{Heinemeyer:2013tqa}
{\bf LHC Higgs Cross Section Working Group} Collaboration, J.~R. Andersen {\em
  et~al.}, {\it {Handbook of LHC Higgs Cross Sections: 3. Higgs Properties}},
  \href{http://arXiv.org/abs/1307.1347}{{\tt arXiv:1307.1347}}.

\bibitem{deFlorian:2016spz}
{\bf LHC Higgs Cross Section Working Group} Collaboration, D.~de~Florian {\em
  et~al.}, {\it {Handbook of LHC Higgs Cross Sections: 4. Deciphering the
  Nature of the Higgs Sector}},  \href{http://arXiv.org/abs/1610.07922}{{\tt
  arXiv:1610.07922}}.

\bibitem{Brivio:2017vri}
I.~Brivio and M.~Trott, {\it {The Standard Model as an Effective Field
  Theory}},  {\em Phys. Rept.} {\bf 793} (2019) 1--98,
  [\href{http://arXiv.org/abs/1706.08945}{{\tt arXiv:1706.08945}}].

\bibitem{David:2020pzt}
A.~David and G.~Passarino, {\it {Use and reuse of SMEFT}},
  \href{http://arXiv.org/abs/2009.00127}{{\tt arXiv:2009.00127}}.

\bibitem{Dawson:2020oco}
S.~Dawson, S.~Homiller, and S.~D. Lane, {\it {Putting standard model EFT fits
  to work}},  {\em Phys. Rev. D} {\bf 102} (2020), no.~5 055012,
  [\href{http://arXiv.org/abs/2007.01296}{{\tt arXiv:2007.01296}}].

\bibitem{Ellis:2020unq}
J.~Ellis, M.~Madigan, K.~Mimasu, V.~Sanz, and T.~You, {\it {Top, Higgs, Diboson
  and Electroweak Fit to the Standard Model Effective Field Theory}},
  \href{http://arXiv.org/abs/2012.02779}{{\tt arXiv:2012.02779}}.

\bibitem{Dedes:2019bew}
A.~Dedes, K.~Suxho, and L.~Trifyllis, {\it {The decay $h\to Z \gamma$ in the
  Standard-Model Effective Field Theory}},  {\em JHEP} {\bf 06} (2019) 115,
  [\href{http://arXiv.org/abs/1903.12046}{{\tt arXiv:1903.12046}}].

\bibitem{Cullen:2019nnr}
J.~M. Cullen, B.~D. Pecjak, and D.~J. Scott, {\it {NLO corrections to $h\to
  b\bar b$ decay in SMEFT}},  {\em JHEP} {\bf 08} (2019) 173,
  [\href{http://arXiv.org/abs/1904.06358}{{\tt arXiv:1904.06358}}].

\bibitem{Dawson:2019clf}
S.~Dawson and P.~P. Giardino, {\it {Electroweak and QCD corrections to $Z$ and
  $W$ pole observables in the standard model EFT}},  {\em Phys. Rev. D} {\bf
  101} (2020), no.~1 013001, [\href{http://arXiv.org/abs/1909.02000}{{\tt
  arXiv:1909.02000}}].

\bibitem{Baglio:2019uty}
J.~Baglio, S.~Dawson, and S.~Homiller, {\it {QCD corrections in Standard Model
  EFT fits to $WZ$ and $WW$ production}},  {\em Phys. Rev. D} {\bf 100} (2019),
  no.~11 113010, [\href{http://arXiv.org/abs/1909.11576}{{\tt
  arXiv:1909.11576}}].

\bibitem{Degrande:2020evl}
C.~Degrande, G.~Durieux, F.~Maltoni, K.~Mimasu, E.~Vryonidou, and C.~Zhang,
  {\it {Automated one-loop computations in the SMEFT}},
  \href{http://arXiv.org/abs/2008.11743}{{\tt arXiv:2008.11743}}.

\bibitem{Dittmaier:1995cr}
S.~Dittmaier and C.~Grosse-Knetter, {\it {Deriving nondecoupling effects of
  heavy fields from the path integral: A Heavy Higgs field in an SU(2) gauge
  theory}},  {\em Phys. Rev. D} {\bf 52} (1995) 7276--7293,
  [\href{http://arXiv.org/abs/hep-ph/9501285}{{\tt hep-ph/9501285}}].

\bibitem{Dittmaier:1995ee}
S.~Dittmaier and C.~Grosse-Knetter, {\it {Integrating out the standard Higgs
  field in the path integral}},  {\em Nucl. Phys. B} {\bf 459} (1996) 497--536,
  [\href{http://arXiv.org/abs/hep-ph/9505266}{{\tt hep-ph/9505266}}].

\bibitem{Gaillard:1985uh}
M.~Gaillard, {\it {The Effective One Loop Lagrangian With Derivative
  Couplings}},  {\em Nucl. Phys. B} {\bf 268} (1986) 669--692.

\bibitem{Chan:1986jq}
L.-H. Chan, {\it {Derivative Expansion for the One Loop Effective Actions With
  Internal Symmetry}},  {\em Phys. Rev. Lett.} {\bf 57} (1986) 1199.

\bibitem{Cheyette:1987qz}
O.~Cheyette, {\it {Effective Action for the Standard Model With Large Higgs
  Mass}},  {\em Nucl. Phys. B} {\bf 297} (1988) 183--204.

\bibitem{DeWitt:1967ub}
B.~S. DeWitt, {\it {Quantum Theory of Gravity. 2. The Manifestly Covariant
  Theory}},  {\em Phys. Rev.} {\bf 162} (1967) 1195--1239.

\bibitem{DeWitt:1980jv}
B.~S. DeWitt, {\it {A gauge invariant effective action}},  in {\em {Oxford
  Conference on Quantum Gravity}}, pp.~449--487, 7, 1980.

\bibitem{tHooft:1975uxh}
G.~'t~Hooft, {\it {The Background Field Method in Gauge Field Theories}},  in
  {\em {12th Annual Winter School of Theoretical Physics}}, pp.~345--369, 1,
  1975.

\bibitem{Boulware:1980av}
D.~G. Boulware, {\it {Gauge Dependence of the Effective Action}},  {\em Phys.
  Rev. D} {\bf 23} (1981) 389.

\bibitem{Abbott:1980hw}
L.~Abbott, {\it {The Background Field Method Beyond One Loop}},  {\em Nucl.
  Phys. B} {\bf 185} (1981) 189--203.

\bibitem{Denner:1994xt}
A.~Denner, G.~Weiglein, and S.~Dittmaier, {\it {Application of the background
  field method to the electroweak standard model}},  {\em Nucl. Phys. B} {\bf
  440} (1995) 95--128, [\href{http://arXiv.org/abs/hep-ph/9410338}{{\tt
  hep-ph/9410338}}].

\bibitem{Denner:2019vbn}
A.~Denner and S.~Dittmaier, {\it {Electroweak Radiative Corrections for
  Collider Physics}},  {\em Phys. Rept.} {\bf 864} (2020) 1--163,
  [\href{http://arXiv.org/abs/1912.06823}{{\tt arXiv:1912.06823}}].

\bibitem{Fuentes-Martin:2016uol}
J.~Fuentes-Martin, J.~Portoles, and P.~Ruiz-Femenia, {\it {Integrating out
  heavy particles with functional methods: a simplified framework}},  {\em
  JHEP} {\bf 09} (2016) 156, [\href{http://arXiv.org/abs/1607.02142}{{\tt
  arXiv:1607.02142}}].

\bibitem{Beneke:1997zp}
M.~Beneke and V.~A. Smirnov, {\it {Asymptotic expansion of Feynman integrals
  near threshold}},  {\em Nucl. Phys. B} {\bf 522} (1998) 321--344,
  [\href{http://arXiv.org/abs/hep-ph/9711391}{{\tt hep-ph/9711391}}].

\bibitem{Smirnov:2002pj}
V.~A. Smirnov, {\it {Applied asymptotic expansions in momenta and masses}},
  {\em Springer Tracts Mod. Phys.} {\bf 177} (2002) 1--262.

\bibitem{Zhang:2016pja}
Z.~Zhang, {\it {Covariant diagrams for one-loop matching}},  {\em JHEP} {\bf
  05} (2017) 152, [\href{http://arXiv.org/abs/1610.00710}{{\tt
  arXiv:1610.00710}}].

\bibitem{Henning:2014wua}
B.~Henning, X.~Lu, and H.~Murayama, {\it {How to use the Standard Model
  effective field theory}},  {\em JHEP} {\bf 01} (2016) 023,
  [\href{http://arXiv.org/abs/1412.1837}{{\tt arXiv:1412.1837}}].

\bibitem{Drozd:2015rsp}
A.~Drozd, J.~Ellis, J.~Quevillon, and T.~You, {\it {The Universal One-Loop
  Effective Action}},  {\em JHEP} {\bf 03} (2016) 180,
  [\href{http://arXiv.org/abs/1512.03003}{{\tt arXiv:1512.03003}}].

\bibitem{Henning:2016lyp}
B.~Henning, X.~Lu, and H.~Murayama, {\it {One-loop Matching and Running with
  Covariant Derivative Expansion}},  {\em JHEP} {\bf 01} (2018) 123,
  [\href{http://arXiv.org/abs/1604.01019}{{\tt arXiv:1604.01019}}].

\bibitem{Ellis:2016enq}
S.~A.~R. Ellis, J.~Quevillon, T.~You, and Z.~Zhang, {\it {Mixed
  heavy\textendash{}light matching in the Universal One-Loop Effective
  Action}},  {\em Phys. Lett. B} {\bf 762} (2016) 166--176,
  [\href{http://arXiv.org/abs/1604.02445}{{\tt arXiv:1604.02445}}].

\bibitem{Ellis:2017jns}
S.~A.~R. Ellis, J.~Quevillon, T.~You, and Z.~Zhang, {\it {Extending the
  Universal One-Loop Effective Action: Heavy-Light Coefficients}},  {\em JHEP}
  {\bf 08} (2017) 054, [\href{http://arXiv.org/abs/1706.07765}{{\tt
  arXiv:1706.07765}}].

\bibitem{Kramer:2019fwz}
M.~Kr\"amer, B.~Summ, and A.~Voigt, {\it {Completing the scalar and fermionic
  Universal One-Loop Effective Action}},  {\em JHEP} {\bf 01} (2020) 079,
  [\href{http://arXiv.org/abs/1908.04798}{{\tt arXiv:1908.04798}}].

\bibitem{Angelescu:2020yzf}
A.~Angelescu and P.~Huang, {\it {Integrating Out New Fermions at One Loop}},
  {\em JHEP} {\bf 01} (2021) 049, [\href{http://arXiv.org/abs/2006.16532}{{\tt
  arXiv:2006.16532}}].

\bibitem{Ellis:2020ivx}
S.~A.~R. Ellis, J.~Quevillon, P.~N.~H. Vuong, T.~You, and Z.~Zhang, {\it {The
  Fermionic Universal One-Loop Effective Action}},  {\em JHEP} {\bf 11} (2020)
  078, [\href{http://arXiv.org/abs/2006.16260}{{\tt arXiv:2006.16260}}].

\bibitem{Criado:2017khh}
J.~C. Criado, {\it {MatchingTools: a Python library for symbolic effective
  field theory calculations}},  {\em Comput. Phys. Commun.} {\bf 227} (2018)
  42--50, [\href{http://arXiv.org/abs/1710.06445}{{\tt arXiv:1710.06445}}].

\bibitem{Cohen:2020fcu}
T.~Cohen, X.~Lu, and Z.~Zhang, {\it {Functional Prescription for EFT
  Matching}},  \href{http://arXiv.org/abs/2011.02484}{{\tt arXiv:2011.02484}}.

\bibitem{Cohen:2020qvb}
T.~Cohen, X.~Lu, and Z.~Zhang, {\it {STrEAMlining EFT Matching}},
  \href{http://arXiv.org/abs/2012.07851}{{\tt arXiv:2012.07851}}.

\bibitem{Fuentes-Martin:2020udw}
J.~Fuentes-Martin, M.~K\"onig, J.~Pag\`es, A.~E. Thomsen, and F.~Wilsch, {\it
  {SuperTracer: A Calculator of Functional Supertraces for One-Loop EFT
  Matching}},  \href{http://arXiv.org/abs/2012.08506}{{\tt arXiv:2012.08506}}.

\bibitem{Boggia:2016asg}
M.~Boggia, R.~Gomez-Ambrosio, and G.~Passarino, {\it {Low energy behaviour of
  standard model extensions}},  {\em JHEP} {\bf 05} (2016) 162,
  [\href{http://arXiv.org/abs/1603.03660}{{\tt arXiv:1603.03660}}].

\bibitem{Schabinger:2005ei}
R.~M. Schabinger and J.~D. Wells, {\it {A Minimal spontaneously broken hidden
  sector and its impact on Higgs boson physics at the large hadron collider}},
  {\em Phys. Rev. D} {\bf 72} (2005) 093007,
  [\href{http://arXiv.org/abs/hep-ph/0509209}{{\tt hep-ph/0509209}}].

\bibitem{Patt:2006fw}
B.~Patt and F.~Wilczek, {\it {Higgs-field portal into hidden sectors}},
  \href{http://arXiv.org/abs/hep-ph/0605188}{{\tt hep-ph/0605188}}.

\bibitem{Bowen:2007ia}
M.~Bowen, Y.~Cui, and J.~D. Wells, {\it {Narrow trans-TeV Higgs bosons and H
  $\to$ hh decays: Two LHC search paths for a hidden sector Higgs boson}},
  {\em JHEP} {\bf 03} (2007) 036,
  [\href{http://arXiv.org/abs/hep-ph/0701035}{{\tt hep-ph/0701035}}].

\bibitem{Pruna:2013bma}
G.~M. Pruna and T.~Robens, {\it {Higgs singlet extension parameter space in the
  light of the LHC discovery}},  {\em Phys. Rev. D} {\bf 88} (2013), no.~11
  115012, [\href{http://arXiv.org/abs/1303.1150}{{\tt arXiv:1303.1150}}].

\bibitem{Kanemura:2015fra}
S.~Kanemura, M.~Kikuchi, and K.~Yagyu, {\it {Radiative corrections to the Higgs
  boson couplings in the model with an additional real singlet scalar field}},
  {\em Nucl. Phys. B} {\bf 907} (2016) 286--322,
  [\href{http://arXiv.org/abs/1511.06211}{{\tt arXiv:1511.06211}}].

\bibitem{Bojarski:2015kra}
F.~Bojarski, G.~Chalons, D.~Lopez-Val, and T.~Robens, {\it {Heavy to light
  Higgs boson decays at NLO in the Singlet Extension of the Standard Model}},
  {\em JHEP} {\bf 02} (2016) 147, [\href{http://arXiv.org/abs/1511.08120}{{\tt
  arXiv:1511.08120}}].

\bibitem{Altenkamp:2018bcs}
L.~Altenkamp, M.~Boggia, and S.~Dittmaier, {\it {Precision calculations for $h
  \to WW/ZZ \to 4$ fermions in a Singlet Extension of the Standard Model with
  Prophecy4f}},  {\em JHEP} {\bf 04} (2018) 062,
  [\href{http://arXiv.org/abs/1801.07291}{{\tt arXiv:1801.07291}}].

\bibitem{Denner:2018opp}
A.~Denner, S.~Dittmaier, and J.-N. Lang, {\it {Renormalization of mixing
  angles}},  {\em JHEP} {\bf 11} (2018) 104,
  [\href{http://arXiv.org/abs/1808.03466}{{\tt arXiv:1808.03466}}].

\bibitem{Fleischer:1980ub}
J.~Fleischer and F.~Jegerlehner, {\it {Radiative Corrections to Higgs Decays in
  the Extended Weinberg-Salam Model}},  {\em Phys. Rev. D} {\bf 23} (1981)
  2001--2026.

\bibitem{Krause:2016oke}
M.~Krause, R.~Lorenz, M.~Margarete, R.~Santos, and H.~Ziesche, {\it
  {Gauge-independent Renormalization of the 2-Higgs-Doublet Model}},  {\em
  JHEP} {\bf 09} (2016) 143, [\href{http://arXiv.org/abs/1605.04853}{{\tt
  arXiv:1605.04853}}].

\bibitem{Denner:2016etu}
A.~Denner, L.~Jenniches, J.-N. Lang, and C.~Sturm, {\it {Gauge-independent
  $\overline{MS}$ renormalization in the 2HDM}},  {\em JHEP} {\bf 09} (2016)
  115, [\href{http://arXiv.org/abs/1607.07352}{{\tt arXiv:1607.07352}}].

\bibitem{Denner:2017vms}
A.~Denner, J.-N. Lang, and S.~Uccirati, {\it {NLO electroweak corrections in
  extended Higgs Sectors with RECOLA2}},  {\em JHEP} {\bf 07} (2017) 087,
  [\href{http://arXiv.org/abs/1705.06053}{{\tt arXiv:1705.06053}}].

\bibitem{deBlas:2014mba}
J.~de~Blas, M.~Chala, M.~Perez-Victoria, and J.~Santiago, {\it {Observable
  Effects of General New Scalar Particles}},  {\em JHEP} {\bf 04} (2015) 078,
  [\href{http://arXiv.org/abs/1412.8480}{{\tt arXiv:1412.8480}}].

\bibitem{Gorbahn:2015gxa}
M.~Gorbahn, J.~M. No, and V.~Sanz, {\it {Benchmarks for Higgs Effective Theory:
  Extended Higgs Sectors}},  {\em JHEP} {\bf 10} (2015) 036,
  [\href{http://arXiv.org/abs/1502.07352}{{\tt arXiv:1502.07352}}].

\bibitem{Chiang:2015ura}
C.-W. Chiang and R.~Huo, {\it {Standard Model Effective Field Theory:
  Integrating out a Generic Scalar}},  {\em JHEP} {\bf 09} (2015) 152,
  [\href{http://arXiv.org/abs/1505.06334}{{\tt arXiv:1505.06334}}].

\bibitem{Brehmer:2015rna}
J.~Brehmer, A.~Freitas, D.~Lopez-Val, and T.~Plehn, {\it {Pushing Higgs
  Effective Theory to its Limits}},  {\em Phys. Rev. D} {\bf 93} (2016), no.~7
  075014, [\href{http://arXiv.org/abs/1510.03443}{{\tt arXiv:1510.03443}}].

\bibitem{Egana-Ugrinovic:2015vgy}
D.~Egana-Ugrinovic and S.~Thomas, {\it {Effective Theory of Higgs Sector Vacuum
  States}},  \href{http://arXiv.org/abs/1512.00144}{{\tt arXiv:1512.00144}}.

\bibitem{Buchalla:2016bse}
G.~Buchalla, O.~Cata, A.~Celis, and C.~Krause, {\it {Standard Model Extended by
  a Heavy Singlet: Linear vs. Nonlinear EFT}},  {\em Nucl. Phys. B} {\bf 917}
  (2017) 209--233, [\href{http://arXiv.org/abs/1608.03564}{{\tt
  arXiv:1608.03564}}].

\bibitem{Jiang:2018pbd}
M.~Jiang, N.~Craig, Y.-Y. Li, and D.~Sutherland, {\it {Complete One-Loop
  Matching for a Singlet Scalar in the Standard Model EFT}},  {\em JHEP} {\bf
  02} (2019) 031, [\href{http://arXiv.org/abs/1811.08878}{{\tt
  arXiv:1811.08878}}].

\bibitem{Haisch:2020ahr}
U.~Haisch, M.~Ruhdorfer, E.~Salvioni, E.~Venturini, and A.~Weiler, {\it
  {Singlet night in Feynman-ville: one-loop matching of a real scalar}},  {\em
  JHEP} {\bf 04} (2020) 164, [\href{http://arXiv.org/abs/2003.05936}{{\tt
  arXiv:2003.05936}}]. [Erratum: JHEP 07, 066 (2020)].

\bibitem{Dawson:2021jcl}
S.~Dawson, P.~P. Giardino, and S.~Homiller, {\it {Uncovering the High Scale
  Higgs Singlet Model}},  \href{http://arXiv.org/abs/2102.02823}{{\tt
  arXiv:2102.02823}}.

\bibitem{Bilenky:1993bt}
M.~S. Bilenky and A.~Santamaria, {\it {One loop effective Lagrangian for a
  standard model with a heavy charged scalar singlet}},  {\em Nucl. Phys. B}
  {\bf 420} (1994) 47--93, [\href{http://arXiv.org/abs/hep-ph/9310302}{{\tt
  hep-ph/9310302}}].

\bibitem{delAguila:2016zcb}
F.~del Aguila, Z.~Kunszt, and J.~Santiago, {\it {One-loop effective lagrangians
  after matching}},  {\em Eur. Phys. J. C} {\bf 76} (2016), no.~5 244,
  [\href{http://arXiv.org/abs/1602.00126}{{\tt arXiv:1602.00126}}].

\bibitem{Stueckelberg:1938zz}
E.~Stueckelberg, {\it {Interaction forces in electrodynamics and in the field
  theory of nuclear forces}},  {\em Helv. Phys. Acta} {\bf 11} (1938) 299--328.

\bibitem{Stueckelberg:1957zz}
E.~Stueckelberg, {\it {Theory of the radiation of photons of small arbitrary
  mass}},  {\em Helv. Phys. Acta} {\bf 30} (1957) 209--215.

\bibitem{Kunimasa:1967zza}
T.~Kunimasa and T.~Goto, {\it {Generalization of the Stueckelberg Formalism to
  the Massive Yang-Mills Field}},  {\em Prog. Theor. Phys.} {\bf 37} (1967)
  452--464.

\bibitem{Lee:1972yfa}
B.~Lee and J.~Zinn-Justin, {\it {Spontaneously Broken Gauge Symmetries Part 3:
  Equivalence}},  {\em Phys. Rev. D} {\bf 5} (1972) 3155--3160.

\bibitem{Denner:1991kt}
A.~Denner, {\it {Techniques for calculation of electroweak radiative
  corrections at the one loop level and results for W physics at LEP-200}},
  {\em Fortsch. Phys.} {\bf 41} (1993) 307--420,
  [\href{http://arXiv.org/abs/0709.1075}{{\tt arXiv:0709.1075}}].

\bibitem{Grojean:2013kd}
C.~Grojean, E.~E. Jenkins, A.~V. Manohar, and M.~Trott, {\it {Renormalization
  Group Scaling of Higgs Operators and $\Gamma(h \to\gamma\gamma)$}},  {\em
  JHEP} {\bf 04} (2013) 016, [\href{http://arXiv.org/abs/1301.2588}{{\tt
  arXiv:1301.2588}}].

\bibitem{Elias-Miro:2013gya}
J.~Elias-Mir\'o, J.~R. Espinosa, E.~Masso, and A.~Pomarol, {\it
  {Renormalization of dimension-six operators relevant for the Higgs decays
  $h\rightarrow \gamma\gamma,\gamma Z$}},  {\em JHEP} {\bf 08} (2013) 033,
  [\href{http://arXiv.org/abs/1302.5661}{{\tt arXiv:1302.5661}}].

\bibitem{Elias-Miro:2013mua}
J.~Elias-Miro, J.~R. Espinosa, E.~Masso, and A.~Pomarol, {\it {Higgs windows to
  new physics through d=6 operators: constraints and one-loop anomalous
  dimensions}},  {\em JHEP} {\bf 11} (2013) 066,
  [\href{http://arXiv.org/abs/1308.1879}{{\tt arXiv:1308.1879}}].

\bibitem{Jenkins:2013zja}
E.~E. Jenkins, A.~V. Manohar, and M.~Trott, {\it {Renormalization Group
  Evolution of the Standard Model Dimension Six Operators I: Formalism and
  lambda Dependence}},  {\em JHEP} {\bf 10} (2013) 087,
  [\href{http://arXiv.org/abs/1308.2627}{{\tt arXiv:1308.2627}}].

\bibitem{Jenkins:2013wua}
E.~E. Jenkins, A.~V. Manohar, and M.~Trott, {\it {Renormalization Group
  Evolution of the Standard Model Dimension Six Operators II: Yukawa
  Dependence}},  {\em JHEP} {\bf 01} (2014) 035,
  [\href{http://arXiv.org/abs/1310.4838}{{\tt arXiv:1310.4838}}].

\bibitem{Alonso:2013hga}
R.~Alonso, E.~E. Jenkins, A.~V. Manohar, and M.~Trott, {\it {Renormalization
  Group Evolution of the Standard Model Dimension Six Operators III: Gauge
  Coupling Dependence and Phenomenology}},  {\em JHEP} {\bf 04} (2014) 159,
  [\href{http://arXiv.org/abs/1312.2014}{{\tt arXiv:1312.2014}}].

\bibitem{Alonso:2014zka}
R.~Alonso, H.-M. Chang, E.~E. Jenkins, A.~V. Manohar, and B.~Shotwell, {\it
  {Renormalization group evolution of dimension-six baryon number violating
  operators}},  {\em Phys. Lett. B} {\bf 734} (2014) 302--307,
  [\href{http://arXiv.org/abs/1405.0486}{{\tt arXiv:1405.0486}}].

\bibitem{Altenkamp:2017ldc}
L.~Altenkamp, S.~Dittmaier, and H.~Rzehak, {\it {Renormalization schemes for
  the Two-Higgs-Doublet Model and applications to $h \to WW/ZZ \to 4$
  fermions}},  {\em JHEP} {\bf 09} (2017) 134,
  [\href{http://arXiv.org/abs/1704.02645}{{\tt arXiv:1704.02645}}].

\bibitem{Altenkamp:2017kxk}
L.~Altenkamp, S.~Dittmaier, and H.~Rzehak, {\it {Precision calculations for $h
  \to WW/ZZ \to 4$ fermions in the Two-Higgs-Doublet Model with Prophecy4f}},
  {\em JHEP} {\bf 03} (2018) 110, [\href{http://arXiv.org/abs/1710.07598}{{\tt
  arXiv:1710.07598}}].

\end{thebibliography}\endgroup

\end{document}